%% file: paper.tex
\documentclass[12pt,preprint]{aastex}
\usepackage{lscape,pdflscape,longtable}
\pdfoutput=1
\newcommand{\conv}{{\ast}}
\newcommand{\ie}{{\em i.e.}}
\newcommand{\uHz}{\mbox{$\mu$Hz}}

\newcommand{\Deg}{^{\rm o}}

\newcounter{mycntr}
\shorttitle{High-Degree Mode Frequencies}
\shortauthors{Korzennik et al.}      
\begin{document}
\title{Accurate Characterization of High-Degree Modes\\ Using MDI Observations}
\author{S.~G.~Korzennik}
\affil{Harvard-Smithsonian Center for Astrophysics, Cambridge, MA 02138, USA}
\author{M.~C.~Rabello-Soares%
\thanks{Now at Physics Department, Universidade Federal de Minas Gerais, Minas Gerais, Brazil}}
\affil{W.~W.~Hansen Experimental Physics Laboratory, Stanford University, 
Stanford, CA 94305-4085, USA}
\author{J.~Schou}
\affil{W.~W.~Hansen Experimental Physics Laboratory, Stanford University, 
Stanford, CA 94305-4085, USA}
\and
\author{T.~P.~Larson}
\affil{W.~W.~Hansen Experimental Physics Laboratory, Stanford University, 
Stanford, CA 94305-4085, USA}
\begin{abstract}
  We present the first accurate characterization of high-degree modes, derived
using the best MDI full-disk full-resolution data set available.  A ninety day
long time series of full-disk two arc-second per pixel resolution dopplergrams
was acquired in 2001, thanks to the high rate telemetry provided by the Deep
Space Network.
  These dopplergrams were spatially decomposed using our best estimate of the
image scale and the known components of MDI's image distortion. A multi-taper
power spectrum estimator was used to generate power spectra for all degrees
and all azimuthal orders, up to $\ell=1000$. We used a large number of tapers
to reduce the realization noise, since at high degrees the individual modes
blend into ridges and thus there is no reason to preserve a high spectral
resolution.
  These power spectra were fitted for all degrees and all azimuthal orders,
between $\ell=100$ and $\ell=1000$, and for all the orders with substantial
amplitude. This fitting generated in excess of $5.2\times10^6$ individual
estimates of ridge frequencies, line-widths, amplitudes and asymmetries
(singlets), corresponding to some 5,700 multiplets ($\ell, n$).
  Fitting at high degrees generates ridge characteristics, characteristics
that do not correspond to the underlying mode characteristics. We used a
sophisticated forward modeling to recover the best possible estimate of the
underlying mode characteristics (mode frequencies, as well as line-widths,
amplitudes and asymmetries).  We describe in detail this modeling and its
validation. The modeling has been extensively reviewed and refined, by
including an iterative process to improve its input parameters to better match
the observations. Also, the contribution of the leakage matrix on the accuracy
of the procedure has been carefully assessed.
  We present the derived set of corrected mode characteristics, that includes
not only frequencies, but line widths, asymmetries and amplitudes. We present
and discuss their uncertainties and the precision of the ridge to mode
correction schemes, through a detailed assessment of the sensitivity of the
model to its input set. The precision of the ridge to mode correction is
indicative of any possible residual systematic biases in the inferred mode
characteristics.
   In our conclusions, we address how to further improve these estimates, and
the implications for other data sets, like GONG+ and HMI.
\end{abstract}
\section{Introduction}

  We have long argued that the inclusion of accurate high-degree modes (\ie,
$300\le\ell\le1000$) has the potential to improve dramatically inferences of
the solar stratification and its dynamics in the outermost 2 to 3\% of the
Sun.  \citet{Rabello-Soares:2000} showed how this can be carried out for the
sound speed and the adiabatic exponent, $\gamma_1$, since the high-$\ell$
modes probe the second and first helium ionization zones. They showed how well
the inclusion of high degree modes helps constrain
$\gamma_1$. \citet{Korzennik+Eff-Darwich:1999} have shown how the inference
of the solar rotation very close to the surface (outer 1\%) can be
dramatically improved by including unbiased high degree rotational
splittings.

  Unfortunately, the determination of the characteristics of the high-degree
modes (\ie, their frequency, line-width, asymmetry and amplitude), remained,
over the past two decades, a challenging task \citep{ Korzennik:1990,
KorzennikEtal:1990, RhodesEtal:1991a, RhodesEtal:1991b, KorzennikEtal:1993,
RhodesEtal:1995, ThompsonEtal:1996, HillEtal:1996, Korzennik:1998,
RhodesEtal:1998, Korzennik+Eff-Darwich:1999, Rabello-SoaresEtal:2001,
RhodesEtal:2001a, RhodesEtal:2001b, ReiterEtal:2002a, ReiterEtal:2002b,
RhodesEtal:2002, KorzennikEtal:2003, ReiterEtal:2003, RhodesEtal:2003,
KorzennikEtal:2004, ReiterEtal:2004, Rabello-SoaresEtal:2006a,
Rabello-SoaresEtal:2006b, KorzennikEtal:2008, Rabello-SoaresEtal:2008a,
Rabello-Soares+Korzennik:2009}.
Indeed, at high degrees, individual modes blend into ridges, causing the
individual mode characteristics to become masked by the ridge. For example, as
we have shown back in \citet{Korzennik:1990}, the ridge frequency is not the
target mode frequency, but in fact corresponds to a value offset by a
significant amount. This frequency offset varies with $n, \ell$ and $m$, and
is determined by the precise contributions from all the modes that blend into
the ridge.

  In global helioseismic data analysis, the individual resonant modes are
isolated as follows, and then fitted. The angular components are separated by
performing a spatial decomposition on each image, projecting the solar surface
onto spherical harmonics. The resulting time series of spherical harmonic
coefficients at a given target $(\ell, m)$ are Fourier transformed,
leading to the separation of the orders of the radial wave function, $n$, in
the frequency domain.
  However, a spherical harmonic decomposition is not orthogonal over a
hemisphere, or, for that matter, the visible solar surface from a single
vantage point of view. This results in what is called spatial leakage, namely,
modes with similar degrees ($\ell'\sim\ell$) and azimuthal orders ($m'\sim
m$), leak into the estimate of a target mode spherical harmonic coefficient at
a given ($\ell,m$).

At low and intermediate degrees, most of these leaks can be separated from the
target mode in the frequency domain and individual modes are resolved and
fitted%
\footnote{In some cases the closest leak, \ie, $\delta m = 2$, $\delta \ell =
0$, depending on the mode FWHM and the spectral resolution, blends with the
target mode, but most fitting methods account for this $m$-blend.}.
However, at high degrees, the spatial leaks lie closer in frequency, due to a
smaller frequency separation, and the modes become wider, as their lifetimes
get smaller. The combination of these two effects results in substantial
overlap of the target mode with its spatial leaks and eventually all the
spatial leaks blend into a ridge. This blending occurs for $\ell\ge300$ for
the f-mode, and for $\ell\ge200$ for the p-modes (\ie, $p_1$ to $p_4$, and at
even lower degrees for the higher orders).
Once modes have blended into ridges, one needs a very good model of the
relative amplitude of all the modes that contribute to the ridge power
distribution to recover the actual mode characteristics. Key elements for this
model are the leakage matrix coefficients and a very good knowledge of the
instrumental properties.

  The high degree mode characteristics presented in this paper are based on
ridge fitting, for degrees up to $\ell=1000$, using one of the longest
available full-disk observations acquired by the Michelson Doppler Imager
(MDI), an instrument on board the Solar and Heliospheric Observatory
(SOHO). The data set we have analyzed is described in
Section~\ref{sec:TheDataSet} as well as the data reduction procedure and the
spectral estimator we used.

  In Section~\ref{sec:TheMethodologyFit}, we describe the fitting procedure we
used to derive ridge characteristics at high degrees.
Section~\ref{sec:TheMethodologyModel} describes the methodology we implemented
to recover the mode characteristics from the fitted ridge values. It
essentially consists of building a sophisticated model of the underlying modes
that contribute to the ridge power distribution. This model generates and fits
a synthetic ridge, producing values that are used to derive a measure of the
bias between the resulting ridge properties and the underlying target mode
used in the modeling.

  Section~\ref{sec:Results} presents the resulting best estimate to date of
high-degree mode characteristics. It includes estimates of the accuracy of the
correction, derived from a comprehensive error budget of the methodology.
Finally, in Section~\ref{sec:Conclusions}, we present our conclusions, how to
further improve our estimates, and any implications for other data sets, like
GONG+ and HMI.

\section{The Data Set}\label{sec:TheDataSet}

  The MDI instrument was launched on December 2nd, 1995 on board the SOHO
spacecraft. The spacecraft, in orbit around the L1 Lagrangian point, is in
constant view of the Sun, providing a near perfect platform for uninterrupted
observations of the solar surface \citep{ScherrerEtal:1995}. While MDI took
high resolution full-disk images nearly all the time, the limited telemetry of
the spacecraft resulted in transmitting to the ground only binned down data,
(from $1024\times1024$ down to $200\times200$ pixels), to fit the data stream
within the telemetry limits. Nevertheless, during limited time periods, but
nearly each year, NASA's Deep Space Network was commissioned to provide
additional telemetry band-width to download unbinned images.  The epochs when
this occurred over the MDI mission, known as Dynamics runs, are listed in
Table~\ref{tab:FullDiskEpochs}.

\input{tableFullDiskEpochs}

   The optimal epoch when full resolution full-disk Dopplergrams are available
was acquired during 2001, with nearly 90 days of continuous observations and a
high duty cycle. In order to derive accurate estimates, using high SNR power
spectra, we focused our efforts on analyzing the 2001 {\em Dynamics} epoch,
combining them both into one time series.
 
\subsection{Data Analysis}\label{sec:DataAnalysis}

  The MDI Dopplergrams analyzed for the study presented here were spatially
decomposed onto spherical harmonics, using our best knowledge of the
instrumental image distortion. \citet{KorzennikEtal:2004} have described in
minute detail the two components of this distortion. One component results
from the characteristics of the MDI instrument optical package itself. It was
estimated from the instrument design optical properties using ray tracing
({\tt ZEMAX}) and validated on the actual in-flight data. The other component
had to be introduced to replicate the ellipticity of the solar limb observed
in the MDI images, an ellipticity much larger than that of the solar
limb. This ellipticity might be the result of a small tilt of the CCD with
respect to the focal plane\footnote{The camera package of MDI was reassembled
shortly before launch to alleviate problems with the CCD package. This
effectively invalidated a lot of the pre-flight tests carried out on the
optical package and is likely to be the source of the small tilt of the CCD
with respect to the focal plane.}. This tilt was estimated to be $\sim
2.6\Deg$ \citep[][and references therein]{KorzennikEtal:2004,
Rabello-SoaresEtal:2008a}.  The improved spatial decomposition we used
incorporates both components of the instrumental image distortion.
  The resulting time series of spherical harmonic coefficients were then gap
filled using a maximum entropy method adapted by Rasmus Munk Larsen and based
on \citet{Fahlman+Ulrych:1982}. The resulting 259,200 minute long time series
have a 97.0\% fill factor (95.8\% prior to gap filling).

  Since the ridges at high degrees have large widths, there is no need for
high spectral resolution. Therefore, we opted to reduce the realization noise
by using a high order sine multi-taper power spectrum estimator. We selected a
61 terms multi-taper, that corresponds to a 90 day long time series to an
effective spectral resolution of 7.8\uHz. This value is approximately the mode
FWHM at $\ell = 500$ and $\ell = 300$ for the $n=0$ and $n=1$ modes,
respectively. But at these degrees the modes blend into ridges, so more
relevant is the ridge FWHM, that is about 14 \uHz\ at $\ell=300$ for the f-mode
and about 17 \uHz\ for $n=1$ at $\ell=200$.
Using this high order sine multi-taper spectral estimator causes some modes at
lower degrees to be substantially wider, while still resolving the ridge
itself.  This widening has the beneficial effect of blending resolved, or
partially resolved, modes into ridges at intermediate degrees ($100 \le\ell\le
300$). This blending results in a range of degrees where we can compare ridge
to mode characteristics, while it eliminates the intermediate case where modes
are only partially resolved.

\begin{figure}[!t]
\centering
\includegraphics[width=.95\textwidth]{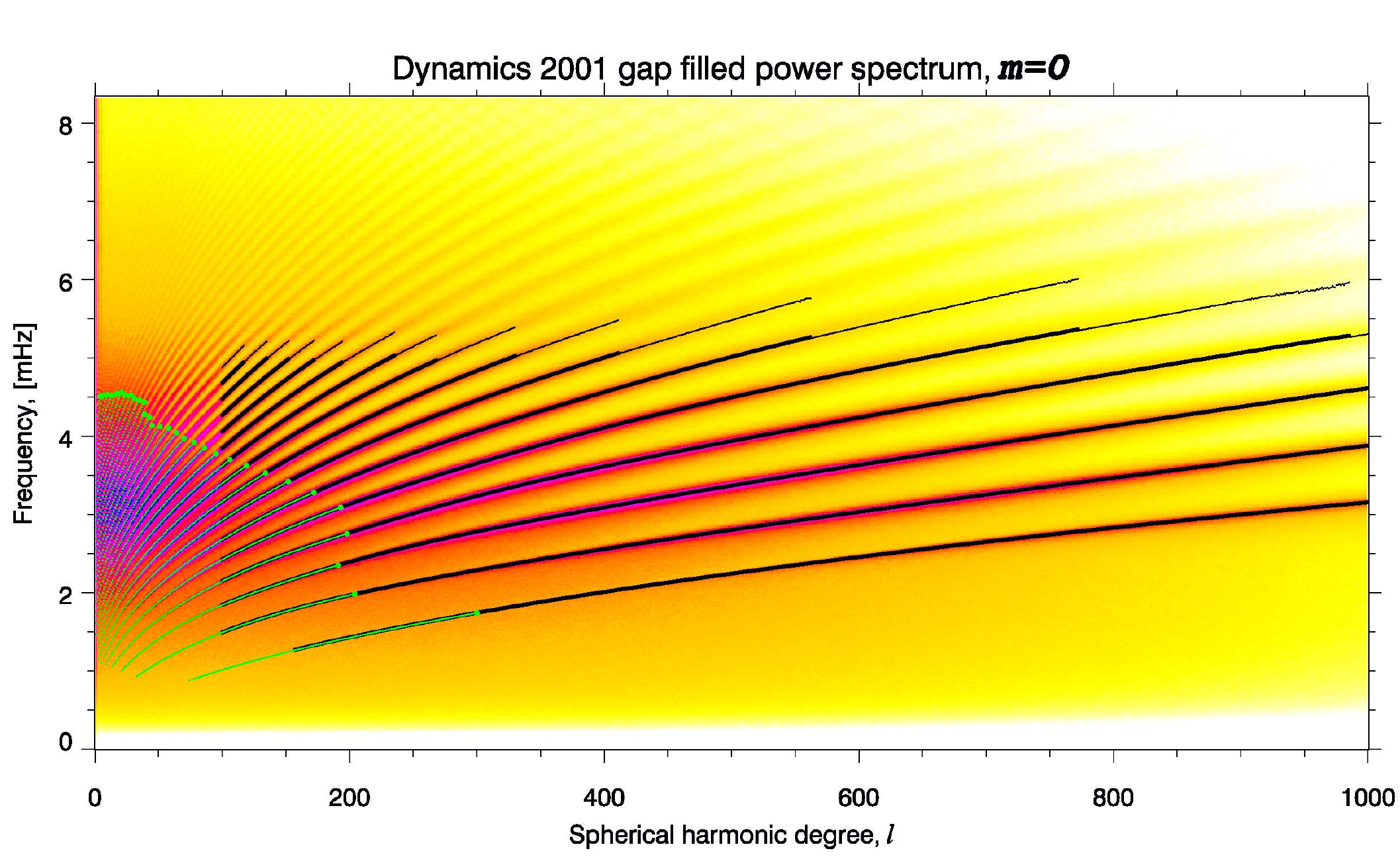}
\caption{Gap-filled {\em Dynamics} 2001 power spectrum, displayed on a
logarithmic scale, for the zonal modes ($m=0$).  The dots represent the fitted
ridges (black) or modes (green), the circles are the remaining fitted ridges
after rejecting the largest fitted order for each degree (see text for
explanation). The largest degree fitted using resolved mode fitting at low and
intermediate degrees is indicated, for each order, by a large green dot.
\label{fig:showPS}}
\end{figure}

  Indeed, the resulting blending of modes at intermediate degrees allows us to
fit ridges at degrees where resolved modes can and have been measured. This
allows us to test whether our methodology to derive mode characteristics from
ridge fitting is correct: can we recover the known mode characteristics from
the measured ridge ones?
  We were able to fit ridges down to $\ell=100$, resulting in an overlap
between mode and ridge fitting covering $100\le\ell\le300$ for the f-modes and
$100\le\ell\le200$ for the p-modes. Figure~\ref{fig:showPS} shows the {\em
Dynamics} 2001 power spectrum for the zonal modes ($m=0$) and the extent of
the ridge fitting.

\section{The Ridge Fitting Methodology}\label{sec:TheMethodologyFit}

  The ridge fitting methodology we used consists of fitting the power
spectra with a sum of modified Lorentzian plus a background term, namely:
\begin{equation}
  P_{\ell,m}(\nu) = \sum_n \tilde{A}_{n,\ell,m} \ 
        {\cal L}_a(\nu,\tilde{\nu}_{n,\ell,m},
                       \tilde{\Gamma}_{n,\ell,m},
                       \tilde{\alpha}_{n,\ell,m})
      + B_{\ell,m}(\nu)
\end{equation}
where ${\cal L}_a$, the modified asymmetric Lorentzian, is defined as (using
$k$ as a shorthand for $n,\ell,m$)
\begin{equation}
{\cal L}_a(\nu, \tilde{\nu}_{k}, \tilde{\Gamma}_{k},\tilde{\alpha}_{k}) =
    \frac{1+\tilde{\alpha}_k\,(\zeta_{k}-\frac{\tilde{\alpha}_k}{2})}
         {1+\zeta^2_{k}}
\label{eq:aLz}
\end{equation}
where $ \zeta_k = \zeta_{n,\ell,m}(\nu)$ is simply
\begin{equation} 
  \zeta_{n,\ell,m} =
  \frac{\nu-\tilde{\nu}_{n,\ell,m}}{\frac{\tilde{\Gamma}_{n,\ell,m}}{2}} 
\end{equation}
and where $B_{\ell,m}$, the background term, is defined as a polynomial
expansion in $\nu$
\begin{equation}
\log B_{\ell,m}(\nu) = \sum_{i=0}^{5} c^b_i(\ell,m)\, \eta_{\nu}^i
\end{equation}
with $\eta_{\nu} =  \frac{\nu-\nu_o}{\nu_s}$.
The parameters $\nu_o$ and $\nu_s$ were both set to 4 mHz, remapping the 0 to
8 mHz frequency range to the $[-1,1]$ interval. This formulation guarantees
the fitted background to be a positive definite quantity.
  The fitted parameters, $\tilde{\nu}_{n,\ell,m}, \tilde{\Gamma}_{n,\ell,m},
\tilde{\alpha}_{n,\ell,m}, \tilde{A}_{n,\ell,m}$, are the resulting ridge
characteristics, namely the ridge frequency, FWHM, asymmetry, and power
amplitude, complemented by the background coefficients, $c^b_i(\ell,m)$.

  The fitting was performed using a least-squares minimization, and carried
out in multiple steps, starting with some good initial guesses. At first, only
the amplitudes and the background coefficients were adjusted, using a
frequency range that covers all the fitted orders. Indeed, the background can
only be adequately constrained when fitting a wide frequency range, since
there is limited spectral range between the ridges.
Next, parameters for each individual order were readjusted, but this time over
a limited frequency range centered on the current estimate of the ridge
frequency for that order. For convergence stability, not all fitted parameters
were initially adjusted. We first adjusted only the amplitude and the FWHM,
then added the central frequency, and only then the asymmetry.
The whole procedure was repeated several times, readjusting the background
term and all the amplitudes and then fitting each order individually again.
The table of initial guesses was compiled by fitting only the zonal spectra,
smoothing the resulting values, either over $\ell$ or $\nu$, as appropriate,
and iterating.

  While one can {\em see} ridges up to the Nyquist frequency in
Fig~\ref{fig:showPS}, the amplitude (or contrast) of the high order ridges get
progressively smaller and smaller.  As a result, fitting the high order ridges
is poorly constrained and becomes numerically unstable. Therefore, the number
of orders that were fitted at each degree had to be limited by some objective
criteria. The criteria we adopted is to fit all the orders whose power
amplitude is at least 0.2\% of the largest observed amplitude at that degree
(\ie, down to a 1:500 ratio). Still, the properties of the highest fitted
order, the one with the smallest amplitude at a given degree, show systematic
effects that are easily understood in term of cross talk between the
background level and contributions from the remainder orders that are not
taken into account. We thus rejected {\em a posteriori} the highest order,
$n$, at each degree.

\begin{figure}[!t]
\centering
\includegraphics[width=.95\textwidth]{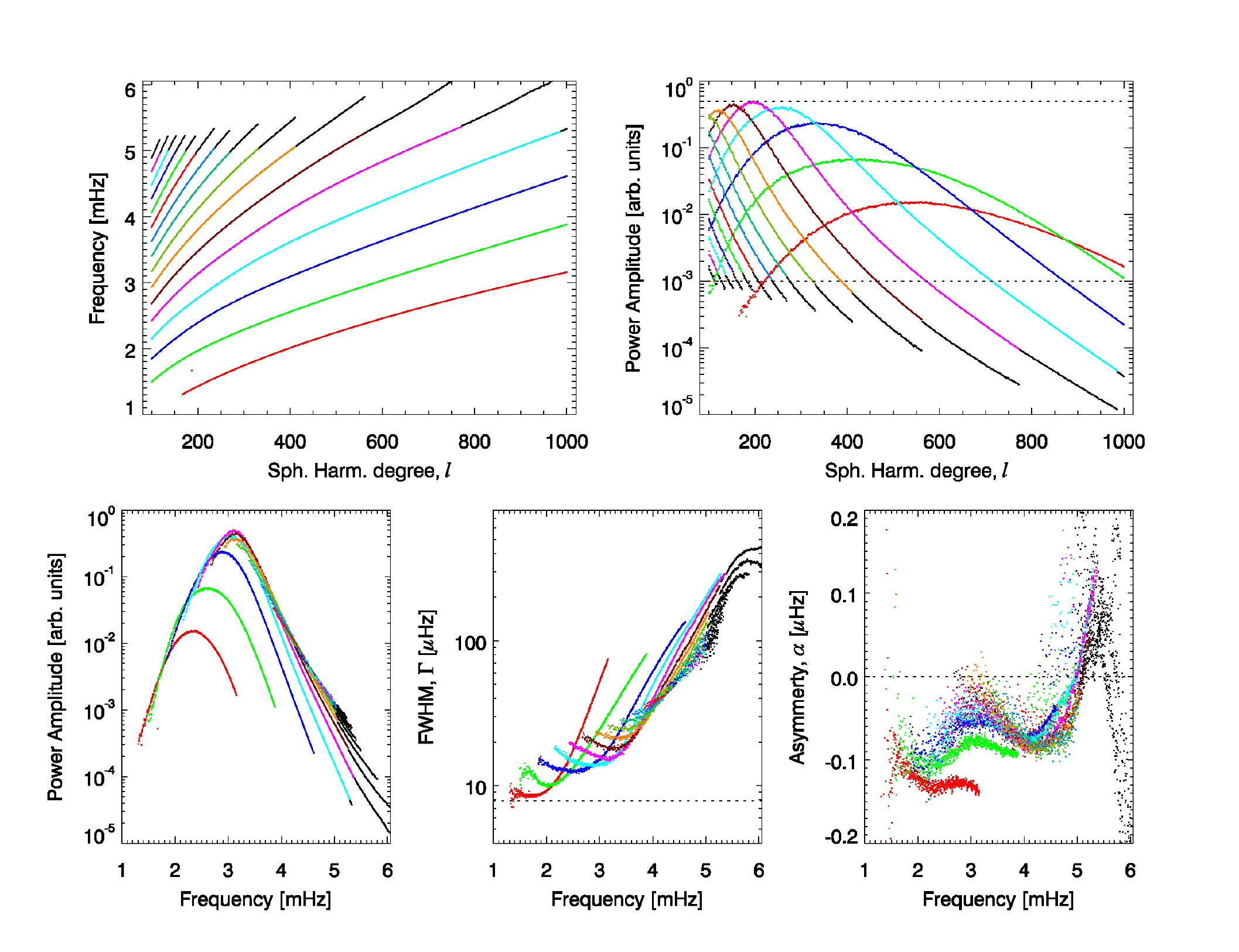}
\caption{Results from ridge fitting, namely ridge frequency, amplitude, FWHM
and asymmetry, once reduced to multiplets. Top left panel: coverage of the
fitting in the $\ell - \nu$ plane. The color corresponds to the value
of the order, $n$, while the black dots are the {\em a posteriori} rejected
values, corresponding to the largest order for each degree (see text for
explanation). The same color scheme is used in all the panels.  Top right
panel: power amplitude as a function of degree, the two horizontal lines are
in the 1:500 ratio. Lower left panel: power amplitude as a function of
frequency. Lower middle panel: FWHM as a function of frequency, the horizontal
dash line corresponds to $W_{N,T}$, the spectral resolution of the $N$th order
sine multi-taper ($N=61$ and $T=90$ days). Lower right panel: asymmetry as a
function of frequency.  Notice how the FWHM and asymmetry of the rejected
values (black dots) are indeed suspicious.
\label{fig:fittedModes}}
\end{figure}

  The fitting has been carried out for all degrees and all azimuthal orders
between $\ell=100$ and $1000$, producing in excess of five million individually
fitted ridges\footnote{In past work, we only fitted some 50 azimuthal orders
at every 10th degree.}. These correspond to 6,681 multiplets ($n,\ell$) that
were further reduced to 5,780 values after rejecting the highest fitted $n$
for each $\ell$.  Figure~\ref{fig:fittedModes} shows the coverage and
properties of the fitted values corresponding to ridge characteristics, once
reduced to multiplets.

  We should point out that the reduction from singlets to multiplets
consists not only of fitting Clebsch-Gordan coefficients to the
frequencies, but also of fitting polynomials in $m/\ell$ for the amplitude,
FWHM and asymmetry. This is explained and justified in the next sections.

\subsection{The Ridge to Mode Correction Methodology}\label{sec:TheMethodologyModel}

  The methodology we implemented to recover mode characteristics from ridge
fitting consists of modeling in detail the contribution of all the modes to a
given ridge. This model is used to generate a synthetic spectrum that is
fitted in a way similar to the data, and produces a correspondence between
ridge characteristics and mode characteristics. This method, initially
presented in \citet{KorzennikEtal:2004}, was further expanded, fine tuned and
refined as described in \citet{Rabello-SoaresEtal:2008a}.

  More recently, we revisited the methodology to not only recover the mode
frequency, but also the FWHM, asymmetry and amplitude. The codes for the
modeling and for the fitting were converted from a 4GL programming language
(IDL) to a compiled language (F90) to make efficient use of the Smithsonian
Institution High Performance Computing cluster, a machine with, to date, a
little over 2,600 compute cores. In the process, the modeling code was also
restructured so as to clearly identify all the input parameters used by the
methodology.

\subsubsection{The Modeling Methodology}\label{sec:modeling.methodology}

\begin{figure}[!t]
\centering
\includegraphics[width=.95\textwidth]{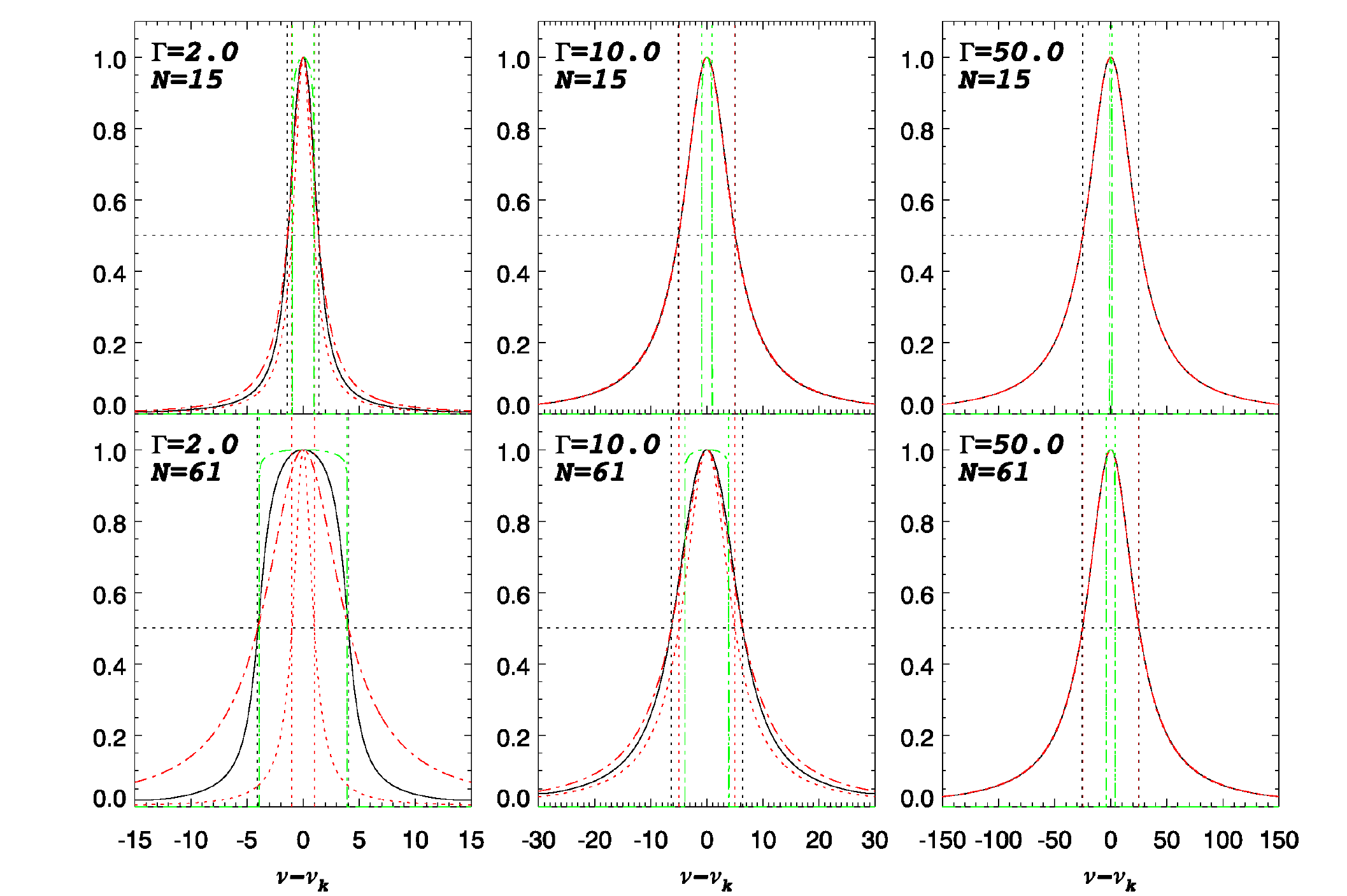}
\caption{Example of convolutions ${\cal L} \otimes {\cal X}$ (see text), for
two values of $N$ (top and bottom panels) and three values of $\Gamma$ (left
to right).  The black curves are the resulting convolutions, the dash red and
green curves represent ${\cal L}$ and ${\cal X}$, respectively, while the red
dot-dash curves correspond to the widened ${\cal L}$, using the effective
FWHM. The vertical lines indicate the corresponding FWHM.  Only for the large
$N$ and narrow peak example does the resulting convolution differ
substantially from a widened Lorentzian. This corresponds to the modeling of
low $\ell$ low $n$ ridges.
\label{fig:xsmt}}
\end{figure}

  The modeling method consists of producing synthetic spectra by overlapping
all the individual modes that leak into a simulated snippet of spectrum around
a given mode. Thus for each $\{n,\ell,m\}$, the model consists in computing
the power spectrum estimated for a limited frequency range, $w$, centered on
$\nu_{n,\ell,m}$, and given by the following superposition:
\begin{equation}
  P_{\ell,m}(\nu) = \left(
    \sum_{\ell'} \sum_{m'} Q_{\ell',m'}^{n,\ell,m}
      {\cal L}_a(\nu, \nu_{n,\ell',m'}, \Gamma_{n,\ell'}, \alpha_{n,\ell'})
    \right) \otimes {\cal X}(\nu, N, W_{T})
               + B_p^{\ell,m}
\label{eq:overlap}
\end{equation}
where $\otimes$ is the convolution operator, ${\cal X}(\nu, N, W_{T})$ is an
empirically derived function, $N$ the number of sine multi-tapers, $W_{T}=1/T$
the spectral resolution corresponding to a time series of length $T$, and $B_p$
represents the power background value.
The frequency range of the model, $w$, is given by
\begin{equation}
w = 15\, (\Gamma_{n,\ell} + \frac{\partial\nu}{\partial\ell}) 
\end{equation}
as to cover the full extent of the ridge.
The empirical convolution function, $\cal X$, is given by: 
\begin{equation}
{\cal X}(\nu, N, W_T) = \left\{
\begin{array}{ll}
  ( 1 - (\frac{2\,\nu}{W_{N,T}\,f_N})^2)^{\frac{1}{N}} 
   & \mbox{if $|\nu| \le \frac{W_{N,T}\,f_N}{2}$} \\
  0 & \mbox{otherwise} \\
\end{array}
\right.
\end{equation}
where $f_N$ is the factor needed to keep the FWHM of $\cal X$ to remain equal
to $W_{N,T} = N\,W_T$, the resolution of the $N$th order sine multi-taper.
Note that the FWHM of the convolution ${\cal L} \otimes {\cal X}$, the
effective FWHM, $\Gamma^E_{n,\ell}$, is
\begin{equation}
\Gamma^E_{n,\ell} = \sqrt{\Gamma^2_{n,\ell} + W_{N,T}^2}
\label{eq:gammaE}
\end{equation}
\ie, the mode FWHM widened, in quadrature, by the resolution of the $N$th
order sine multi-taper.
For cases where $\Gamma_{n,\ell} > W_{N,T}$, the result of the convolution is
nearly equivalent to an asymmetric Lorentzian with that effective FWHM. But
for cases where $\Gamma_{n,\ell} < W_{N,T}$ (narrow modes, the low $\ell$ and
low $n$ cases) the shape of ${\cal L} \otimes {\cal X}$ is no longer an
asymmetric Lorentzian, as illustrated in Fig.~\ref{fig:xsmt}. As a result, the
ridge produced by the superposition described by Eq.~\ref{eq:overlap}, for
intrinsically narrow modes, is narrower than the result of superposing widened
Lorentzians.

The mode profiles are represented by asymmetric Lorentzians, ${\cal L}_a$, as
defined in Eq.~\ref{eq:aLz} while the frequencies, $\nu_{n,\ell,m}$ and
$\nu_{n,\ell',m'}$, are estimated using
\begin{equation}
  \nu_{n,\ell,m}   = \nu_{n, \ell,m=0} +  \delta\nu_{n,\ell}(m) 
\end{equation}
and
\begin{equation}
 \nu_{n,\ell',m'} = \nu_{n, \ell',m=0} +  \delta\nu_{n,\ell'}(m')
 \label{eq:dnu_approx}
\end{equation}
and where
\begin{equation}
 \delta\nu_{n,\ell}(m) = \sum_{i=1}^6 a_i(n,\ell) P_i(\frac{m}{L})
 \label{eq:dnu_ai_approx}
\end{equation}

The amplitude of the leaked power, $Q$, is given by:
\begin{equation}
 Q_{\ell',m'}^{n,\ell,m} = (\tilde{C}_{\ell',m'}^{n,\ell,m})^2 R_p(\ell,\ell')
\end{equation}
where $\tilde{C}$ corresponds to the distorted spatial leak coefficients, the
sum of the radial and horizontal components:
\begin{equation}
 \tilde{C}_{\ell',m'}^{n,\ell,m} =  \tilde{C}_r(\ell',m';\ell,m)
                 + \beta_{n,\ell}\, \tilde{C}_h(\ell',m';\ell,m)
\end{equation}
with $\beta_{n,\ell}$ being the ratio of the horizontal to the vertical
displacement components.

  The power ratio term, $R_p$, is introduced to represent the power
attenuation with degree resulting from the instrumental PSF%
\footnote{We recognize that this is very much an empirical approach, but we
          would need a reliable model of the instrumental PSF to do it
          differently.}, and is parametrized by a
polynomial expansion as follows:
\begin{equation}
  \log\left[R(\ell,\ell')\right] = \sum_{k=0}^2 c^R_k(\ell) \, (\ell'-\ell)^k 
\end{equation}

  The distortion of the leaks by the differential rotation is given by
\begin{equation}
\tilde{C}_x(\ell',m';\ell,m) = 
   \sum_{\ell''} G^{\ell''}_{\ell',m'} C_x(\ell'',m';\ell,m)
\label{eq:mixing}
\end{equation}
for $x=r,h$, \ie, the radial and horizontal components, respectively.
The mixing coefficients, $G$, derived by \citet{Woodard:1989}, are:
\begin{equation}
    G^{\ell'}_{\ell,m} = G^{\ell'}_{\ell,m}(b_2, b_4, \frac{\partial\nu}{\partial\ell}) 
     = (-)^p \int_{-\pi}^{\pi} \cos(p\Theta + \delta \Phi(\Theta)) d\Theta
\end{equation}
where
\begin{eqnarray}
 p &=& \frac{\ell-\ell'}{2} \\
 \delta \Phi(\Theta) &=&  c_0 (c_1\sin(\Theta) + c_2\sin(2\Theta)) \\
 c_0 &=& -\frac{m}{2}\,\frac{1}{\frac{\partial\nu}{\partial\ell}} \\
 c_1 &=& -\frac{1}{2} (b_2\, y^2 + b_4\, y^4)              \\
 c_2 &=& \frac{1}{16} b_4\, y^4                          \\
 y^2 &=& {1 - (\frac{m}{L})^2}
\end{eqnarray}
and where $b_2$ and $b_4$ represent the solar differential rotation, when
parametrized as
\begin{equation}
\frac{\Omega(\theta)}{2\pi} = b_0 + b_2 \cos^2\theta + b_4 \cos^4\theta
\end{equation}
with $\theta$ being the co-latitude and $\Omega(\theta)$ the surface rotation
rate.

The undistorted leakage coefficients, $C$, are given by the following
integrals \citep[see][]{KorzennikEtal:2004}:
\begin{eqnarray}
   C_r({\ell, m; \ell', m'}) & = & \hspace*{1.15em}
   \oint W  Y^{m\conv}_{\ell}  Y^{m'}_{\ell'} 
        \sin\theta \cos\phi \, d\Omega
   \label{eq:theoleakr} \\
   C_{\theta}({\ell, m; \ell', m'}) & = &
  \!\!\!\! -\frac{1}{L} \oint W  Y^{m\conv}_{\ell} 
  \partial_{\theta}  Y^{m'}_{\ell'} \cos\theta \cos\phi \, d\Omega
   \label{eq:theoleakt} \\
   C_{\phi}({\ell, m; \ell', m'}) & = & 
     \frac{1}{L} \oint W  Y^{m\conv}_{\ell} 
   \partial_{\phi}  Y^{m'}_{\ell'} \frac{1}{\sin\theta} \sin\phi\, d\Omega
   \label{eq:theoleakp}
\end{eqnarray}
where $^{\conv}$ represents the complex conjugate operator, $Y^{m}_{\ell}$ the
spherical harmonic of degree $\ell$ and azimuthal order $m$, $\theta$ and
$\phi$ the co-latitude and longitude respectively, and $W(\theta,\phi)$ the
spatial window function of the observations -- \ie, the function that delimits
the angular span of the observations and that includes any additional
attenuation like the spatial apodization.  The two components of the leakage
matrix are the radial component $C_r$ and the horizontal components $C_h =
C_{\theta} + C_{\phi}$.
The complete list of the input parameters for our modeling is summarized in
Table~\ref{tab:inputParams}.

\begin{table}[!t]
\centering 
\begin{tabular}{||l|l||}
\hline
$\nu_{n,\ell}$    & mode frequency        \\
$\Gamma_{n,\ell}$ & mode FWHM             \\
$\alpha_{n,\ell}$ & mode asymmetry        \\
$B_p(\ell)$       & power background      \\
$a_i(n,\ell)$     & frequency splittings coefficients \\
$c^R_k(\ell)$     & coefficients defining the power ratio wrt $\ell$ \\
$b_2, b_4$         & parametrization of the surface differential rotation \\
$N_{\delta\ell'},N_{\delta m'}$  & extent of sum on $\ell'$ and $m'$  in Eq.~\ref{eq:overlap} \\
$N_{\delta\ell''}$ & extent of sum on $\ell''$  in Eq.~\ref{eq:mixing} \\
$C_r, C_h$         & radial and horizontal leakage matrix coefficients \\
\hline
$\beta_{n,\ell}$  & horizontal to vertical ratio, $\beta_{n,\ell} = (\frac{\nu_{n,\ell}}{\nu_{0,\ell}})^2$ \\
$\frac{\partial\nu}{\partial\ell}$ & tabulated values, derived from $\{\nu_{n,\ell}\}$ \\
\hline\hline
\end{tabular}
\caption{List of input parameters, and {\em secondary} input parameters
(bottom, whose values are derived from the primary parameters), for the ridge
modeling.\label{tab:inputParams}}
\end{table}

\subsubsection{Model Derivation}

  A model was run at first using some initial input parameters. We ran it
for every degree between $\ell=100$ and $\ell=1000$, and for 51 values of $m$,
spanning uniformly the $[-\ell,+\ell]$ interval.
That initial input set was derived using results from ridge fitting as well as
results from mode fitting at low and intermediate degrees, namely:
\begin{list}{--}{\setlength{\topsep}{0pt}\setlength{\itemsep}{0pt}}
 \item The values for the mode frequencies, $\nu_{n,\ell}$, were derived from
 ridge fitting results, after smoothing them using a bivariate polynomial in
 $n$ and $\ell$.
  \item The FWHM values, $\Gamma_{n,\ell}$, were derived from values obtained
  by fitting individual modes at low and intermediate degrees\footnote{We used
  the result from fitting a 12.5 year long time series of MDI observations
  (Korzennik, {\em in prep.}).}  combined with values derived from ridge
  fitting, corrected in a somewhat crude way for the mode to ridge widening.

  \item The asymmetry, $\alpha_{n,\ell}$, was set to be a bivariate polynomial
  in $\nu$ and $\ell$, based on ridge fitting results (since, as shown below,
  the ridge asymmetry is a good estimate of the mode asymmetry).
  \item The background values, $B_p(\ell)$, were set by a polynomial in
  $\ell$, derived from the fit to the zonal power spectra.
  \item The frequency splitting coefficients, $a_i(n,\ell)$, were determined
   from a polynomial in $\log(\nu/L)$, with only even non-zero coefficients.
   These coefficients were derived by fitting low and intermediate frequency
   splittings, estimated by fitting resolved modes.
  \item The power ratio coefficients, $c^R_k(\ell)$, were derived by fitting
  $\frac{\partial\log \tilde{A}}{\partial\ell}$, where the power amplitudes,
  $\tilde{A}$, correspond to measured zonal ridge power amplitudes, themselves
  smoothed using a bivariate polynomial in $\nu$ and $\ell$.
  \item The ratio $\beta_{n,\ell}$ was set to
  $(\nu_{n,\ell}/\nu_{n=0,\ell})^2$, after fitting a cubic polynomial in
  $\sqrt{L}$ to the f-mode frequencies derived from ridge fitting.
  \item We adopted $b_2=-69.6\pm1.7$ nHz and $b_4= -58.8\pm2.9$ nHz, values
  derived from a rotation inversion of a co-eval epoch using individual
  resolved modes at low and intermediate degrees \citep{EffDarwich+Korzennik:2012}.
  \item The derivative $\frac{\partial\nu}{\partial\ell}$ was computed
  directly from the set of $\{\nu_{n,\ell}\}$.
 \item The extent of the summations were set to
\begin{equation}
N_{\delta\ell'} = \left\{ \begin{array}{ll}
 10               & \ell\le 400 \\
 2 (1 + \ell/100) & \mbox{otherwise}\end{array}\right.
\end{equation}
\begin{equation}
N_{\delta\ell''} = \left\{ \begin{array}{ll}
 12        & \ell\le 600 \\
 2\ell/100 & \mbox{otherwise}\end{array}\right.
\end{equation}
while $N_{\delta m} = 10$. These values were determined to be optimum in
\citet{Rabello-SoaresEtal:2008a}.
 \item The coefficients of the leakage matrix, $C_r, C_h$, were computed by
 one of us (JS) to correspond to the spatial decomposition of the MDI {\em
 Dynamics} observations.
\end{list}

\begin{figure}[!t]
\centering
\includegraphics[width=.95\textwidth]{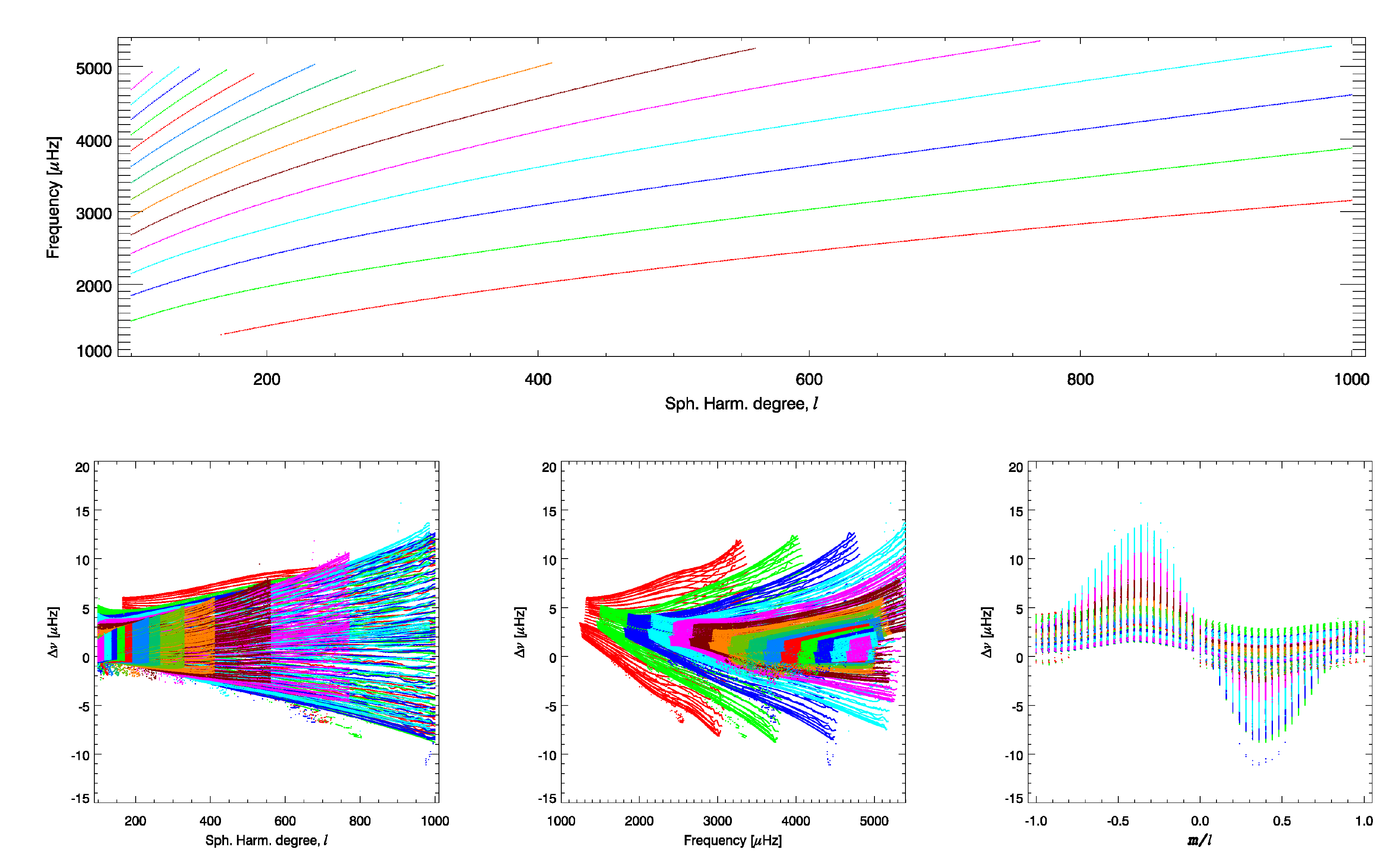}
\caption{Top panel: coverage in the $\ell-\nu$ of the modeling (shown for
$m=0$ only), the color corresponds to different values of $n$. Bottom
panels: frequency offsets, $\Delta^{\nu}$, between mode and ridge frequency as a
function of degree, $\ell$ (left), frequency, $\nu$ (middle), and relative
azimuthal order, $m/\ell$ (right). \label{fig:ridgeModel1}}
\end{figure}

\begin{figure}[!t]
\centering
\includegraphics[width=.95\textwidth]{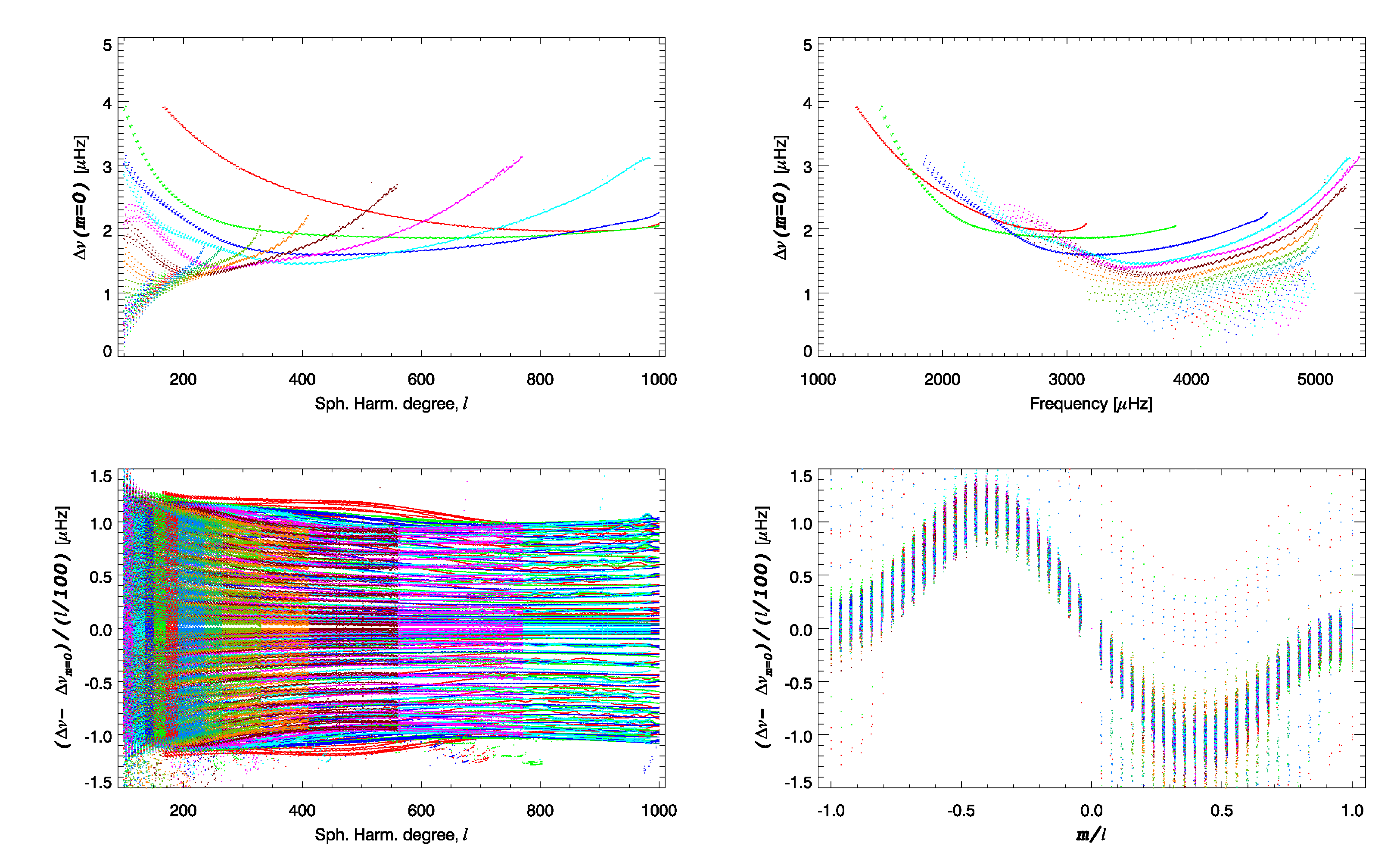}
\caption{Top panels: frequency offsets, $\Delta^{\nu}$, between mode and ridge
frequency (as in Fig.~\ref{fig:ridgeModel1}) but for $m=0$ only, plotted as a
function of degree or frequency.  Bottom panels: frequency offsets with
respect to the corresponding zonal value and divided by $\ell/100$, also
plotted versus degree or frequency.  Note how $\Delta^{\nu}_{m=0}$ is primarily a
function of $\nu$, while $(\Delta^{\nu} -\Delta^{\nu}_{m=0})/\ell$ is primarily a
function of $m/\ell$. The color corresponds to the value of the order, $n$.
\label{fig:ridgeModel2}}
\end{figure}

\begin{figure}[!t]
\centering
\includegraphics[width=.95\textwidth]{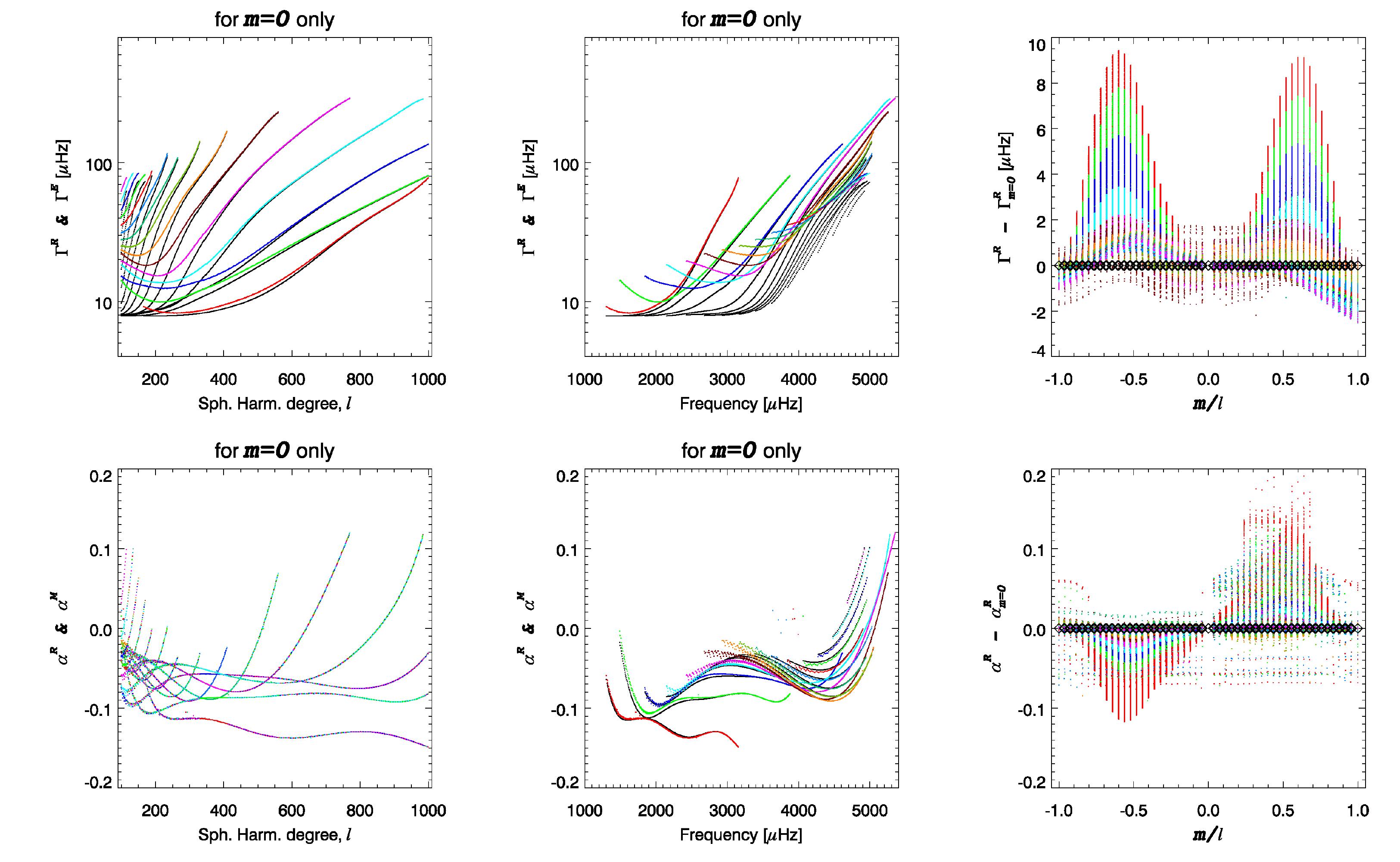}
\caption{Top panels: comparison of the model's ridge FWHM (colored dots) to
the input effective FWHM (black dots), versus degree, frequency or $m/\ell$
(left to right respectively). Bottom panels: comparison of the model's ridge
asymmetry (colored dots) to the input asymmetry (black dots), versus degree,
frequency or $m/\ell$ (left to right respectively). The left two panels in
both rows show only the $m=0$ values. The right most panels show the variation
with respect to the corresponding zonal values as a function of the relative
azimuthal order.  The diamond symbols in these panels correspond to the model
mode input values: both $\Gamma^E_{n,\ell}$ and $\alpha_{n,\ell}$ are constant
with $m$. The variation seen here results primarily from the variation of the
leakage with $m$.  The color corresponds to the value of the order, $n$.
\label{fig:ridgeModel3}}
\end{figure}

Figures~\ref{fig:ridgeModel1} to \ref{fig:ridgeModel3} present the key
properties of this model. Figure~\ref{fig:ridgeModel1} shows the raw frequency
offset (\ie, the difference between the target mode and the ridge frequency),
while Fig.~\ref{fig:ridgeModel2} shows on the one hand the zonal frequency offset,
$\Delta^{\nu}_{n,\ell,m=0}$, and how it scales primarily with frequency, and
on the other, the frequency offset with respect to its corresponding zonal
value, $\Delta^{\nu}_{n,\ell,m}-\Delta^{\nu}_{n,\ell,m=0}$ divided by the
$\ell$, since this quantity scales linearly with $\ell$. Once scaled, it is
mostly a function of the relative azimuthal order, $\frac{m}{\ell}$.

  Figure~\ref{fig:ridgeModel3} shows the mode to ridge widening, for the zonal
ridges, and also how the ridge width varies with the relative azimuthal order,
$\frac{m}{\ell}$, even though the mode width itself, an input parameter to the
model, is constant with $m$. Similarly, that figure shows that the ridge
asymmetry changes with $\frac{m}{\ell}$, while the corresponding input model
mode asymmetry is also constant with $m$.  The distortion of the eigenvalues
by the differential rotation changes the observed power distribution with a
strong dependence on $m$ -- hence it not only changes the ridge central
frequency but its width and asymmetry.

\subsubsection{The Modeling Iterative Process}

  Some of the input parameters are not precisely known {\em a priori} since
they are the parameters we want to estimate accurately, and are thus
approximated by values derived from ridge fitting. While we have shown that
the ridge to mode frequency correction sensitivity to these input parameters
is small \citep{Rabello-SoaresEtal:2008a}, we recognize that improving the
input parameters whenever possible can only improve the accuracy of such
corrections.
In our continuing effort to improve our determination of unbiased mode
parameters at high-degrees, we implemented an iterative procedure to adjust
some of the input parameters so as to produce a ridge model whose
characteristics match the measured ridge values.

  We adjusted iteratively the following parameters of the model input set:
$\nu_{n,\ell}$, $\Gamma_{n,\ell}$, $\alpha_{n,\ell}$, $a_i({n,\ell})$ and
$c_k^R(\ell)$. It consisted of the following adjustments at each iteration
step:
\begin{list}{--}{\setlength{\topsep}{0pt}\setlength{\itemsep}{0pt}}
\item Input frequencies, $\nu_{n,\ell}$:
\begin{equation}
  \nu_{n,\ell} \leftarrow \nu_{n,\ell,m} \leftarrow \tilde{\nu}_{n,\ell,m} - \Delta^{\nu}_{n,\ell,m}
\end{equation}
where $\tilde{\nu}$ is the observed ridge frequency, estimated by fitting the
observations, and $\Delta^{\nu}_{n,\ell,m}$ is the ridge to mode frequency
offset predicted by the model --\ie, the model resulting ridge frequency minus
the model input mode frequency-- interpolated at each $m$, using a polynomial
fit in $\frac{m}{\ell}$. The singlets frequencies, $\nu_{n,\ell,m}$, are then
reduced to multiplets, $\nu_{n,\ell}$, by fitting Clebsch-Gordan coefficients.
\item The FWHM, $\Gamma_{n,\ell}$, is adjusted at each iteration as follows:
\begin{equation}
\Gamma_{n,\ell} \leftarrow
\sqrt{(\tilde{\Gamma}^M_{n,\ell}\, f^\Gamma_{n,\ell})^2-W_{N,T}^2}
\end{equation}
where the factor $f^\Gamma$ is given by
\begin{equation}
  f^\Gamma_{n,\ell} = <\tilde{\Gamma}_{n,\ell,m}/\tilde{\Gamma}_{n,\ell,m}^M>
\end{equation}
and $\tilde{\Gamma}_{n,\ell,m}$ is the ridge FWHM estimated by fitting the
observations, while $\tilde{\Gamma}_{n,\ell,m}^M$ is the ridge FWHM measured
when fitting the model, interpolated at each $m$ by a polynomial fit in
$\frac{m}{\ell}$. The brackets represent averaging over $m$.
The quantity $W_{N,T}$ is the frequency resolution, set to the resolution of
the observations ($N=61$ and $T=90$ days).
\item Input asymmetry, $\alpha_{n,\ell}$, similarly is adjusted at each
 iteration as follows:
\begin{equation}
\alpha_{n,\ell} \leftarrow \alpha^M_{n,\ell}\, f^\alpha_{n,\ell}
\end{equation}
where the factor $f^\alpha$ is given by
\begin{equation}
f^\alpha_{n,\ell} = <\tilde{\alpha}_{n,\ell,m}/\tilde{\alpha}_{n,\ell,m}^M>
\end{equation} 
and $\tilde{\alpha}_{n,\ell,m}$ is the ridge asymmetry estimated by fitting
the observations, while $\tilde{\alpha}_{n,\ell,m}^M$ is the ridge asymmetry
measured when fitting the model, interpolated at each $m$ by a polynomial fit
in $\frac{m}{\ell}$. The brackets represent averaging over $m$.
\item Amplitudes and $c_k^R(\ell)$ are also re-adjusted at each iteration as
  follows:
\begin{equation}
A_{n,\ell} \leftarrow A^M_{n,\ell}/(\sum_k f^A_k \ell^k)
\end{equation}
where the polynomial $\sum_k f^A_k \ell^k$ is determined by fitting the ratio
$\tilde{A}^M_{n,\ell,m=0}/\tilde{A}_{n,\ell,m=0}$ with respect to
$\ell$. $\tilde{A}^M$ and $\tilde{A}$ represent the model and the observed
ridge power amplitudes, respectively, for the zonal modes.

The $c_k^R(\ell)$ coefficients are then derived by fitting $\frac{\partial\log
A}{\partial\ell}$. 
\item The following input variables are then readjusted or, for the secondary
  input parameters, recomputed as follows:
\begin{list}{---}{\setlength{\topsep}{0pt}\setlength{\itemsep}{0pt}}

  \item The FWHM, $\Gamma_{n,\ell}$, and the asymmetry, $\alpha_{n,\ell}$, are
  smoothed using a bivariate polynomial in $\nu$ and $\ell$.
  \item The splittings coefficients, $a_i(n,\ell)$, are recomputed using the
  new input frequencies.
  \item The horizontal to vertical ratio, $\beta$, is recomputed using the
  same prescription as in the preceding section, but using the new input
  frequencies. 
  \item The derivative, $\frac{\partial\nu}{\partial\ell}$ is recomputed
  using the new input frequencies.
\end{list}
\end{list}

  This procedure was run for some 10 iterations, where the models were
computed only every 5th degree.  The resulting input set was then used to
compute a final model at each degree.  After a handful of iterations, the
characteristics of the model barely changed and the procedure converged, with
relative changes at the last steps of the iterative process at the 0.1\%
level.

\subsubsection{Sensitivity of the Model}

 The values for $\nu_{n,\ell}$, $\Gamma_{n,\ell}$, $\alpha_{n,\ell}$, $c^R_k$
and $a_i({n,\ell})$ have been fine tuned using an iterative process. Since the
resulting ridge model values when using these parameters match the observed
ridge values, their uncertainty and thus the corresponding model sensitivity
can be estimated by the changes at the last iteration (0.1\%).  A more
conservative alternative approach is to perturb each value by its
corresponding measured ridge uncertainty.

  We computed additional models where some input parameters were increased by
their corresponding observational uncertainty. 

\begin{figure}[!t]
\centering
\includegraphics[width=.95\textwidth]{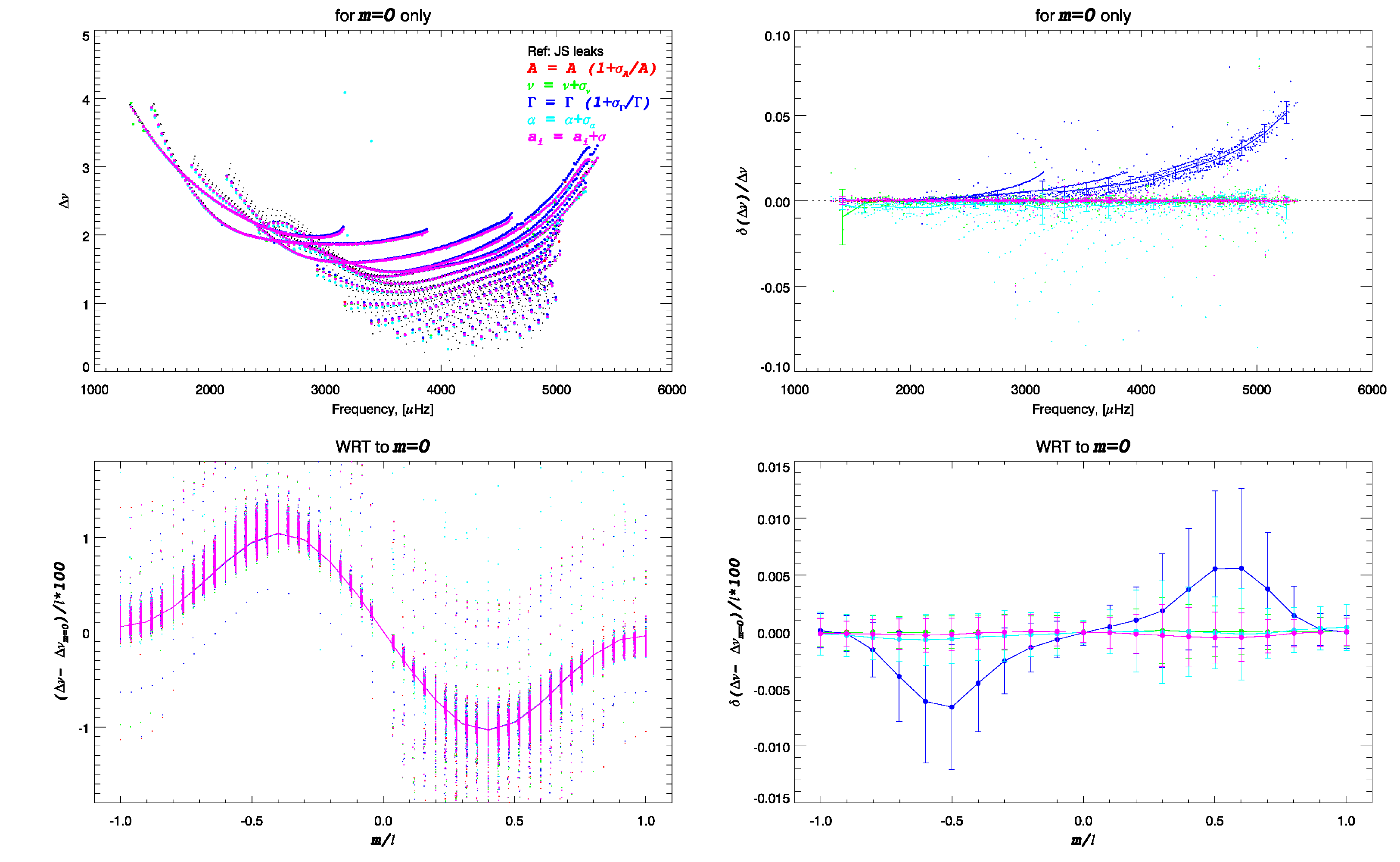}
\caption{Comparison of the ridge to mode frequency offset, produced by
different models. The top panels show the zonal modes only as a function of
frequency, while the bottom panels show the scaled offset, relative to the
zonal values, as a function of $m/\ell$. The left panels compare the offset,
the right panels show the relative difference (top right) or the difference
(bottom right). The lines with error bars represent binned values and the
scatter within each bin.
The models correspond to perturbing the modes amplitudes, $A$, frequencies, $\nu$, FWHM,
$\Gamma$, asymmetry, $\alpha$, and, rotational splittings, $a_i$, by one-sigma.
\label{fig:compareCase1}}
\end{figure}

The effect of perturbing the modes amplitudes, $A$, frequencies, $\nu$, FWHM,
$\Gamma$, asymmetry, $\alpha$, and, rotational splittings, $a_i$, by one-sigma
are illustrated in Fig.~\ref{fig:compareCase1}, where the relative change of
the frequency offsets between ridge and mode, $\Delta^\nu$, is shown
as the change of the zonal offset,
$\Delta^\nu_{n,\ell,m=0}$, and as the change of the scaled frequency offset
with respect to the zonal value, $(\Delta^\nu_{n,\ell,m} - \Delta^\nu_{n,\ell,m=0})/\ell$.
Note how the zonal offset is barely affected, except when perturbing the FWHM.
The azimuthal signature is also small with just the FWHM perturbation showing a
systematic and specific change.

  The input parameters $\beta_{n,\ell}$, the horizontal to vertical ratio,
$b_2, b_4$, the coefficients corresponding to the parametrization of the
surface differential rotation, and $C_r, C_h$, the leakage matrix
coefficients, were not iterated upon. The sensitivity of the model to their
accuracy needs also to be estimated.

\begin{figure}[!t]
\centering
\includegraphics[width=.95\textwidth]{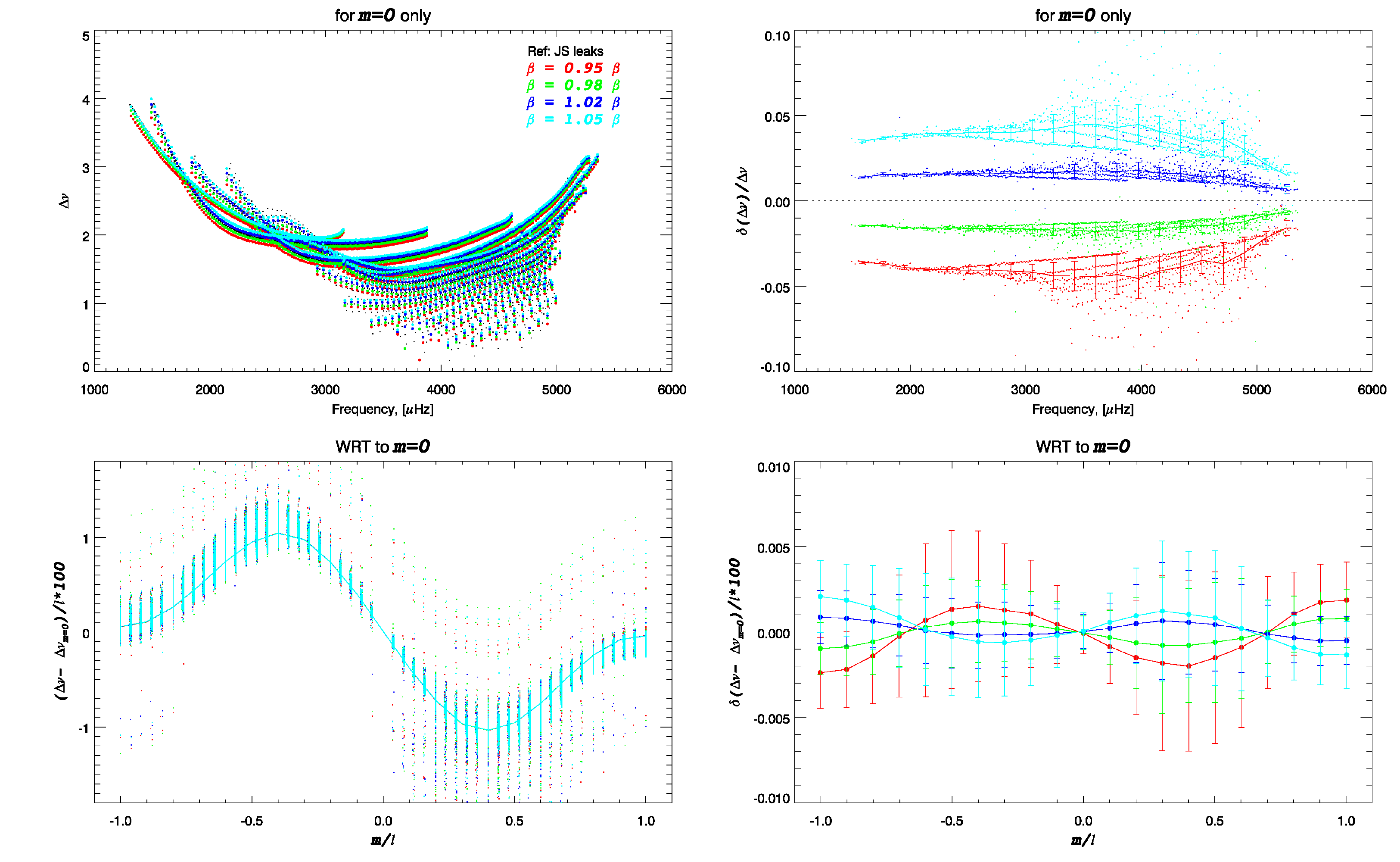}
\caption{Comparison of the ridge to mode frequency offset, produced by
different models, in the same format as Fig.~\ref{fig:compareCase1}.
The effect of perturbing the horizontal to vertical ratio, $\beta$, is
illustrated for positive and negative perturbation of 2 and 5\%.
\label{fig:compareCase2}}
\end{figure}

  The value of $\beta_{n,\ell}$ is well constrained by the values of
$\nu_{n,\ell}$ under the prescription we used. Still, this prescription does
not force $\beta_{n=0,\ell}$ to be exactly unity\footnote{In fact it varies
within 0.9970 and 1.0011.}. This prescription is also based on estimating
that ratio using a formula resulting from a specific outer boundary condition,
namely that the Lagrangian pressure perturbation vanish ($\delta p = 0$), in
the small amplitude oscillation equations for the adiabatic and non-magnetic
case. While this is most likely a very good approximation, the conditions near
the surface are not adiabatic, and the ratio $\beta$ might not be exactly the
one predicted by the adiabatic case. \citet{RhodesEtal:2001a} estimated that
$\beta$ is $0.995\pm0.004$ of the theoretical value.  We ran models with
perturbed values of $\beta$ to assess their sensitivity to this
parameter.
The effect of perturbing $\beta$ is illustrated in
Fig.~\ref{fig:compareCase2}. That figure shows that the zonal frequency offset
changes proportionally to the perturbation, while the change of the scaled
offset with respect to the zonal values displays a small azimuthal signature
that peaks at the 1 and 2\% levels, respectively (in terms of relative change
of the offset).

\begin{figure}[!t]
\centering
\includegraphics[width=.95\textwidth]{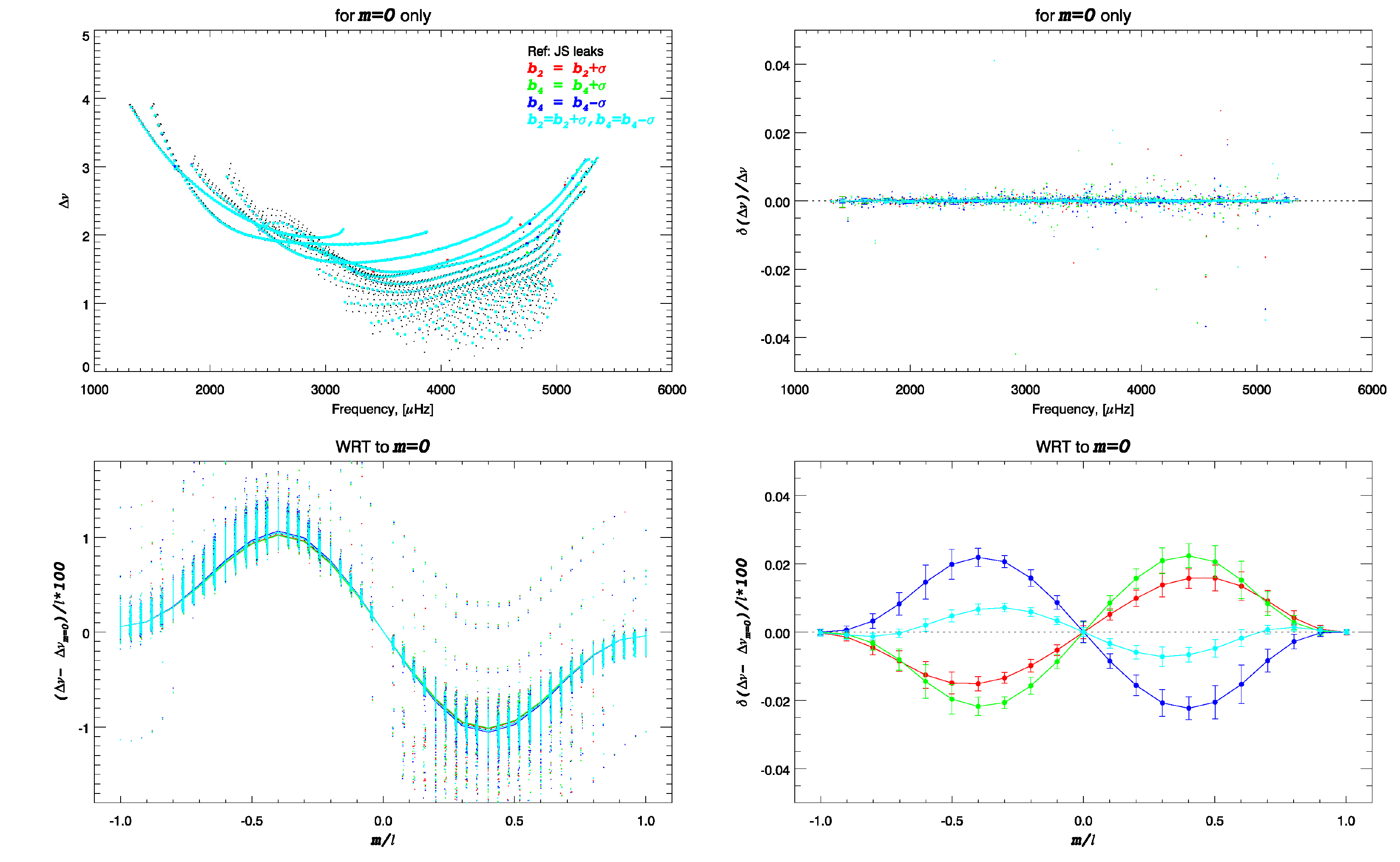}
\caption{Comparison of the ridge to mode frequency offset, produced by
different models, in the same format as Figs.~\ref{fig:compareCase1} and
\ref{fig:compareCase2}. 
The figures show the effect of perturbing the parametrization of the surface
differential rotation ($b_2, b_4$) used to compute the mode mixing resulting
from the distortion of the eigenfunctions by the differential rotation. 
\label{fig:compareCase3}}
\end{figure}

  We selected our best estimate for $b_2=-69.6\pm1.7, b_4=-58.8\pm2.9$ nHz,
using our own inversion profile derived from measuring rotational frequency
splittings at low and intermediate modes for a co-eval epoch
\citep{EffDarwich+Korzennik:2012}.  We have computed models using one-sigma
perturbation in $b_2$ and $b_4$, as shown in Fig.~\ref{fig:compareCase3}, to
assess the effect of perturbing the parametrization of the surface
differential rotation used to compute the mode mixing resulting from the
distortion of the eigenfunctions by the differential rotation.
The change of the zonal frequency offset is negligible (at the 0.1\% level),
while the change of the scaled offset with respect to the zonal values
displays a clear azimuthal signature at the 2\% level. This illustrates the
interdependency between knowing the rotation profile to properly correct the
frequency splittings and deriving the rotation profile from the splittings.

  Finally, to quantify the sensitivity of our result to the leakage matrix
coefficients, we computed models that use an independent estimate of the
leakage coded by one of us (SK). The leakage was computed by spatially
decomposing images of the three components of the line of sight velocity
signal (see Eqs.~\ref{eq:theoleakr} to \ref{eq:theoleakp}). The sampling of
these images and their spatial apodization were set so as to correspond to the
MDI Dynamics observations.
This was carried out for the same subset of $(\ell,m)$, as for the reference
leakage matrix coefficients (computed by JS).

  We computed a set of leakage matrix coefficients for $B_o=0$, where $B_o$ is
the heliographic latitude at disk center, and for $B_o=-5.68^{\rm o}$, the
mean value of $B_o$ for the 2001 Dynamics epoch.
  We computed another set for that same value of $B_o$, but including an
estimate of the PSF of MDI. That estimate was derived from running the
procedure {\tt HGEOM}, a procedure that is part of the GONG reduction and
analysis software package.
{\tt HGEOM} returns an estimate of the azimuthaly averaged MTF \citep[\ie, the
Fourier transform of the PSF, see][and references therein]{KorzennikEtal:2004}.
  We thus also included an axisymmetric PSF, whose profile was set by the MTF
estimated by {\tt HGEOM}. 

\begin{figure}[!t]
\centering
\includegraphics[width=.95\textwidth]{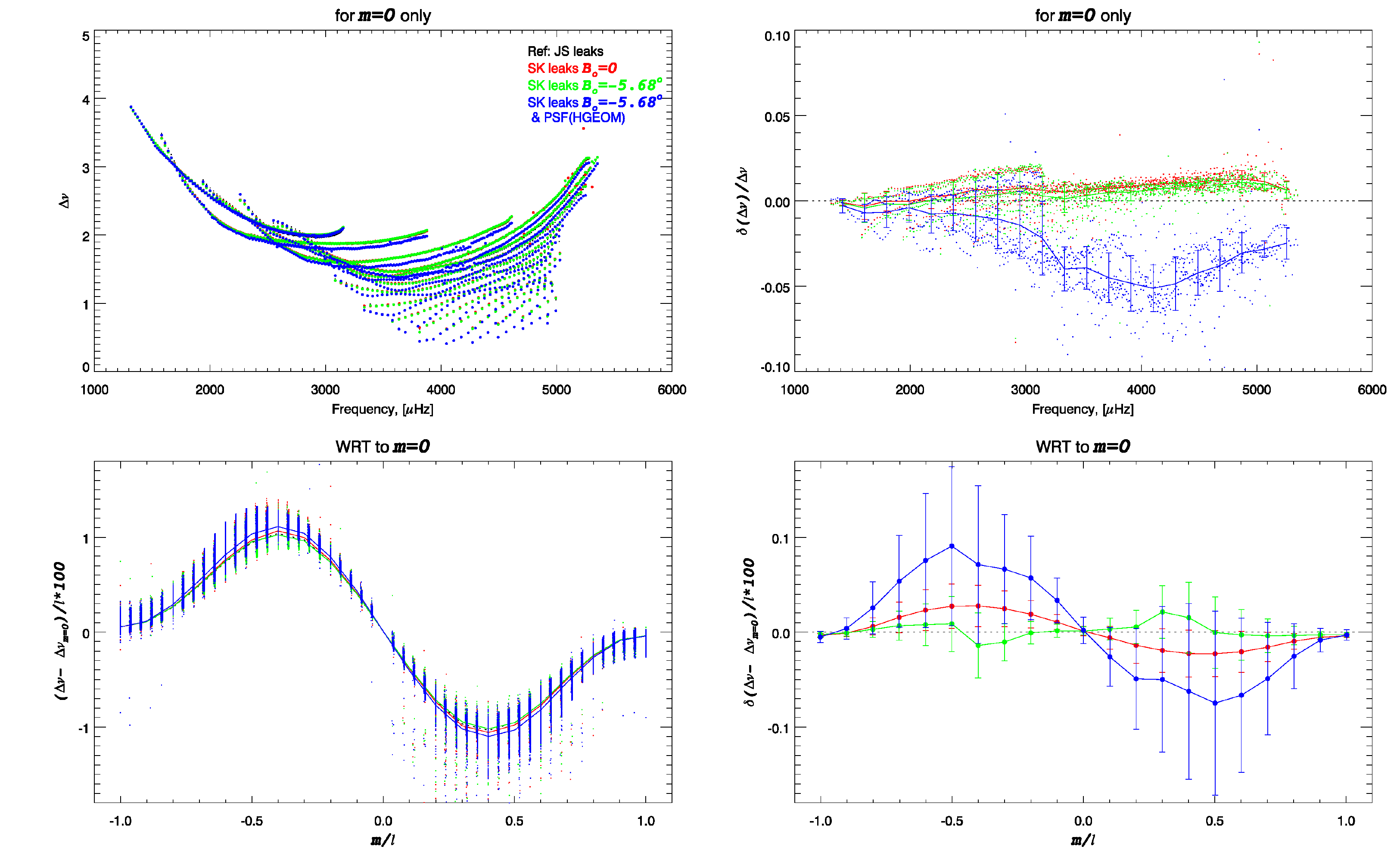}
\caption{Comparison of the ridge to mode frequency offset, produced by
different models, in the same format as Figs.~\ref{fig:compareCase1} to
\ref{fig:compareCase3}.
The effect of using different leakage matrix coefficients (see text for
description) is illustrated, using two different and independent
implementations (black versus colors), including the actual mean value of
$B_o$ (in red and green), as well as including an axisymmetric instrumental
PSF (in blue).
\label{fig:compareCase4}}
\end{figure}

 The changes resulting from using different leakage matrices are presented in
Fig.~\ref{fig:compareCase4}.
This figure shows that using a different leakage matrix computation produces a
relative change in the zonal frequency offset at the 1\% level, except when
including a PSF, where that change peaks at 5\%. Similarly, the largest change
in the scaled offset with respect to the zonal offset is seen when including a
PSF.
These results corroborate our previous conclusions \citep{KorzennikEtal:2004}
that the largest contribution to the uncertainty of the high degree mode
properties is our limited knowledge of the PSF of the MDI instrument.

\begin{table}[!t]
\centering
\begin{tabular}{||c|c||c|c||}\hline
          &              &
          \multicolumn{2}{c||}{Frequency offset relative sensitivity} \\
Parameter & Perturbation & zonal & azimuthal \\\hline
frequency, $\nu$    & $1\ \sigma$      & $0.001 \pm 0.002$ (0.009) & 0.001 ($0.001 \pm 0.002$) \\
FWHM,      $\Gamma$ & $1\ \epsilon$    & $0.012 \pm 0.012$ (0.052) & 0.007 ($0.001 \pm 0.005$) \\
asymmetry, $\alpha$ & $1\ \sigma$      & $0.003 \pm 0.006$ (0.004) & 0.001 ($0.001 \pm 0.002$) \\
amplitude, $A$      & $1\ \epsilon$    & $0.001 \pm 0.001$ (0.001) & 0.001 ($0.001 \pm 0.001$) \\
splittings, $a_i$   & $1\ \sigma$      & $0.001 \pm 0.001$ (0.001) & 0.001 ($0.001 \pm 0.002$) \\
ratio, $\beta$      & 2\%              & $0.016 \pm 0.005$ (0.018) & 0.001 ($0.001 \pm 0.002$) \\
rot.\, $b_2,b_4$    & $1\ \sigma$      & $0.001 \pm 0.001$ (0.001) & 0.022 ($0.001 \pm 0.014$) \\
leakage matrix      & $B_o=0^{\rm o}$  & $0.008 \pm 0.006$ (0.013) & 0.028 ($0.001 \pm 0.023$) \\
leakage matrix   & $B_o=-5.68^{\rm o}$ & $0.005 \pm 0.008$ (0.011) & 0.021 ($0.001 \pm 0.020$) \\
leakage matrix & $B_o=-5.68^{\rm o}$ \& PSF & $-0.029 \pm 0.022$ (0.051) & 0.091 ($0.004 \pm 0.076$) \\\hline

\end{tabular}
\caption{Sensitivity of the ridge to mode frequency offset with respect to the
model's input parameters. The zonal values are characterized by the mean and
RMS of the differences (and within parenthesis the maximum of the absolute
value of the binned differences), while the azimuthal values are characterized
by the maximum of the absolute value of the binned differences (and within
parenthesis the mean and RMS of the raw differences).
\label{tab:sensitivity}}
\end{table}

  The relative change of $\Delta^\nu$ for all the perturbed models, presented
in Figs.~\ref{fig:compareCase1} to \ref{fig:compareCase4}, are also summarized
in Table~\ref{tab:sensitivity}, where the zonal values are characterized by
the mean and RMS of the differences, while the azimuthal values are
characterized by the maximum of the absolute value of the binned differences.

  The sensitivity of the ridge to mode frequency offset to a $1\,\sigma$
perturbation is indeed at the 0.1\% level, but for the FWHM and for changes of
$\beta$ and the leakage matrix. In the case of the FWHM, the $1\,\sigma$
perturbation is a large perturbation because the ridge fitting precision is
low. Such a perturbation produces a model whose ridge width is, on average,
off by 4\% (versus 0.2\% for the reference model) when compared to the
observed ridge width. So the model precision is constrained by producing a
model that matches the observations and not by the uncertainty on the ridge
width. As for the sensitivity with $\beta$ we see that it is proportional to
the perturbation, so if we consider that $\beta$ may be off by as much as
0.5\% from the theoretical value \citep{RhodesEtal:2001a}, the model
uncertainty is at most at the 0.4\% level.  As for the sensitivity of our
model to a particular leakage matrix, we see large changes associated to our
limited knowledge of the PSF of the MDI instrument.
A more conservative estimate of the model precision is thus more at the 0.5\%
level than the 0.1\% level, but could be as large as a few percent if the
leakage matrix we use is off. We opted, somewhat arbitrarily, to use 0.1\% and
1\% as basic and conservative values for the precision of $\Delta^\nu$, the
ridge to mode frequency offset.

\section{Results}\label{sec:Results}

  The approach we used to derive the characteristics of high degree modes is
to fit and characterize the power ridges and compute a sophisticated model of
the ridge power to estimate the bias between the ridge characteristics and the
underlying characteristics of the mode. While the most useful and sought after
characteristic is the mode frequency, we have also expanded our efforts to
derive estimates of the mode width, its asymmetry and its amplitude.

\subsection{Frequencies \& Frequency Splittings}

   Mode blending causes the ridge frequency to be offset with respect to the
underlying target mode frequency. Our ridge model allows us to estimate this
offset, and the sensitivity of that estimate on the model input parameters. We
have thus computed this offset for every degree between $\ell=100$ and
$\ell=1000$, and each order for which the ridge was fitted, but only for 51
values of $m$ spanning the $[-\ell,+\ell]$ range. We used a polynomial fit in
$m/\ell$ to re-sample the offset at each $m$.

  An estimate of the mode frequency is computed by subtracting from the
measured ridge frequency the ridge to mode frequency offset:
\begin{equation}
  \nu_{n,\ell,m} = \tilde{\nu}_{n,\ell,m} - \Delta^{\nu}_{n,\ell,m}
\label{eq:corrFreq}
\end{equation}
where $\tilde{\nu}$ is the ridge frequency and $\Delta^{\nu}$ the ridge to
mode offset:
\begin{equation}
  \Delta^{\nu}_{n,\ell,m} = \tilde{\nu}^M_{n,\ell,m} - {\nu}^M_{n,\ell,m} 
\label{eq:defDeltaFreq}
\end{equation}
${\nu}^M$ being the mode frequency and $\tilde{\nu}^M$ the resulting ridge
frequency in our model. The frequency multiplets, $\nu_{n,\ell,m}$, are then
reduced to singlets, $\nu_{n,\ell}$, by performing the usual Clebsch-Gordan
coefficients expansion in $m$.

  The uncertainty on the resulting mode frequency estimate is determined by
the uncertainty on the ridge estimate and the uncertainty on the correction,
the latter being either the precision of the iterative process (0.1\%) or a
more conservative estimate, as discussed in the previous section (1\%). These
two uncertainties are presumed independent and therefore combined in
quadrature, \ie:
\begin{equation}
\sigma^2_{\nu_{n,\ell,m}} = \tilde{\sigma}^2_{\nu_{n,\ell,m}} +
 \sigma^2_{\Delta^{\nu}_{n,\ell,m}}
\label{eq:errFreq}
\end{equation}

\begin{figure}[!t]
\centering
\includegraphics[width=.95\textwidth]{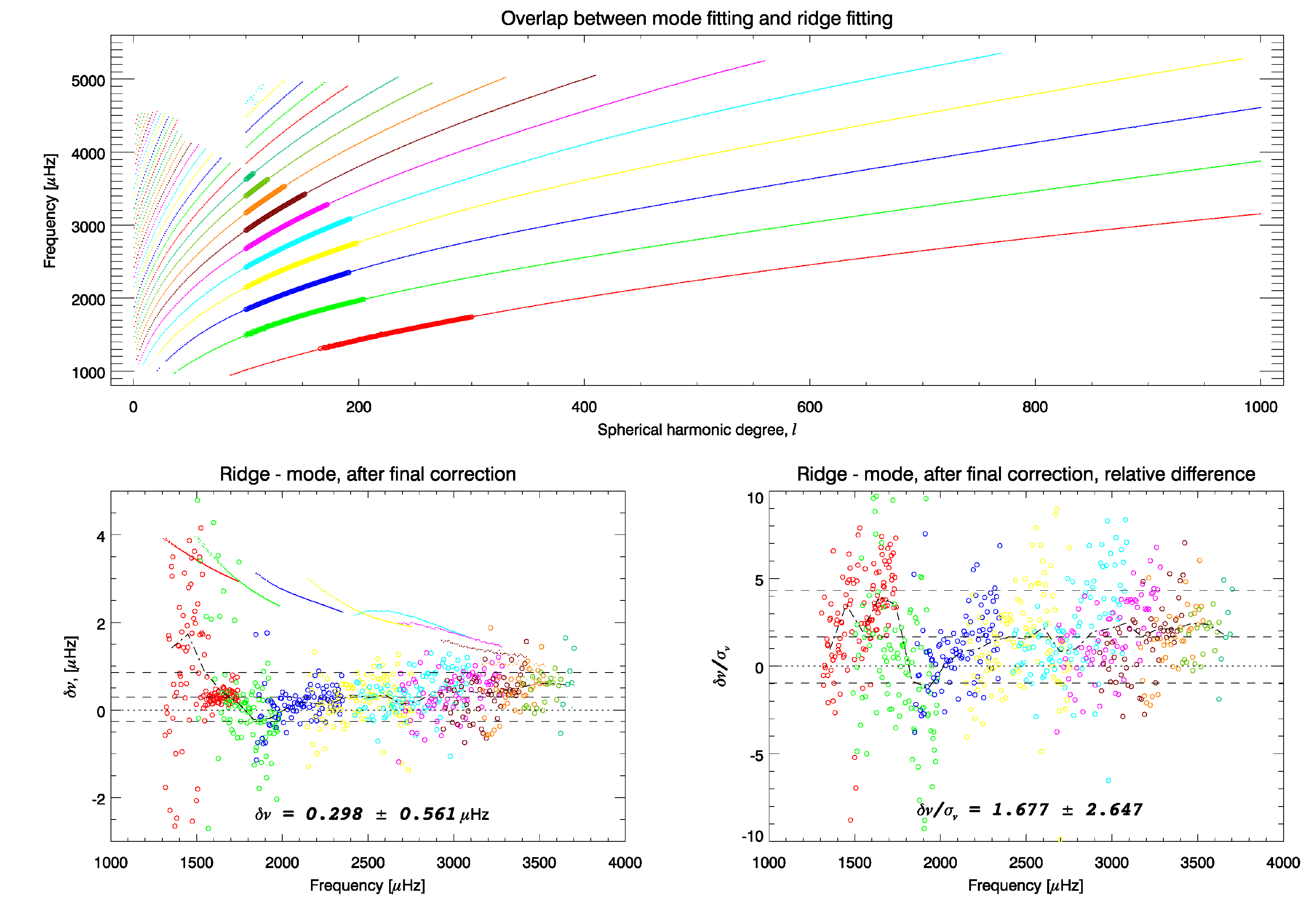}
\caption{Comparison of estimate of mode multiplet frequencies derived from
ridge fitting to actual mode frequency computed by fitting resolved modes at
intermediate degrees. The top panel shows the overlap between both methods in
the $\ell-\nu$ plane (filled circles). The bottom left panel shows the raw
frequency differences (open circles) as a function of frequency, with the dot
dash line representing the binned raw differences. The dots on that figure
correspond to the ridge to mode frequency correction. The bottom right panel
shows the scaled frequency differences, \ie, the frequency differences divided
by their respective uncertainties (open circles) as a function of frequency,
with the dot dash line representing the binned scaled differences. A dotted
line is drawn at 0, while the dotted lines are drawn at the mean and plus or
minus one RMS of the raw differences. The color corresponds to the value of
the order, $n$.
\label{fig:compareFrequency}}
\end{figure}

\begin{figure}[!t]
\centering
\includegraphics[width=.95\textwidth]{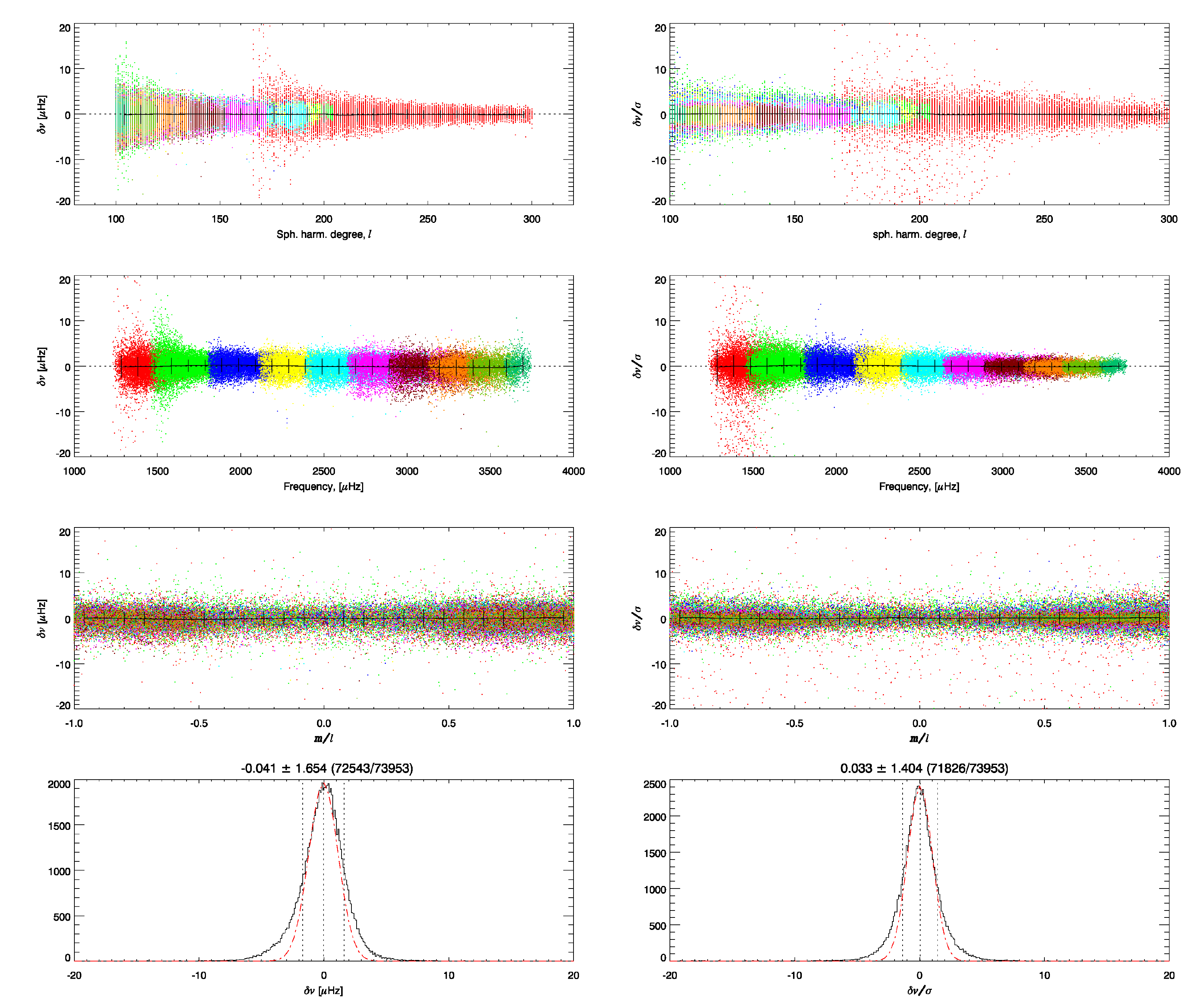}
\caption{Comparison of estimate of mode singlet frequencies derived from ridge
fitting to actual mode frequency computed by fitting resolved modes at
intermediate degrees. The frequency differences (left panels), and the scaled
differences (differences divided by their respective uncertainties, right
panels) are plotted as a function of degree, frequency, and $m/\ell$, and as a
histogram (top to bottom).  In the top three rows the color corresponds to the
value of the degree, $n$, while the black lines represent binned values with
the RMS within each bin shown as error bars. These differences show no trend
with degree, frequency or relative azimuthal order, but for the scatter.
The red curves in the histograms are the Gaussians corresponding to the mean
and standard deviation of the distributions (indicated by the vertical dotted
lines). Note the excess and skewness in the wings of the actual distribution.
\label{fig:compareSingletFrequency}}
\end{figure}

  As described in Sec.~\ref{sec:DataAnalysis}, we have some overlap between
mode and ridge fitting covering $100\le\ell\le300$ for f-modes and
$100\le\ell\le200$ for p-modes.  Figure~\ref{fig:compareFrequency} shows how
well the estimate of the mode multiplet frequencies derived from ridge fitting
match the actual mode frequency computed by fitting resolved
modes\footnote{The resolved modes singlets used for the comparisons are the
ones estimated by Korzennik ({\em in prep.})
using a 144 day long nearly co-eval epoch.}.
The residual bias is nearly uniform, about 0.3 \uHz\, or at the $1.67\,\sigma$
level.
Figure~\ref{fig:compareSingletFrequency} compares the mode singlet
frequencies, derived from ridge fitting, to the frequencies computed by
fitting resolved modes\footnote{The methodology used by Korzennik fit all the
individual singlets in a given multiplet.}. The singlets agree at the
$0.1\,\sigma$ level, with no trends with degree, frequency of azimuthal order.
The scatter of the differences shows some systematic fluctuations with
frequency (or order).
The distribution of the raw differences, although nearly randomly distributed,
show a departure from a Gaussian distribution in the wings (\ie, for the less
frequent larger differences), that is asymmetric. This skewness might explain
why the comparison of the multiplets shows a larger bias. Note also that the
RMS of the scaled differences is 1.4, or a little larger than unity.

\begin{figure}[!t]
\centering
\includegraphics[width=.95\textwidth]{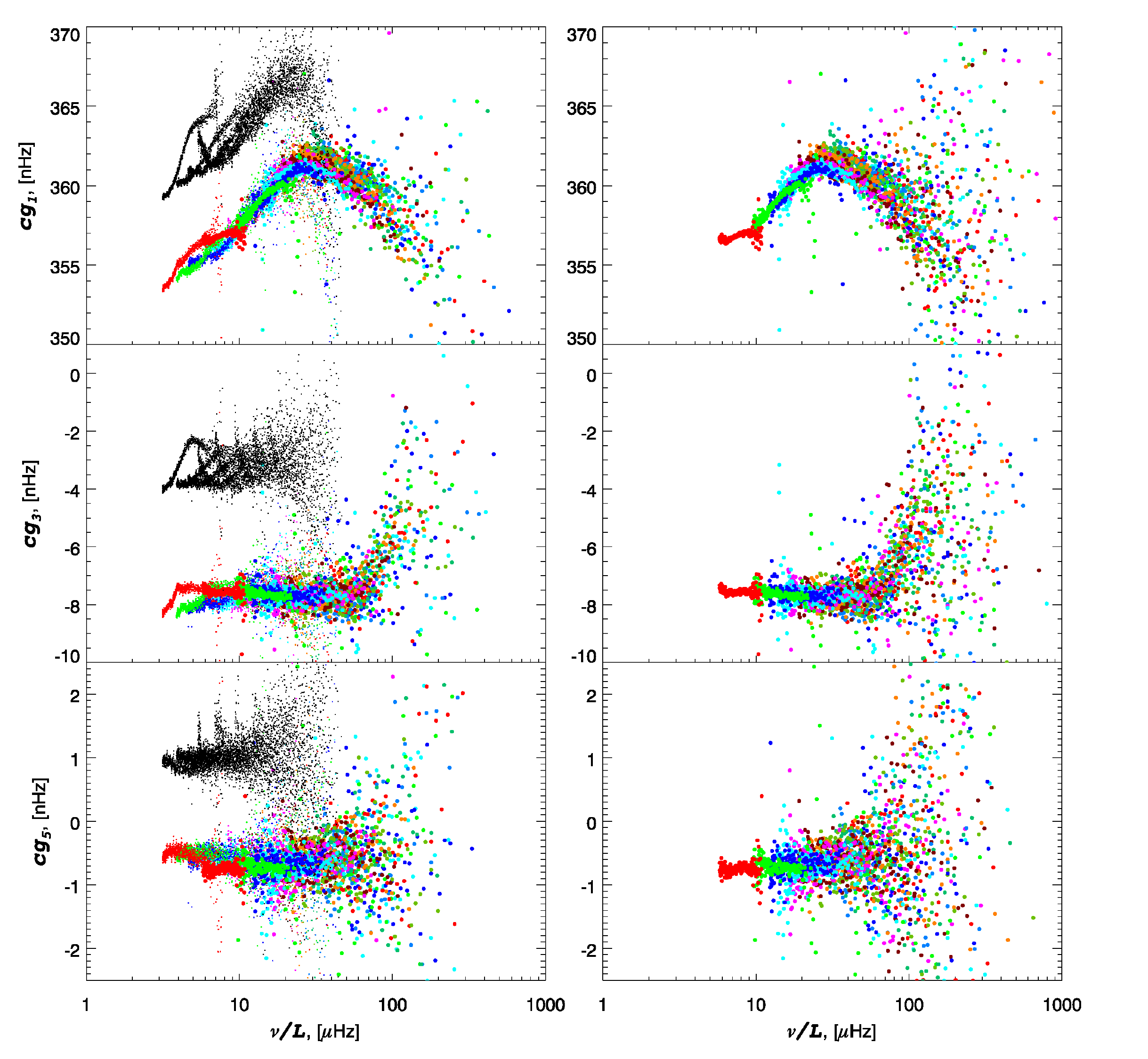}
\caption{Left panels: comparison of estimates of mode frequencies splittings,
parametrized as Clebsch-Gordan coefficients, plotted as a function of the
ratio $\nu/L$, a proxy for the mode inner turning point, derived from ridge
fitting after correcting for the ridge to mode offset (small dots) to
frequency splitting coefficients computed by fitting resolved modes (large
dots).
The black dots are the frequency splitting coefficients computed from
uncorrected ridge frequencies. The corrected splittings match the estimates
derived from fitting resolved modes.
Right panels: same Clebsch-Gordan coefficients computed by fitting resolved
modes, only (same large dots). 
The color corresponds to the value of the order, $n$.  Top to bottom panels
correspond to different odd coefficients.
\label{fig:compareSplittings}}
\end{figure}

  Figure~\ref{fig:compareSplittings} shows how well our estimates of the
frequency splittings derived from corrected ridge frequencies, once
parametrized in terms of Clebsch-Gordan coefficients, match the frequency
splittings computed by fitting resolved modes.  The substantial offsets
present without that correction are canceled by our correction. One is
hard-pressed in these plots to see the transition from mode fitting to ridge
fitting along a given order.

\begin{figure}[!t]
\centering
\includegraphics[width=.95\textwidth]{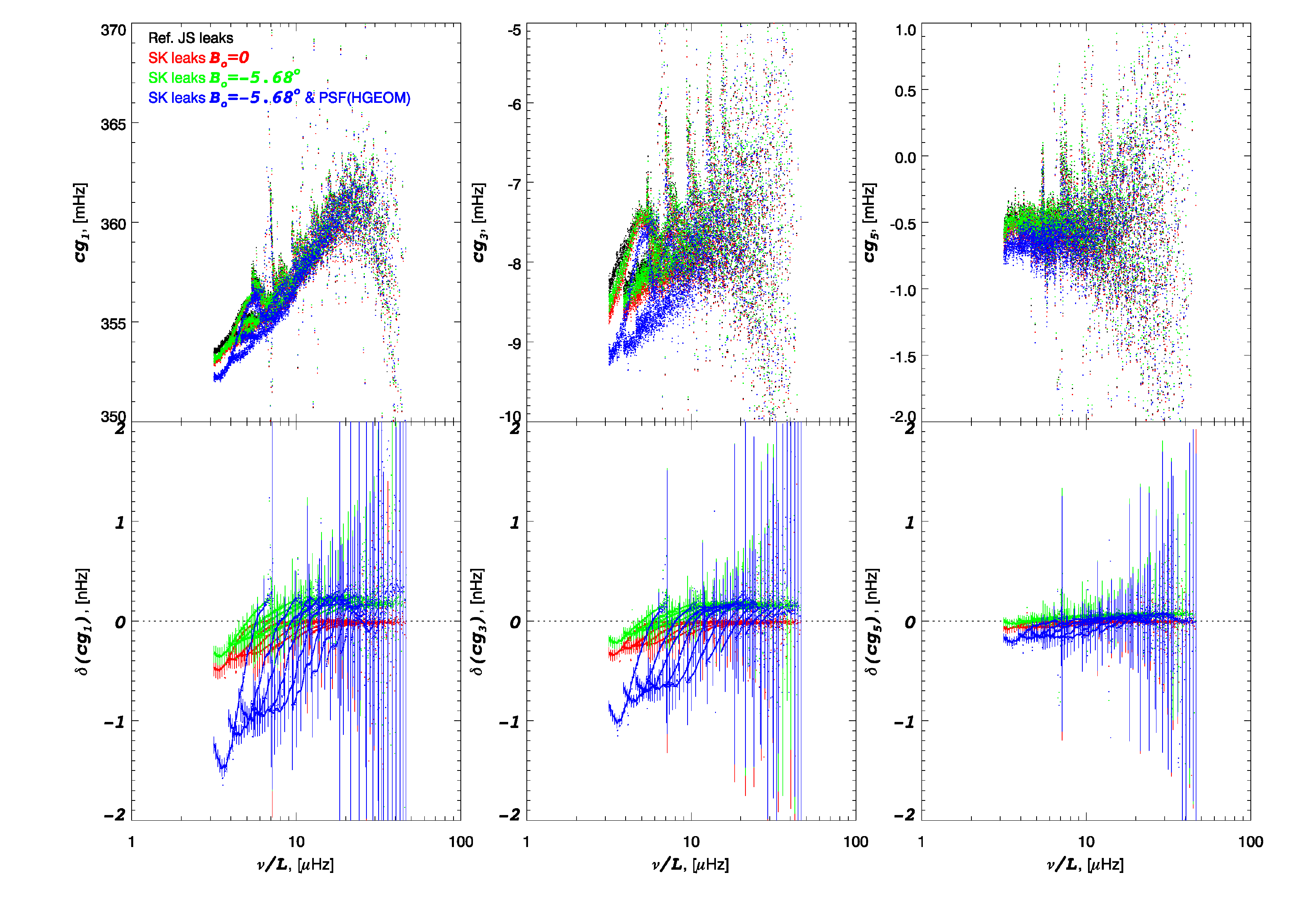}
\caption{Comparison of estimates of mode frequencies splittings, parametrized
as Clebsch-Gordan coefficients, plotted as a function of the ratio $\nu/L$,
and derived from ridge fitting after correcting for the ridge to mode offset,
using ridge-to-mode offsets computed for the same four different models
presented in Fig.~\ref{fig:compareCase4}.
The bottom panels show the differences with respect to the reference model, and the
coefficients error bars (not shown for each value for clarity).  Left to right
panels correspond to different odd coefficients.
The changes are, for the shallow modes, significantly larger than their
uncertainties.
\label{fig:compareSplittingsWRTModel}}
\end{figure}

Figure~\ref{fig:compareSplittingsWRTModel} shows how these Clebsch-Gordan
coefficients change when correcting the ridge frequencies using mode-to-ridge
offsets estimated from models using different leakage matrices. This figure,
which presents the same effect than the one presented in
Fig.~\ref{fig:compareCase4}, but in terms of splitting coefficients, shows
significant changes in the parametrization of the rotational splittings,
primarily at the lowest values of $\nu/L$ (shallow modes). As we stated
earlier, this results from our limited knowledge of the PSF of the MDI
instrument.

\subsection{Widths, Asymmetries \& Amplitudes}

  The most sought after characteristics of solar oscillations are the mode
frequencies, since they are the most precise and best understood
properties. They have allowed us to infer interesting properties of the Sun.
We also realize that the modes FWHM, asymmetry and amplitude are useful
properties to characterize, and important quantities when modeling the near
surface acoustic field.  In a manner similar to the case of the frequencies,
the FWHM, amplitude and to a lesser degree the asymmetry estimated by fitting
the ridges are not the FWHM, amplitude and asymmetry of the underlying modes.
With the assistance of the ridge modeling, we devised specific methodologies
to derive estimates of the mode values from the ridge estimates.

\subsubsection{FWHM}

\begin{figure}[!t]
\centering
\includegraphics[width=.95\textwidth]{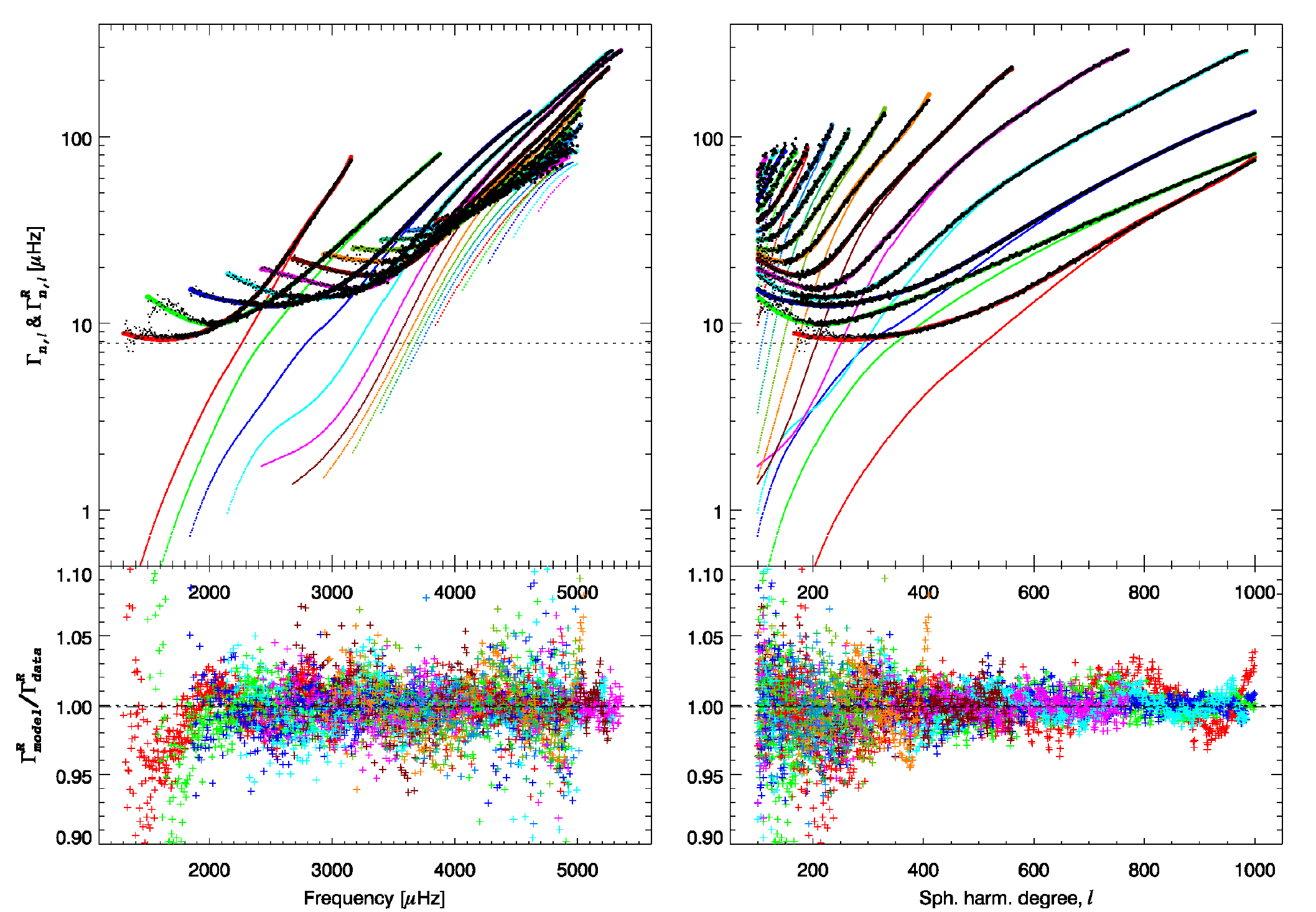}
\caption{Top panels: measured ridge FWHM (black dots) as a function of
frequency or spherical harmonic degree (left and right panels, respectively),
compared to our best model for the ridge FWHM (colored circles), derived from
an optimized model for mode FWHM estimates (colored dots), for $m=0$. 
Note how, for a substantial fraction of modes, the resulting ridge FWHM is
much larger than the mode FWHM.  The small black dots correspond to cases
where the 3-$\sigma$ uncertainty on the ridge width is larger than the mode
width estimates.
The horizontal line corresponds to $W_{N,T}$, the resolution of the $N$th
order sine multi-taper used as spectral estimator.
Bottom panels: ratio of model to measured ridge FWHMs, indicating how well our
predicted ridge FWHMs agree with the observed ones, but for some residual
disagreement at the lowest frequencies, where the mode FWHM is more than an
order of magnitude smaller then the resulting ridge FWHM.
The different colors correspond to the order, $n$.
\label{fig:compareFWHM}}
\end{figure}

\begin{figure}[!t]
\centering
\includegraphics[width=.95\textwidth]{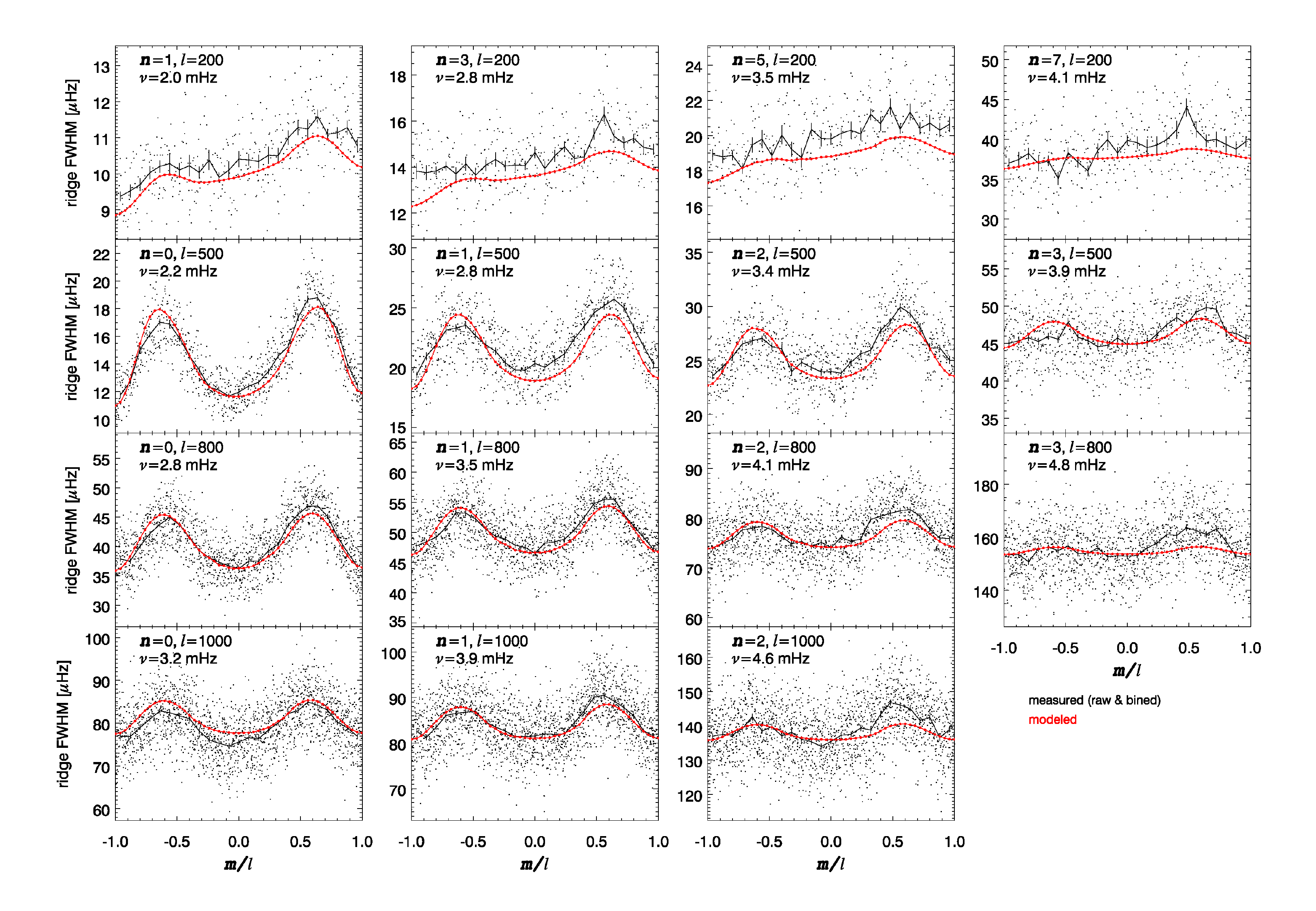}
\caption{Comparison of ridge FWHM for a selection of multiplets, as a function
of the ratio $m/\ell$. The dots correspond to individual ridge width
measurements, and the black curves the corresponding binned values. The red
curves correspond to our best model of the ridge width, as evaluated at 51
equispaced $m$ values. In most cases, our model reproduces rather well the
variation of ridge width with the azimuthal order, $m$, although the input
mode width of the model is constant with $m$.
\label{fig:compareFWHMvsMOL}}
\end{figure}

  Figure~\ref{fig:compareFWHM} compares the ridge FWHM, $\tilde{\Gamma}$,
resulting from ridge fitting to the predicted ridge width from our model, for
the zonal modes. The iterative procedure produced a model for the ridge FWHM
that matches, but for the very low frequencies, the measured values. The
intrinsic mode width estimates used in the model are also shown on that
figure. These are, for a substantial fraction of modes, a lot smaller than the
ridge widths. In fact, the uncertainty on the fitted ridge width becomes, in
some cases, larger than the mode width estimate itself, especially at the low
degree end of the low order ridges.  For these modes, recovering the
intrinsic mode width from ridge fitting is poorly constrained and in some
cases too poorly constrained to derive a meaningful estimate.

  Figure~\ref{fig:compareFWHMvsMOL} also compares the measured ridge widths to
the model predictions, but as a function of the ratio $m/\ell$, and for a
selection of multiplets. This figure illustrates how well our model reproduces
the variation of the ridge width with the azimuthal order, although the mode
width in the model is constant with $m$.

  The zonal ridge FWHM is by no means the mode FWHM. Its value reflects the
mode FWHM, the spectral resolution, and as a result of the mode blending, the
slope of the ridge with respect to the degree (\ie,
$\frac{\partial\nu}{\partial\ell}$), since it is that slope that sets the
separation in frequency of the leaks that contribute to the ridge.  We devised
a simple two step widening model to estimate the mode width from the ridge
width. It is based on the Gaussian convolution analogy: the convolution of
Gaussian by a Gaussian is also a Gaussian, whose width is the square root of
the sum of the square of the convolved Gaussian widths. So by analogy, we
relate the width of the ridge to the width of the mode using the following
simple and convenient model:
\begin{equation}
\tilde{\Gamma}^2_{n,\ell} = {(\Gamma^E_{n,\ell})}^2 +  {\cal W}_{n,\ell}^2
\label{eq:fwhm_convol}
\end{equation}
where $\Gamma^E_{n,\ell}$ is the effective mode width (as defined in
Eq.~\ref{eq:gammaE}) and ${\cal W}$ is a mode to ridge widening term, that we
estimate from our modeling procedure.  To account for the ridge fitting
uncertainty, we rewrite Eq.~\ref{eq:fwhm_convol} as:
\begin{equation}
( \tilde{\Gamma}_{n,\ell} \pm \tilde{\sigma}_{n,\ell})^2 = (\Gamma^E_{n,\ell} \pm \sigma_{n,\ell})^2 +  {\cal W}^2_{n,\ell}
\label{eq:fwhm_convol2}
\end{equation}
The mode to ridge widening term,  ${\cal W}$, is estimated from the model
\begin{equation} 
{\cal W}^2_{n,\ell} = (\tilde{\Gamma}^{M}_{n,\ell} )^2 - (\Gamma^{E,M}_{n,\ell})^2
\label{eq:estW}
\end{equation}
where $\Gamma^{E,M}$ is the input model's effective mode width, \ie, the mode
width widened by the spectral resolution, and $\tilde{\Gamma}^{M}$ the model's
resulting ridge width.

This can be rewritten as (after dropping the $n,\ell$ subscripts):
\begin{eqnarray}
\tilde{\Gamma}^2 + \tilde{\sigma}^2 + 2\, \tilde{\Gamma}\ \tilde{\sigma} & = & 
  (\Gamma^E)^2 + \sigma^2 +  2\, \Gamma^E\ \sigma + {\cal W}^2 \\
\tilde{\Gamma}^2 + \tilde{\sigma}^2 - 2\, \tilde{\Gamma}\ \tilde{\sigma} & = & 
  (\Gamma^E)^2 + \sigma^2 -  2\, \Gamma^E\ \sigma + {\cal W}^2
\end{eqnarray}
and simple arithmetic leads to the following two equations:
\begin{eqnarray} 
\tilde{\Gamma}\  \tilde{\sigma}     & = &   \Gamma^E\ \sigma \\
\tilde{\Gamma}^2 + \tilde{\sigma}^2 & = & (\Gamma^E)^2 + \sigma^2 + {\cal W}^2
\end{eqnarray}
from which we derive $\Gamma^E$ and  $\sigma$ as follows:
\begin{eqnarray} 
\sigma & = & \tilde{\sigma} \frac{\tilde{\Gamma}}{\Gamma^E}
\label{eq:estSigmaGamma} \\
\Gamma^E & = & \sqrt{\tilde{\Gamma}^2 - {\cal W}^2 + g\,\tilde{\sigma}^2}
\label{eq:estGammaE} 
\end{eqnarray}
where 
\begin{equation} 
 g = 1 - (\frac{\tilde{\Gamma}}{\Gamma^E})^2
\end{equation}
This correction factor, $g$, depends on $\Gamma^E$ the variable we solve for,
but can be re-evaluated iteratively, starting by setting $g=0$, to solve
Eq.~\ref{eq:estGammaE}. Only 3 to 4 iterations are needed to reach a $10^{-6}$
precision.

Evaluating of $\Gamma^E$ becomes poorly constrained when ${\cal W}^2$ is
commensurable with $\tilde{\Gamma}^2 + g\,\tilde{\sigma}^2$, since the result of
their subtraction must to be a positive quantity. But since $\Gamma^E$ must be
larger than the spectral resolution, $W_{N,T}$, we can force $\Gamma^E$, when
derived from Eq.~\ref{eq:estGammaE}, to always be at least $W_{N,T}$.

\begin{figure}[!t]
\centering
\includegraphics[width=.95\textwidth]{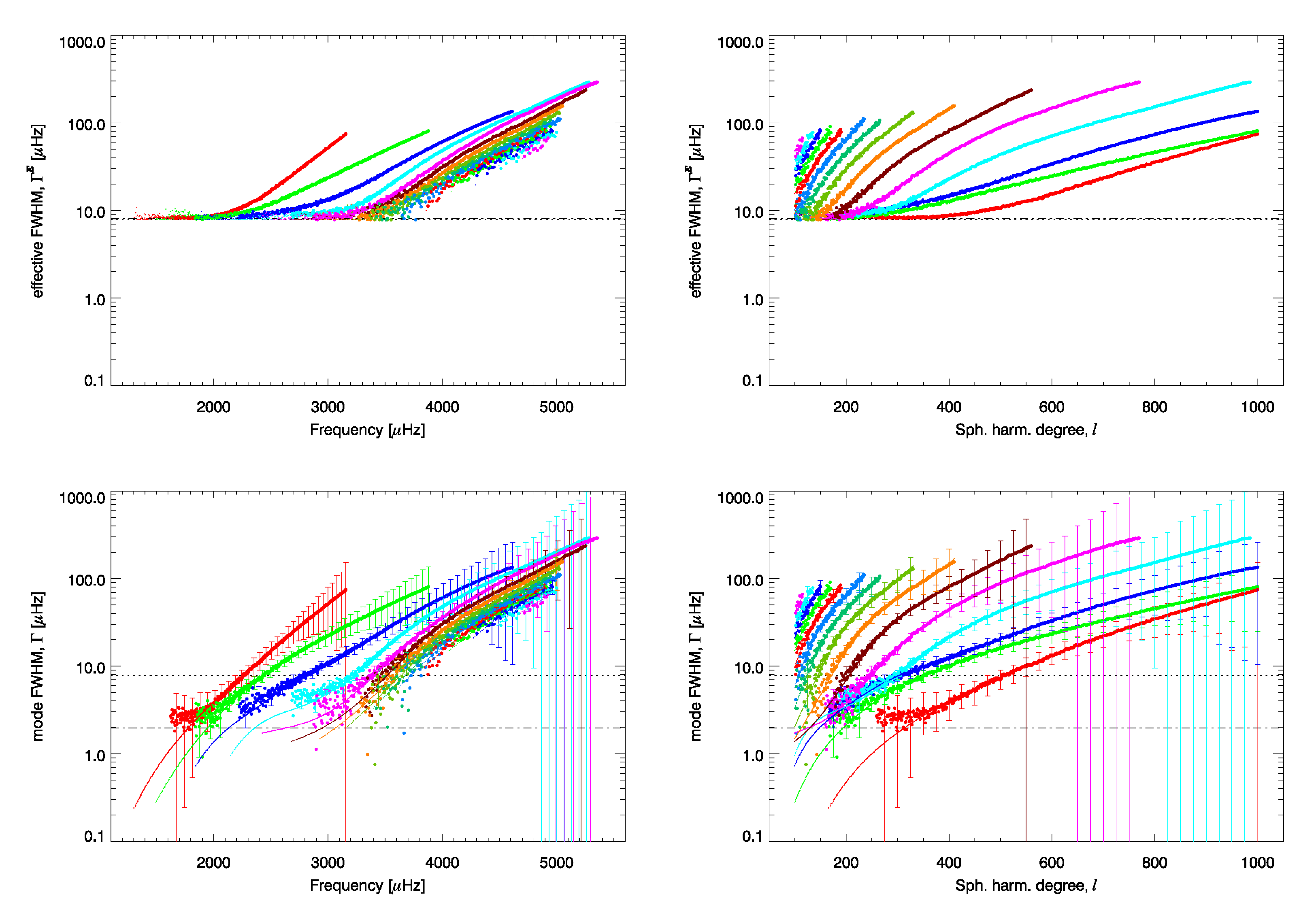}
\caption{Top panels: effective FWHM, estimated from the ridge width, using the
  widening model (see description in text), as a function of frequency or
  degree (left and right panels, respectively). The dots correspond to
  effective widths below the reliability threshold. Bottom panels: mode FWHM,
  estimated from the effective width, using a 1/4 threshold (see explanation
  in text), and their associated uncertainty, as a function of frequency or
  degree (left and right panels, respectively). The dots in these panels are
  the mode widths used in our model of the ridge. 
  The color corresponds to the value of the order, $n$.
  The horizontal lines are drawn at the respective reliability limits (see
  text for explanation).
\label{fig:correctFWHM}}
\end{figure}

An estimate of the mode FWHM is then derived from the effective width by
solving Eq.~\ref{eq:gammaE}, hence
\begin{equation} 
\Gamma  = \sqrt{{\Gamma^E}^2 - W_{N,T}^2}
\label{eq:estGamma}
\end{equation}
Results of this equation are guaranteed to be greater or equal to zero if
$\tilde{\Gamma}^E$ is set to always be at least $W_{N,T}$. But estimates of
$\Gamma$ much smaller than $W_{N,T}$, derived from $\Gamma^E$ according to
Eq.~\ref{eq:estGamma}, are in practice meaningless. To alleviate this, we
introduce a reliability threshold, $f_r$, as to only infer $\Gamma$ from
$\Gamma^E$ when $\Gamma \ge W_{N,T}\,f_r$.  We used $f_r=1/4$, namely we
derived an estimate of $\Gamma$ using Eq.~\ref{eq:estGamma} only when
$\Gamma^E \ge 1.0308\, W_{N,T}$.  This means that for a given $n$ the mode
width correction scheme cannot always be extended to the lowest fitted degrees.
The results of this correction scheme for the effective width and the mode
width are plotted for zonal modes in Fig.~\ref{fig:correctFWHM}.

\subsubsection{Asymmetry}

\begin{figure}[!t]
\centering
\includegraphics[width=.95\textwidth]{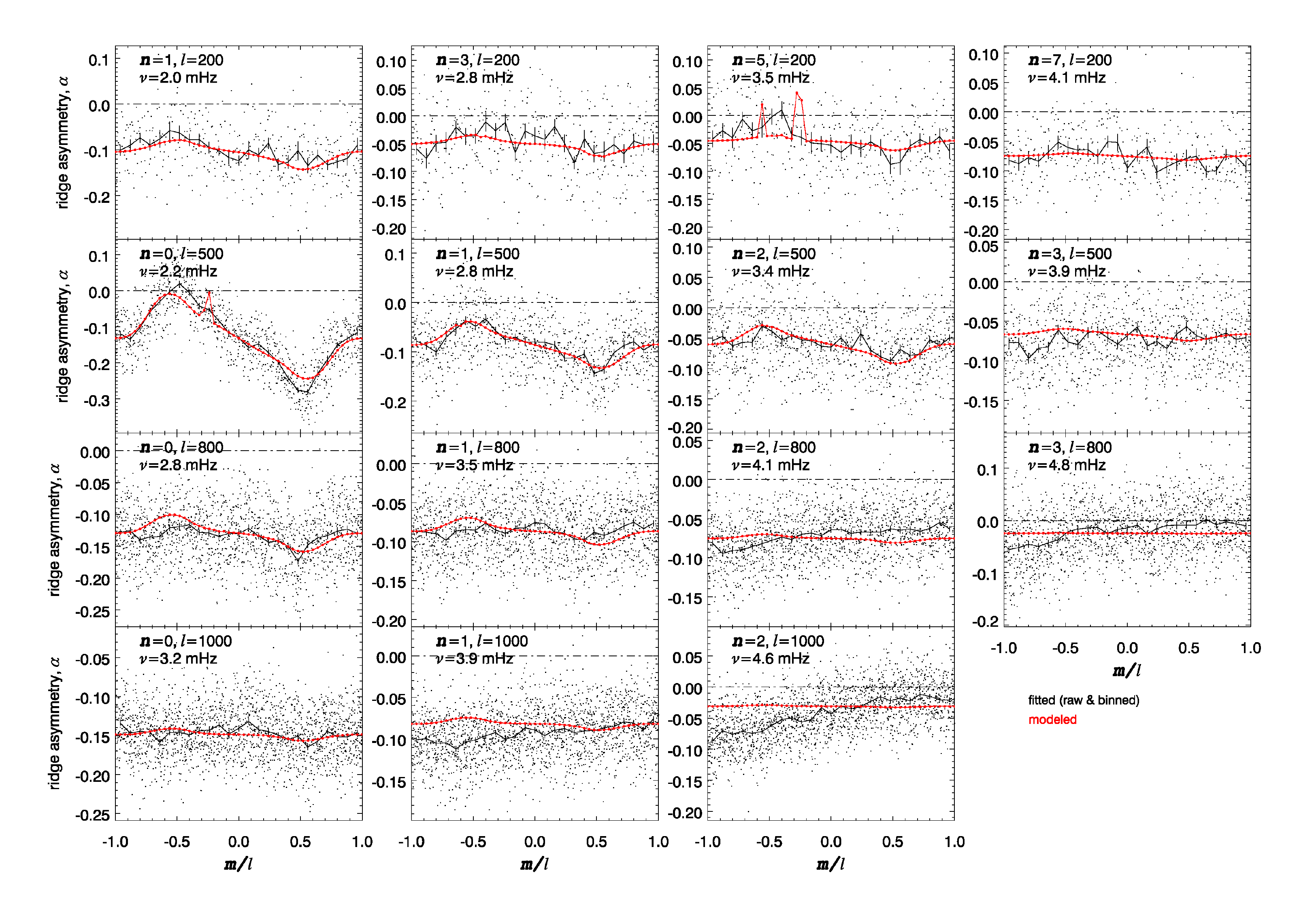}
\caption{Comparison of ridge asymmetry, for a selection of multiplets (same as
in Fig.~\ref{fig:compareFWHMvsMOL}), as a function of the ratio $m/\ell$. The
dots correspond to individual ridge asymmetry measurements, and the black
curves the corresponding binned values. The red curves correspond to our best
model of the ridge asymmetry. Our model reproduces some of the variation of
ridge asymmetry with the azimuthal order, $m$, although the mode asymmetry in
the model is set to be constant with $m$.
\label{fig:compareAlphavsMOL}}
\end{figure}

  Figure~\ref{fig:compareAlphavsMOL} shows the measured ridge asymmetry as a
function of $\frac{m}{\ell}$, for a selection of multiplets (the same set as
in Fig.~\ref{fig:compareFWHMvsMOL}). The figure shows that our model
reproduces the observed variation of the ridge asymmetry with $m$, although
the mode asymmetry in the model is constant with $m$. For the selected modes
in that figure, the variation of the asymmetry with $m$ is large and well
modeled at $\ell=500$ for $n=0$ and $n=1$.  The change of that variation with
$n$ and $\ell$ is properly modeled, but our model fails to reproduce in detail
the observed variation at all $n$ and $\ell$, especially the asymmetry of the
observed variation with respect to $m$.

\begin{figure}[!t]
\centering
\includegraphics[width=.95\textwidth]{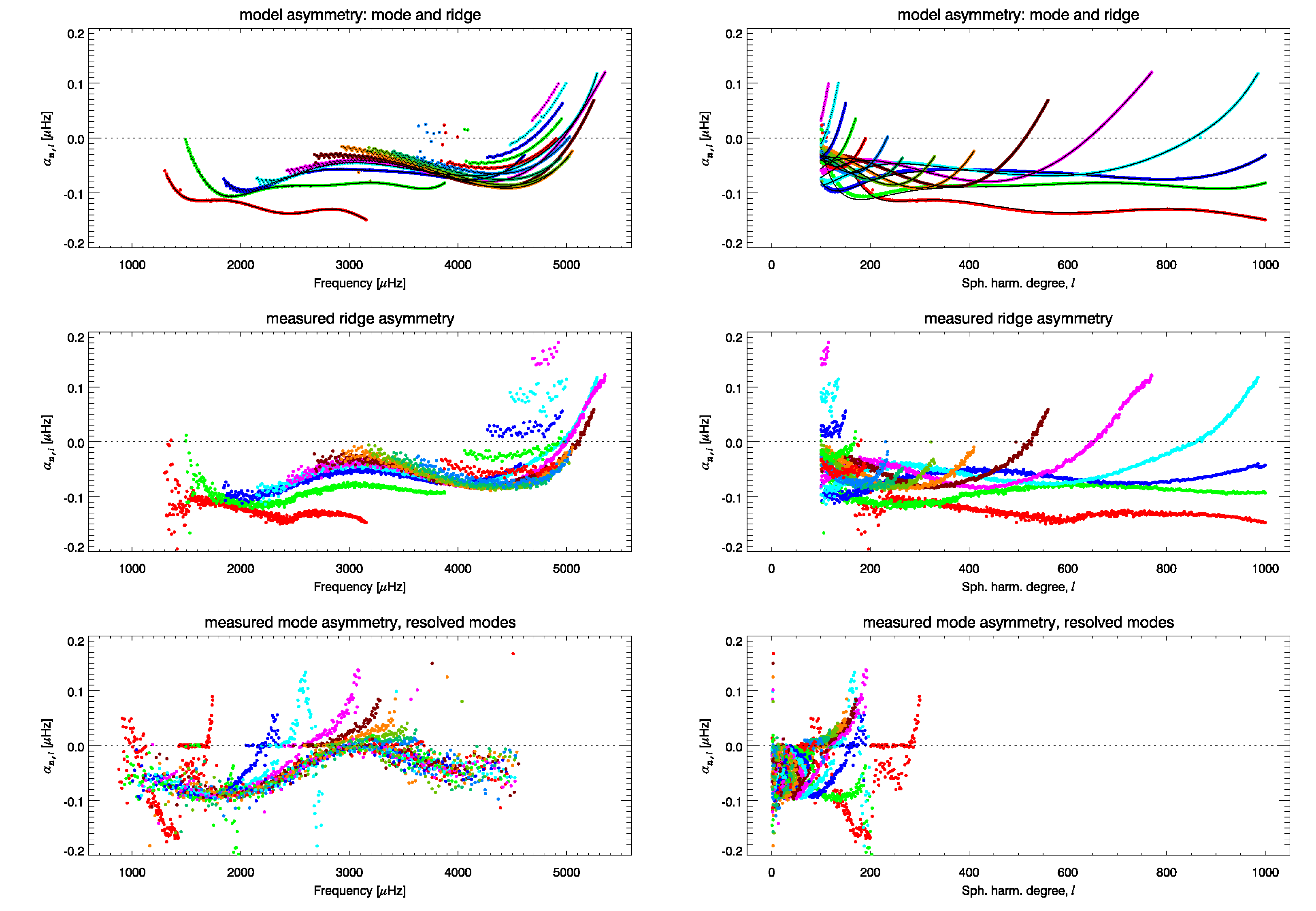}
\caption{%
Ridge asymmetry, for zonal modes only, as a function of frequency or degree,
left and right panels, respectively.
The top panels show the model asymmetry: the input mode asymmetry (black dots)
and the resulting ridge asymmetry (colored dots). Our model predicts that, for
the zonal modes, the asymmetry of the ridge is essentially the mode asymmetry.
The middle panels show the measured ridge asymmetry for intermediate and high
degrees. The model asymmetry produced by our iterative process matches (by
construction) the observed ridge asymmetry.
The bottom panels show the mode asymmetry as measured from fitting resolved
modes at low and intermediate degrees. It shows a qualitative similarity -- a
similar dependence of $\alpha$ on $\nu$ -- but a poor quantitative agreement.
The color corresponds to the value of the order, $n$.
\label{fig:compareAlpha}}
\end{figure}

  Figure~\ref{fig:compareAlpha} shows, for zonal modes, the ridge asymmetry as
predicted by our model as well as the measured ridge asymmetry. These two sets
match, as a direct result of the iterative process on the input values. The
figure also shows that the ridge asymmetry of the zonal modes is a good
estimate of the mode asymmetry.

\begin{figure}[!t]
\centering
\includegraphics[width=.95\textwidth]{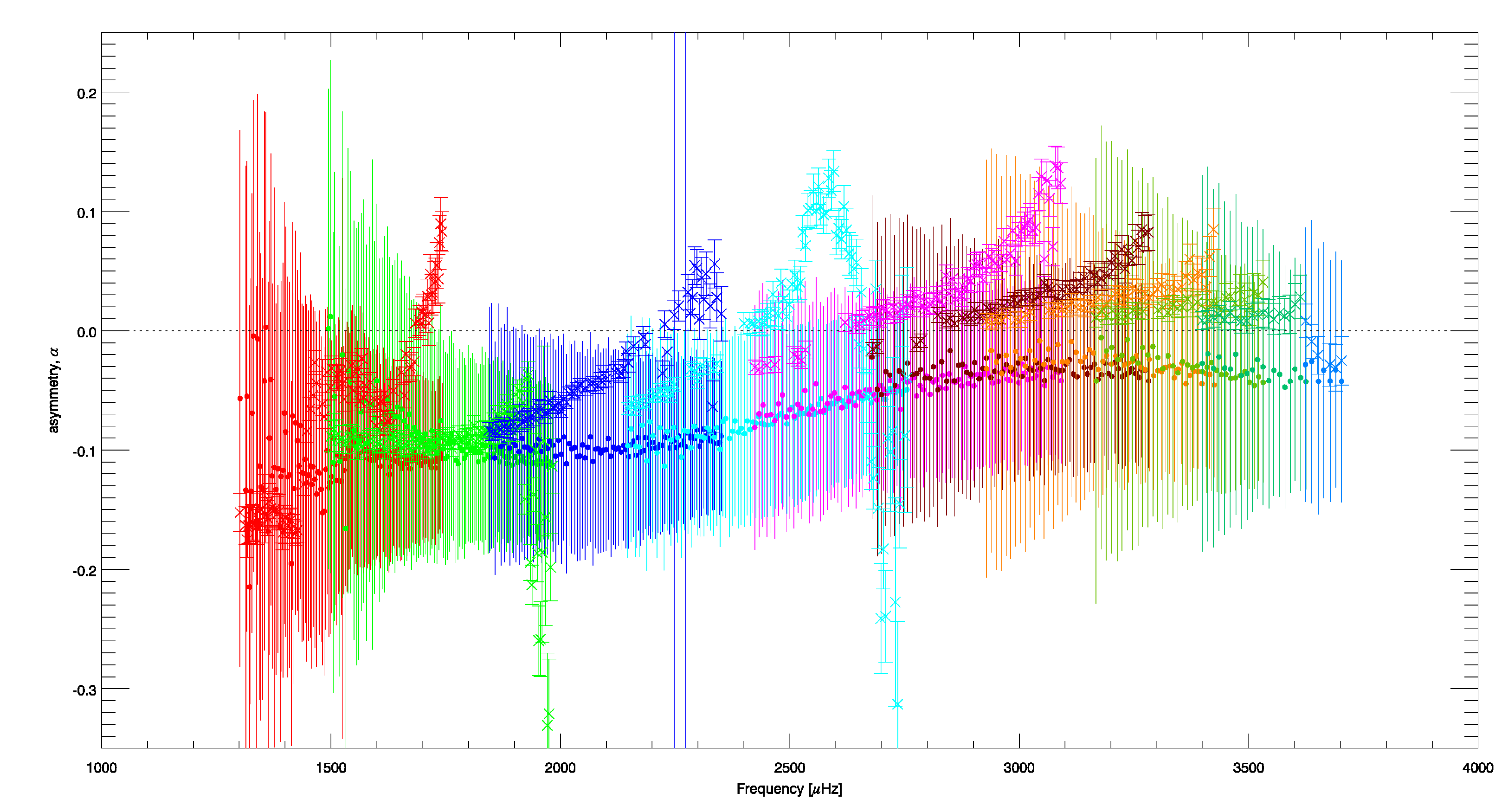}
\caption{Comparison between mode asymmetry determined from fitting resolved
modes at low and intermediate degrees (crosses and error bars with serifs) and
mode asymmetry estimated from ridge fitting at high degrees (dots), for the
overlapping region (see Fig.~\ref{fig:compareFrequency}).  The color
corresponds to the value of the order, $n$. Note how, but for some f-modes and
most of the $n=1$ modes, the ridge fitting estimates are systematically lower
than the individual mode fitting values, although that offset is mostly within
$1\,\sigma$.
\label{fig:compareAlphaX}}
\end{figure}

 The comparison of the asymmetry, as measured from resolved mode fitting at
low and intermediate degrees\footnote{These are fitting results using the
methodology developed by Korzennik ({\em in prep.})  and a very long time
series (12.5 year) of MDI observations.} shows a qualitative similarity -- a
similar dependence of $\alpha$ on $\nu$. The direct comparison of the
asymmetry for the overlapping fitting range between high degree and resolved
modes is shown in Fig.~\ref{fig:compareAlphaX}. The estimates derived from
ridge fitting are systematically lower than the estimates derived from fitting
resolved modes, although the differences are mostly within the uncertainties
on the ridge fitting asymmetry. The mean difference is $0.059 \pm 0.049$, the
mean scaled difference (difference divided by the respective uncertainty) is
$0.68 \pm 0.64$.
Note that the asymmetry is a somewhat tricky parameter to fit. It has a strong
cross-talk with the local slope of the background (\ie, the local background
asymmetry). As a result, the initial guess of the asymmetry can affect the
resulting fitted value. This explains the few discontinuties seen in the model
asymmetries in some panels of Fig.~\ref{fig:compareAlpha} and the set of
values clustered around zero in the bottom panel of
Fig.~\ref{fig:compareAlphaX} (asymmetry estimates of resolved modes).

\subsubsection{Amplitude}
\label{sec:amp}

\begin{figure}[!t]
\centering
\includegraphics[width=.9\textwidth]{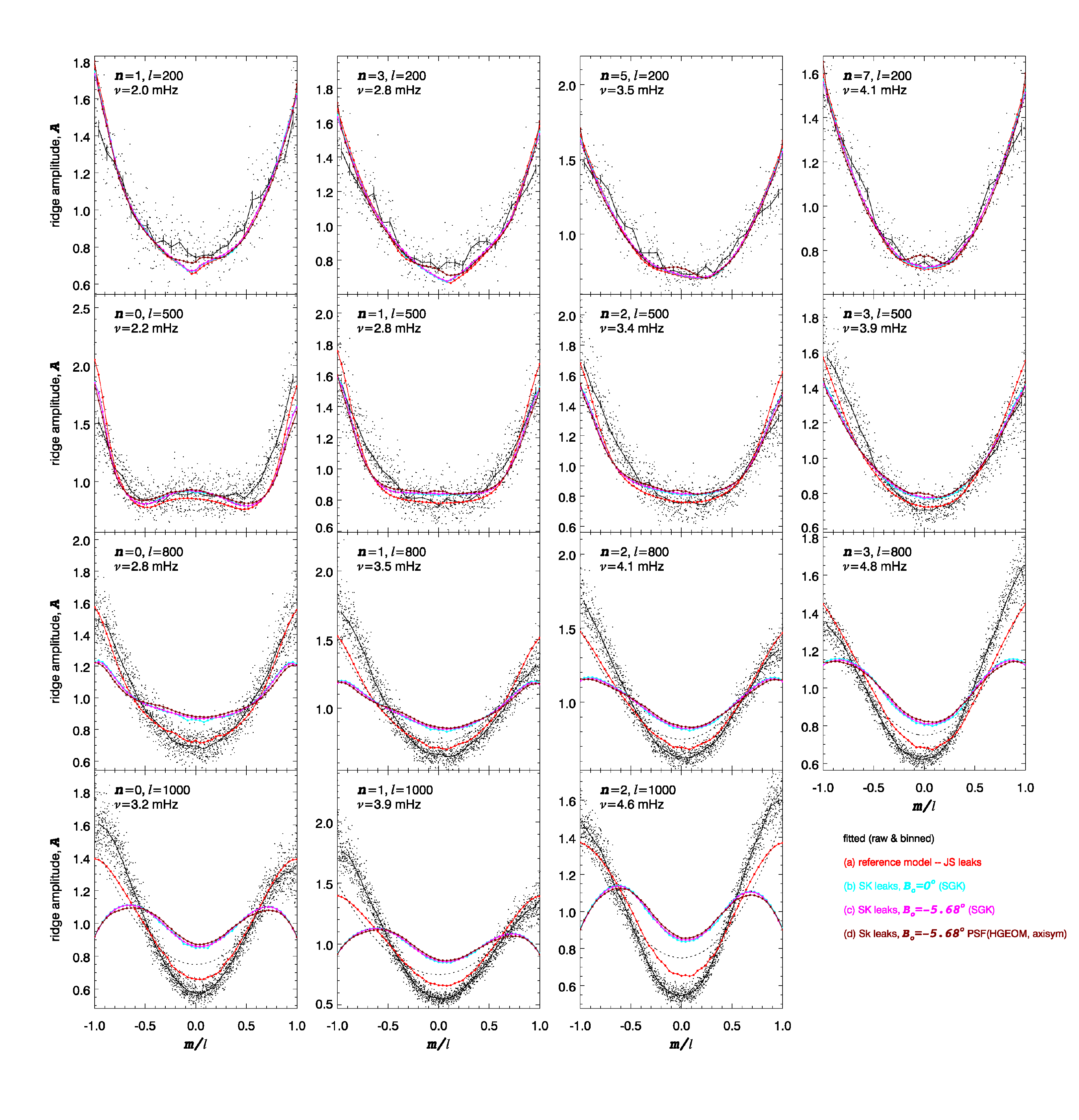}
\caption{Comparison of ridge amplitude, for a selection of multiplets (as in
Figs.~\ref{fig:compareFWHMvsMOL} and \ref{fig:compareAlphavsMOL}), as a
function of the ratio $m/\ell$. The dots correspond to individual ridge power
amplitude measurements, and the black curves the corresponding binned
values. The colored  curves correspond to different models of the ridge
amplitude, resulting from different models for the leakage matrix.
The dotted line corresponds to $1+(\frac{m}{\ell})^2$.
As for the asymmetry, the models reproduce over all the variation of ridge
amplitude with the azimuthal order, while the mode amplitude in the model is
constant with $m$. The reference model shows the best match with the
observations and the discrepancies between the various model appear at high
degrees and high frequencies. Note also how all of the models fail to
reproduce the observed asymmetry of the ridge power amplitude.
\label{fig:compareAmpXvsMOL}}
\end{figure}

Figure~\ref{fig:compareAmpXvsMOL} shows how our model reproduces the observed
variation of the ridge amplitude with azimuthal order, although the mode
amplitude is constant with $m$. The dominant variation, close to
$1+(\frac{m}{\ell})^2$, is well reproduced, but the observed values show a
departure from the reference model that gets progressively larger at higher
degrees. The model also fails to reproduce the asymmetry of the variation of
the ridge power amplitude with respect to the azimuthal order, $m$.

\begin{figure}[!t]
\centering
\includegraphics[width=.95\textwidth]{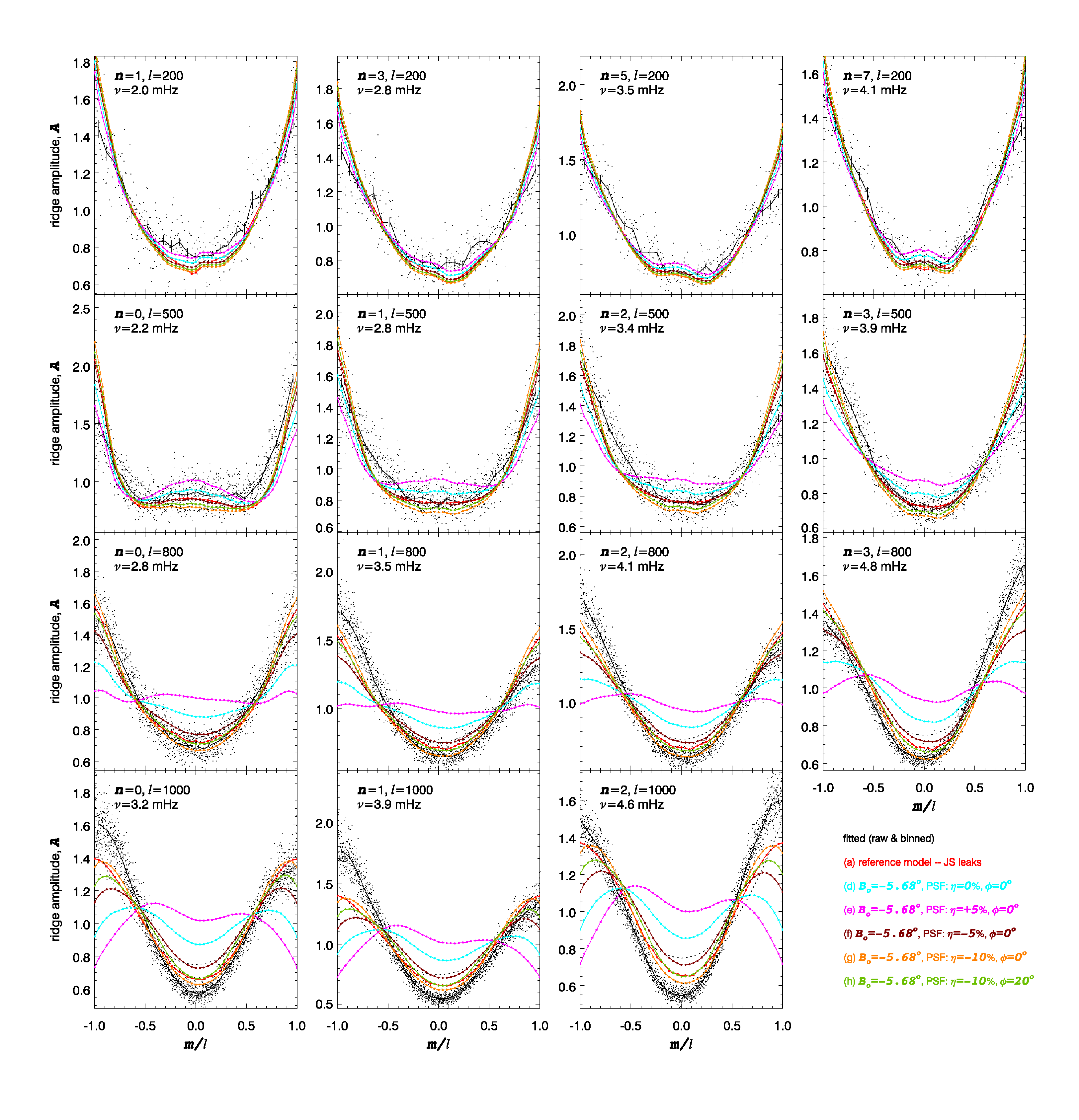}
\caption{Comparison of scaled ridge amplitude, as a function of the ratio
$m/\ell$, scaled as in Fig.~\ref{fig:compareAmpXvsMOL}, for models using
different leakage matrices (see text for the description of the models). The
dots correspond to individual ridge power amplitude measurements, while the
black curves are the corresponding binned values. The colored curves
correspond to different models.
\label{fig:compareAmpYvsMOL}}
\end{figure}

That figure also illustrates the dependence of the observed variation of the
ridge amplitude with $m$ on the leakage matrix used in the modeling. Results
from using an independent leakage matrix computation by one of us (SK) are
also shown in that figure. Figure~\ref{fig:compareAmpYvsMOL} shows additional
models, where by including various PSFs, we explored the dependence of the
ridge amplitude variation with $m$ on the PSF included in the leakage matrix
computation.

We computed a set of models using leakage matrices evaluated by including
various PSFs. We used the PSF profile determined by {\tt HGEOM} but instead of
using an axisymmetric PSF, we derived an elliptic PSF, $\cal F$, with
\begin{equation}
  {\cal F}(x,y) = F(r)
\end{equation}
where $F(r)$ is the PSF profile determined by {\tt HGEOM} and where the
mapping between $r$, the radius, and the position $(x,y)$ is a rotated
ellipse:
\begin{equation}
r = \sqrt{(\frac{x'}{a})^2 + (\frac{y'}{b})^2}
\end{equation}
and
\begin{eqnarray}
  x' &=& x \cos \phi - y \sin \phi \\
  y' &=& x \sin \phi + y \cos \phi
\end{eqnarray}
while $a=1-\eta$ and $b=1+\eta$.
 Figures~\ref{fig:compareAmpXvsMOL} and~\ref{fig:compareAmpYvsMOL} present the
variation of the ridge power amplitude with $m$ resulting from the following 8
cases:
\begin{list}{(\alph{mycntr})}%
{\usecounter{mycntr}\setlength{\topsep}{0pt}\setlength{\itemsep}{0pt}}
\item Standard leakage matrix (computed by JS)
\item Independent leakage matrix calculation, for $B_o=0^{\rm o}$ (computed by  SK) 
\item Independent leakage matrix calculation, for $B_o=-5.68^{\rm o}$
  (computed by SK) 
\item As above, plus axisymmetric PSF ($\eta=0,\ \phi=0^{\rm o}$)
\item As (c), plus elliptical PSF $\eta=+0.05,\ \phi=0^{\rm o}$ 
\item As (c), plus elliptical PSF $\eta=-0.05,\ \phi=0^{\rm o}$ 
\item As (c), plus elliptical PSF $\eta=-0.10,\ \phi=0^{\rm o}$ 
\item As (c), plus elliptical PSF $\eta=-0.10,\ \phi=20^{\rm o}$
\end{list}
with the first two cases repeated in both figures.

While in none of these models does the amplitude variation with $m$ match the
observations, the figures illustrate how the precise variation of $\tilde{A}$
with $m$ at various $\ell$ is quite sensitive to the PSF and suggest that the
PSF of MDI is likely to be non-axisymmetrical.

\begin{figure}[!t]
\centering
\includegraphics[width=.95\textwidth]{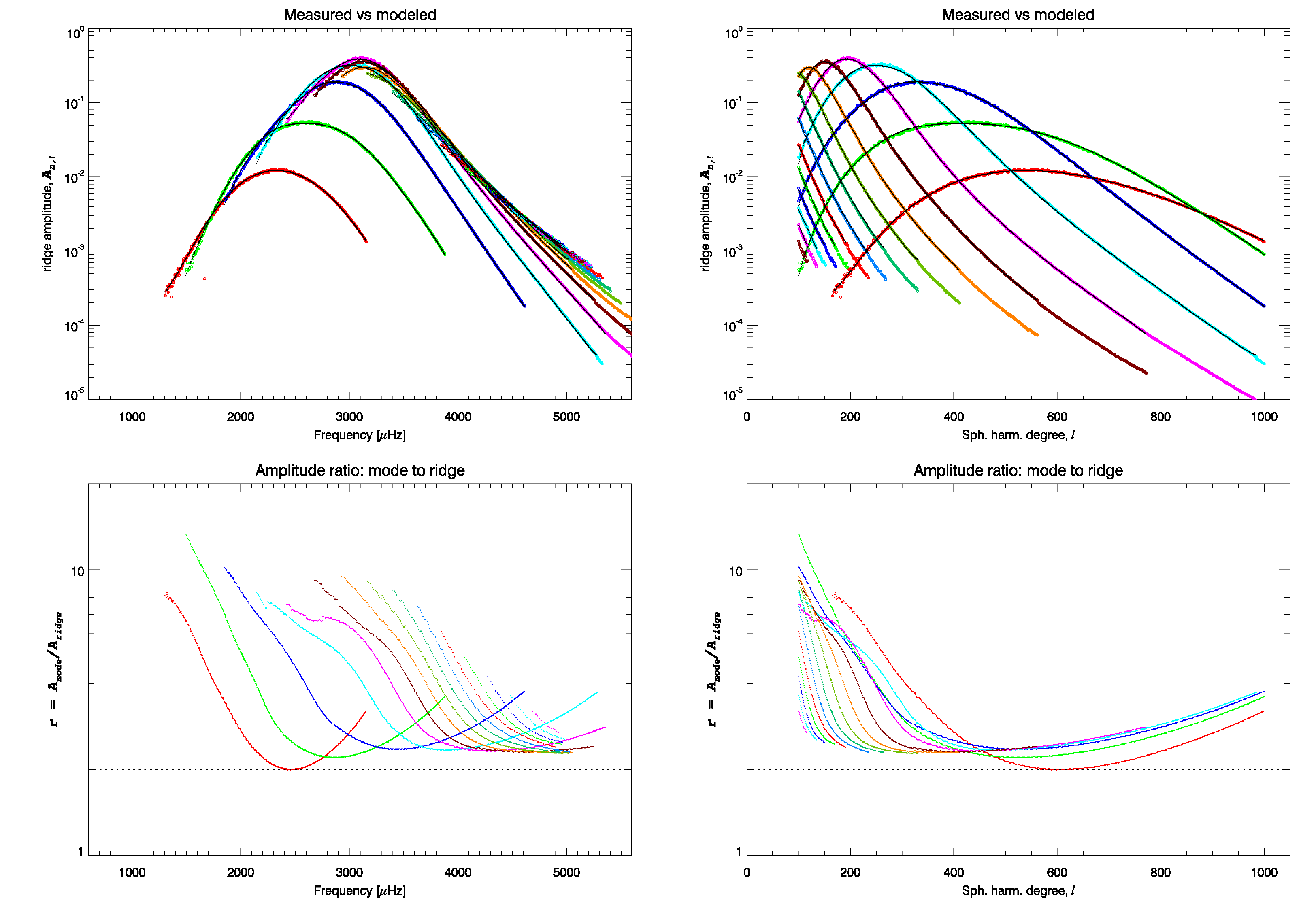}
\caption{Top panels: comparison of the ridge power amplitude, as a function of
frequency (panels on the left) or degree (panels on the right), for zonal
modes. The model values (black dots) are compared to the measured ridge
amplitude (colored dots). These agree rather well -- by construction -- since
the model input mode amplitudes were adjusted iteratively to achieve such a
match.
The color corresponds to the value of the order, $n$.
Bottom panels: ratio between the mode amplitude and the ridge amplitude
estimated from our ridge model.
\label{fig:compareAmp}}
\end{figure}

  A comparison of the zonal amplitudes is shown in Fig.~\ref{fig:compareAmp},
where we show the measured zonal ridge power amplitude and compare it to the
ridge amplitude predicted by our model. The ridge zonal amplitudes agree
rather well -- by construction -- since the model input mode amplitudes were
adjusted iteratively to achieve such a match. The figure also shows the ratio
between the mode amplitude and the ridge amplitude. While below $\ell=300$ the
ridge to mode amplitude ratio is a complex function of order and degree, at
the higher degrees, that ratio is mostly a function of $\ell$. A reasonable
estimate of the mode amplitude can thus be derived using the ratio predicted
by our model.

\subsection{Final Correction}

  The steps used to estimate mode parameters from ridge parameters are
summarized as follow. First, for all the singlets we corrected the ridge
frequency, by subtracting the ridge to mode offset determined by our model,
and propagated the correction uncertainty to the fitting uncertainty. Namely
we use Eqs.~\ref{eq:corrFreq} and \ref{eq:errFreq}:
\begin{eqnarray}
\nu_{n,\ell,m}            & = & \tilde{\nu}_{n,\ell,m} - \Delta^\nu_{n,\ell,m}  \nonumber \\
\sigma^2_{\nu_{n,\ell,m}}    & = & \sigma^2_{\tilde{\nu}_{n,\ell,m}} +\sigma^2_{\Delta^\nu_{n,\ell,m}}\nonumber
\end{eqnarray}
where $\Delta^\nu_{n,\ell,m}$ is defined in Eq.~\ref{eq:defDeltaFreq}
and
where $\sigma_{\Delta^{\nu}_{n,\ell,m}}$ is taken to be 1\% or 0.1\% of $\Delta^\nu_{n,\ell,m}$.
The zonal frequency for a given multiplet, $\nu_{n,\ell}$, is computed as
usual, by fitting Clebsh-Gordan coefficients to the corrected singlets
frequencies, $\nu_{n,\ell,m}$.

For the FWHM, asymmetry and amplitude, singlets values for a given multiplet
are fitted to a polynomial in $m/\ell$. The zonal value is then estimated as
the value at $m=0$ of the fitted polynomial, with the uncertainty associated
to the fit.
The resulting zonal estimates for the ridge FWHM, asymmetry and amplitude are
then corrected to produce zonal mode estimates.

These corrections are:
\begin{eqnarray}
\Gamma^2_{n,\ell}          & = & (\Gamma^E_{n,\ell})^2 - W_{N,T}^2  \nonumber\\
     & = & \tilde{\Gamma}_{n,\ell}^2 + g\,\tilde{\sigma}_{n,\ell}^2 - {\cal W}_{n,\ell}^2 - W_{N,T}^2 \nonumber\\
\sigma_{\Gamma_{n,\ell}}   & = & \sigma_{\tilde{\Gamma}_{n,\ell}}  \tilde{\Gamma}_{n,\ell} /{\cal W}_{n,\ell} \nonumber\\
\alpha_{n,\ell}            & = & \tilde{\alpha}_{n,\ell} - \Delta^\alpha_{n,\ell} \nonumber\\
\sigma^2_{\alpha_{n,\ell}} & = & \tilde{\sigma}^2_{\alpha_{n,\ell}} + \sigma^2_{\Delta^\alpha_{n,\ell,m}} \nonumber\\
A_{n,\ell}                 & = & \tilde{A}_{n,\ell} / r^A_{n,\ell} \nonumber\\
\sigma_{A_{n,\ell}}        & = & \tilde{\sigma}_{r^A_{n,\ell}} / r^A_{n,\ell} \nonumber
\end{eqnarray}
following Eqs.~\ref{eq:estW}, \ref{eq:estGammaE}, \ref{eq:estSigmaGamma} and
\ref{eq:estGamma}, where $\Delta^\alpha_{n,\ell}$ is the small difference
between ridge and mode asymmetry predicted by the model of the ridge. The
precision on the asymmetry correction, $\sigma_{\Delta^\alpha_{n,\ell,m}}$, is
negligible and thus is neglected, as the correction itself is minute.
The quantity $r^A_{n,\ell}$ is the ratio between the ridge and mode amplitude
in the model.

  These corrections have been carried out using models computed with different
leakage matrices and different mode asymmetries. We computed various models
since on the one hand, we were unable to produce a good model of the ridge
amplitude variation with $m$ that matches the observations, and because on the
other hand our estimate of the asymmetry based on ridge fitting does not match
the estimates resulting from fitting resolved modes at low and intermediate
degrees.

  We considered five leakage matrices and two asymmetry models. For the
asymmetry, we either allowed it be a bivariate polynomial of frequency and
degree, $f_1(\nu,\ell)$, and adjusted that function iteratively to match the
ridge fitting at intermediate and high degree (as described earlier), or we
fixed it to be solely a function of frequency, $f_2(\nu)$, set to a polynomial
that matches the estimate of the asymmetry resulting from mode fitting at low
and intermediate degrees. As for the five leakage matrices, we considered the
five cases (a), (b), (c), (d) and (h) described in section~\ref{sec:amp}.

\begin{table}[!t]
\caption{Models Properties: comparison of various models ridge FWHM, amplitude
         and asymmetry to the corresponding observed values (for $m=0$).  The
         tabulated values are the average and standard deviation of either
         ratios or differences.} 
 ~ \\
\centering
\begin{tabular}{ll|ccc|}
  Leakage  & Asymmetry    &                                       &                            &                     \\
  matrix   & type         &  $\tilde{\Gamma}_M/\tilde{\Gamma}_O$  & $\tilde{A}_M/\tilde{A}_O$  & $\tilde{\alpha}_M-\tilde{\alpha}_O$ \\
\hline
      &                 &                    &                      &                    \\
  (a) & $f_1(\nu,\ell)$ & 0.999 $\pm$ 0.015  & 1.001 $\pm$ 0.018    & 0.000 $\pm$ 0.010  \\
  (b) & $f_1(\nu,\ell)$ & 0.999 $\pm$ 0.015  & 0.993 $\pm$ 0.020    & 0.000 $\pm$ 0.010  \\
  (c) & $f_1(\nu,\ell)$ & 1.000 $\pm$ 0.015  & 0.999 $\pm$ 0.019    & 0.000 $\pm$ 0.010  \\
  (d) & $f_1(\nu,\ell)$ & 1.012 $\pm$ 0.021  & 1.000 $\pm$ 0.731    & 0.000 $\pm$ 0.009  \\
  (h) & $f_1(\nu,\ell)$ & 1.013 $\pm$ 0.022  & 1.001 $\pm$ 0.781    & 0.000 $\pm$ 0.009  \\
      &                 &                    &                      &                    \\
  (a) & $f_2(\nu)$      & 0.999 $\pm$ 0.015  & 1.001 $\pm$ 0.018    & 0.049 $\pm$ 0.052  \\
  (b) & $f_2(\nu)$      & 0.999 $\pm$ 0.015  & 0.992 $\pm$ 0.019    & 0.049 $\pm$ 0.052  \\
  (c) & $f_2(\nu)$      & 1.000 $\pm$ 0.015  & 0.998 $\pm$ 0.019    & 0.049 $\pm$ 0.052  \\
  (d) & $f_2(\nu)$      & 1.013 $\pm$ 0.021  & 0.998 $\pm$ 0.728    & 0.048 $\pm$ 0.052  \\
  (h) & $f_2(\nu)$      & 1.014 $\pm$ 0.021  & 0.999 $\pm$ 0.778    & 0.048 $\pm$ 0.053  \\
  \\
\end{tabular}
\label{tab:modelComp}
\end{table}

\begin{table}[!t]
\caption{Frequency match, singlets: comparison of mode frequencies, at
  intermediate degree for individual singlets ($\nu_{n,\ell,m}$), determined
  either by fitting resolved modes or by fitting ridges and correcting for the
  ridge-to-model offset estimated for various models. The mean and RMS,
  after a $3\,\sigma$ rejection, are tabulated for raw and scaled differences,
  as well as the number of overlapping modes and the number of modes kept
  after the $3\,\sigma$ rejection.}
 ~ \\
\centering
\begin{tabular}{ll|cc|cc|}
  Leakage  & Asymmetry    &  $\Delta\nu_{n,\ell,m}$ & & $\frac{\Delta\nu_{n,\ell,m}}{\sigma_{\nu_{n,\ell,m}}}$ & \\
  matrix   & type         &  [\uHz] & ($N, N_k$) & & ($N, N_k$)\\
\hline
(a) & $f_1(\nu,\ell)$ &  0.098 $\pm$ 1.656 & (73858, 72498) & 0.135 $\pm$ 1.409 & (73858, 71740) \\
(b) & $f_1(\nu,\ell)$ &  0.094 $\pm$ 1.656 & (73862, 72503) & 0.130 $\pm$ 1.408 & (73862, 71739) \\
(c) & $f_1(\nu,\ell)$ &  0.093 $\pm$ 1.654 & (73864, 72499) & 0.131 $\pm$ 1.406 & (73864, 71737) \\
(d) & $f_1(\nu,\ell)$ &  0.118 $\pm$ 1.660 & (73870, 72535) & 0.151 $\pm$ 1.410 & (73870, 71748) \\
(h) & $f_1(\nu,\ell)$ &  0.118 $\pm$ 1.659 & (73868, 72529) & 0.151 $\pm$ 1.410 & (73868, 71743) \\
    &                 &                    &                &                   &                \\
(a) & $f_2(\nu)$      &  0.086 $\pm$ 1.655 & (73860, 72495) & 0.123 $\pm$ 1.408 & (73860, 71743) \\
(b) & $f_2(\nu)$      &  0.083 $\pm$ 1.654 & (73860, 72490) & 0.119 $\pm$ 1.407 & (73860, 71739) \\
(c) & $f_2(\nu)$      &  0.096 $\pm$ 1.739 & (65602, 64473) & 0.141 $\pm$ 1.387 & (65602, 63630) \\
(d) & $f_2(\nu)$      &  0.108 $\pm$ 1.658 & (73869, 72528) & 0.143 $\pm$ 1.408 & (73869, 71741) \\
(h) & $f_2(\nu)$      &  0.108 $\pm$ 1.658 & (73869, 72526) & 0.144 $\pm$ 1.408 & (73869, 71739) \\
\end{tabular}
\label{tab:modelCompS}
\end{table}

\begin{table}[!t]
\caption{Frequency match, multiplets: comparison of mode frequencies,
  as in Table~\ref{tab:modelCompS}, but for multiplets ($\nu_{n,\ell}$).}
 ~ \\
\centering
\begin{tabular}{ll|cc|cc|}
  Leakage  & Asymmetry    &  $\Delta\nu_{n,\ell}$ & & $\frac{\Delta\nu_{n,\ell}}{\sigma_{\nu_{n,\ell}}}$ & \\
  matrix   & type         &  [\uHz] & ($N, N_k$) & & ($N, N_k$)\\
\hline
(a) & $f_1(\nu,\ell)$ & 0.152 $\pm$ 0.543 & (678, 636) & 0.806 $\pm$ 2.216 & (678, 649) \\
(b) & $f_1(\nu,\ell)$ & 0.136 $\pm$ 0.543 & (678, 636) & 0.739 $\pm$ 2.210 & (678, 649) \\
(c) & $f_1(\nu,\ell)$ & 0.137 $\pm$ 0.537 & (678, 635) & 0.731 $\pm$ 2.201 & (678, 649) \\
(d) & $f_1(\nu,\ell)$ & 0.122 $\pm$ 0.571 & (678, 638) & 0.682 $\pm$ 2.312 & (678, 651) \\
(h) & $f_1(\nu,\ell)$ & 0.121 $\pm$ 0.573 & (678, 638) & 0.684 $\pm$ 2.320 & (678, 651) \\
    &                 &                   &           &                    &            \\
(a) & $f_2(\nu)$      & 0.167 $\pm$ 0.542 & (678, 636) & 0.871 $\pm$ 2.215 & (678, 649) \\
(b) & $f_2(\nu)$      & 0.153 $\pm$ 0.542 & (678, 636) & 0.807 $\pm$ 2.205 & (678, 649) \\
(c) & $f_2(\nu)$      & 0.118 $\pm$ 0.541 & (596, 560) & 0.550 $\pm$ 2.087 & (596, 576) \\
(d) & $f_2(\nu)$      & 0.134 $\pm$ 0.568 & (678, 638) & 0.725 $\pm$ 2.293 & (678, 651) \\
(h) & $f_2(\nu)$      & 0.135 $\pm$ 0.569 & (678, 638) & 0.732 $\pm$ 2.296 & (678, 651) \\
\end{tabular}
\label{tab:modelCompM}
\end{table}

 Table~\ref{tab:modelComp} shows for each of these 10 cases some properties of
the resulting models, namely how the widths, amplitudes, and, asymmetries in
our modeled ridges match the observations. The widths and asymmetries agree
with observations as a result of the iterative process on the model input
parameters. By contrast, the modeled asymmetries do not agree with the
observations when they are kept as the preset function of frequency,
$f_2(\nu)$, based on low and intermediate degrees estimates.

Tables~\ref{tab:modelCompS} and \ref{tab:modelCompM} show how well the mode
frequencies estimated from ridge fitting, when corrected using these 10
models, match the values measured by fitting resolved modes, for that same
overlap range between ridge fitting and resolved mode fitting.
The tabulated values are the average and standard deviation of the frequency
differences, and the scaled differences ($\Delta\nu/\sigma_\nu$), after a
$3\,\sigma$ rejection. The number of overlapping modes and the number of these
kept after the $3\,\sigma$ rejection are also listed.

\begin{figure}[!t]
\centering
\includegraphics[width=.95\textwidth]{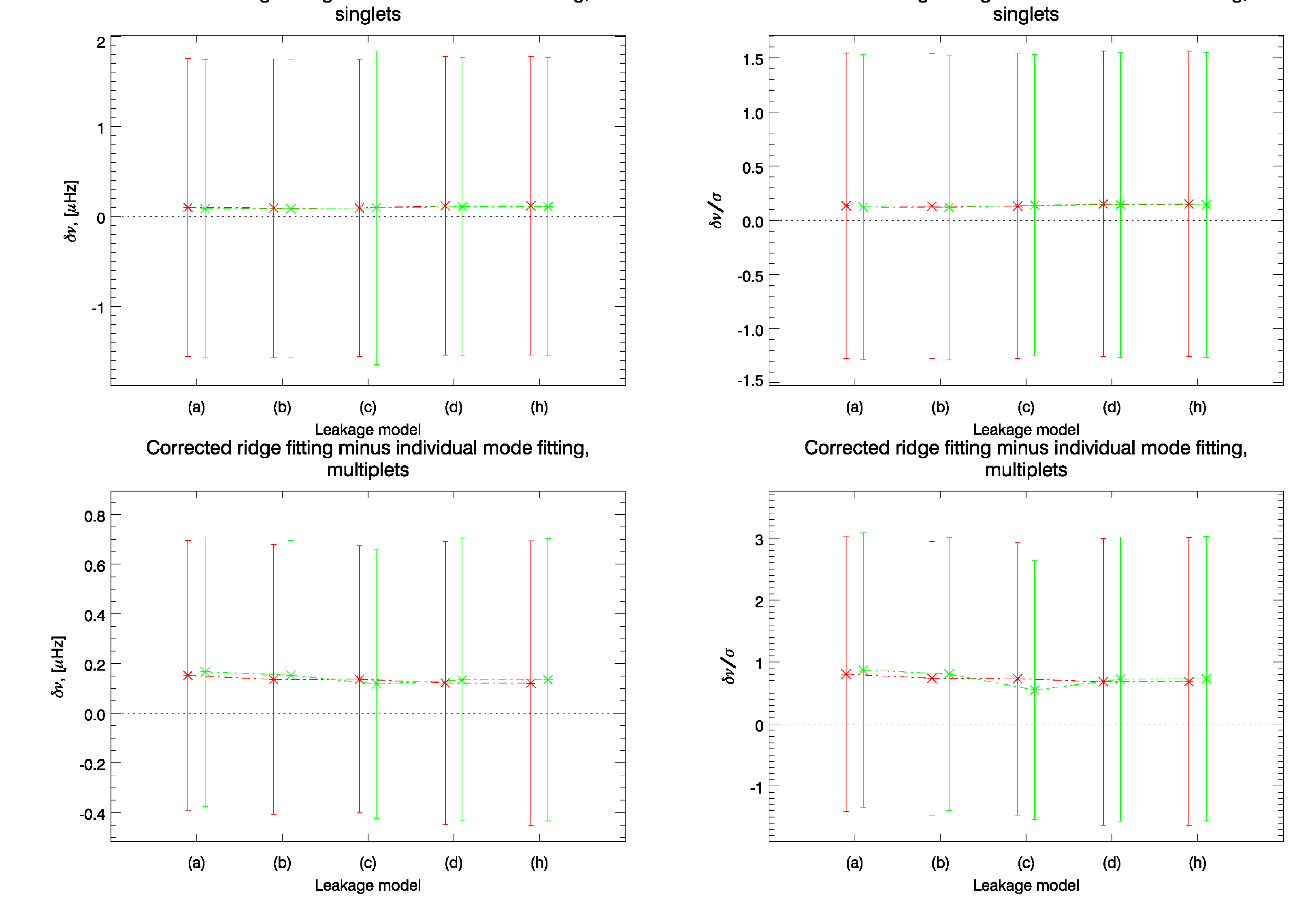}
\caption{Mean and RMS for frequency differences (left panels)
and scaled differences (right panels), between mode frequencies estimated from
ridge fitting and values derived from fitting resolved modes, after correcting
for the ridge-to-mode offset using different models, as presented in
Tables~\ref{tab:modelCompS} and \ref{tab:modelCompM}.
The top panels correspond to singlets, while the bottom panels correspond to
multiplets. The red and green curves are models with the $f_1(\nu,\ell)$ or
$f_2(\nu)$ asymmetry laws, respectively.
No model is significanly better, when using this metric.
\label{fig:modelCompX}}
\end{figure}

Table~\ref{tab:modelCompS} compares singlets, while Table~\ref{tab:modelCompM}
compares multiplets.
These tabulated values are plotted in Fig.~\ref{fig:modelCompX}, where the
error bars on the plot are the tabulated RMS.
These comparisons show that no model is significantly better amongst these ten
when comparing estimate of mode frequencies in the overlapping range
between ridge fitting and resolved mode fitting. 

  For illustration, a fraction of the resulting parameters are tabulated in
the Appendix in Tables~\ref{tab:multiplets} to \ref{tab:cgCoefs}, namely some
singlets $(n,\ell,m)$, multiplets $(n,\ell)$, and corresponding Clebsh-Gordan
coefficients. The complete tables are available in digital form at \\
{\tt 
https://www.cfa.harvard.edu/$\sim$sylvain/research/tables/HiL/}.

\subsection{Remaining Issues}

One of the two major remaining issues is our inability to model the MDI
instrument PSF. This inability results in incorporating an inadequate leakage
matrix, that in turn produces a model of the variation of the ridge amplitude
with $m$ that does not match the observations, especially at the higher
degrees. It may also explain the mismatch of the variation with $m$ of the
ridge width and asymmetry, although at a smaller scale.

Since we have co-eval full disk observations between MDI and HMI, we should be
able to derive the MDI PSF from these data. Indeed, on the one hand, the HMI
instrument optical characteristics have been carefully measured prior to
launch, while on the other hand the higher spatial resolution of HMI will
allow us to characterize precisely the PSF of the MDI instrument even without
knowing very precisely the PSF of the HMI instrument.

The other remaining issue is the mismatch between the estimates of the mode
asymmetry derived from low and intermediate degrees and from intermediate and
high degrees. While the introduction of a better PSF for the MDI instrument
might slightly change the offset between the mode and ridge asymmetry, the
zonal values are unlikely to be much affected. This mismatch is not
significant, since the estimates from ridge fitting have large error bars, but
it is systematic. We speculate that the difficulty in constraining the
background might explain this systematic offset. The background term is likely
not symmetric across the ridge, but it is poorly constrained when fitting at
intermediate and high degrees since there is almost no free spectral range
between the ridges. It is thus conceivable that the measured ridge asymmetry
might be contaminated by the local slope of the unconstrained background
across the ridge.

Other remaining issues are whether our model of the image distortion and its
orientation (the effective $P$ angle) are taken into account at the required
precision at the spatial decomposition step. These may be responsible at some
level for the mismatch between our models and the observations of the
variation of the ridge amplitude with $m$.

\section{Conclusions}\label{sec:Conclusions}

We have successfully derived and implemented a procedure to estimate mode
parameters for high degree modes, using ridge fitting. We are able to derive
not only frequencies, but widths, asymmetries and amplitudes. For a range of
degrees we have overlapping mode determination derived from ridge fitting and
from resolved mode fitting. Thus we could quantitatively assess the precision
of our method and found no residual significant differences.

Our extensive analysis of the precision of our correction scheme shows that
the correction precision is substantially better than the fitting precision,
but for two caveats: one is the residual, although not significant,
discrepancy between the determination of the asymmetry from ridge fitting and
from mode fitting that prevents us from asserting that our model of the mode
asymmetry is well constrained.  The other is the contribution of the
instrumental PSF, through the leakage matrix, that for MDI we are unable to
constrain from the data and for which we do not have a valid pre flight
determination.

In both cases, our inability to constrain these properties results in some
potential remaining systematic errors in the corrected parameters. We are
unable to find an objective metric to select a {\em best} model.
The PSF ``problem'' can, and we hope will, be solved by analyzing co-eval
data acquired by HMI and MDI.

As for implications on other data sets, the optimistic view is to expect that
when analyzing HMI data, an instrument that delivers full resolution full disk
observations at all times, the instrument PSF will be shown to be well
known. Applying our methodology to these data would allow us to confirm this.
For the GONG data, we have to be realistic and recognize that the {\em
effective} PSF is likely to also be a problem, since the contribution of the
atmosphere needs to be taken into consideration, and one has to merge data
from six different instruments.

\acknowledgments
\section*{Acknowledgments}

 The Solar Oscillations Investigation - Michelson Doppler Imager project on
SOHO is supported by NASA grant NAG5--8878 and NAG5--10483 at Stanford
University.  SOHO is a project of international cooperation between ESA and
NASA.
 SGK was supported by Stanford contract PR--6333 and NASA grants NAG5--9819
and NNG05GD58G.

\clearpage 

\appendix
\section*{Appendix: Tables}

This appendix presents a subset of the mode characteristics resulting from
ridge fitting after correcting them according to the procedure described in this
paper.
We show in Table~\ref{tab:singlets} a selection of singlets $(n,\ell,m)$ (\ie,
for 11 equispaced values of $m$, 3 three values of $n$ and for values of
$\ell$ that are mutiples of 100). The blank entries correspond to cases where
the fitting failed. A selection of multiplets $(n,\ell)$ is presented in
Table~\ref{tab:multiplets}, for all $n$ and for values of $\ell$ that are
mutiple of 100; and the corresponding Clebsh-Gordan first three even
coefficients in Table~\ref{tab:cgCoefs}.
The complete tables are available in digital form at  \\
{\tt 
https://www.cfa.harvard.edu/$\sim$sylvain/research/tables/HiL/}.

\input tableSAll

\newpage
\input tableMAll

\newpage
\input tableCGAll

\end{document}

%% file: tableFullDiskEpochs.tex
\begin{table}[!ht]
{\small
\begin{tabular}{ccc|c|l}
 Year & Start & End  & Length & Name \\%
      & \multicolumn{2}{c|}{month/day} & [day]  &      \\\hline
 1996 & 05/23 & 07/24 & 61.3 & {\em Dyn.} I    \\\hline\hline
 1997 & 04/13 & 06/12 & 60.4 & {\em Dyn.} IIa  \\
 1997 & 06/13 & 06/19 &  5.2 & {\em Dyn.} IIb  \\
 1997 & 06/19 & 06/30 & 11.9 & {\em Dyn.} IIc  \\
 1997 & 06/30 & 07/14 & 13.2 & {\em Dyn.} IId  \\\hline
      &       &       & 90.7 &                 \\\hline\hline
 1998 & 01/09 & 03/03 & 52.9 & {\em Dyn.} IIIa \\
 1998 & 03/05 & 04/10 & 36.0 & {\em Dyn.} IIIb \\\hline
      &       &       & 88.9 &                 \\\hline\hline
 1999 & 02/28 & 03/12 & 11.6 & {\em Dyn.} IVa  \\
 1999 & 03/13 & 05/28 & 76.6 & {\em Dyn.} IVb  \\\hline
      &       &       & 88.2 &                 \\\hline\hline
 2000 & 04/02 & 04/24 & 21.5 & {\em Dyn.} Va   \\
 2000 & 05/09 & 07/11 & 62.5 & {\em Dyn.} Vb   \\
 2000 & 08/15 & 08/29 & 13.9 & {\em Dyn.} Vc   \\\hline
      &       &       & 97.9 &                 \\\hline\hline
 2001 & 02/28 & 05/14 & 75.1 & {\em Dyn.} VIa  \\
 2001 & 05/14 & 05/28 & 13.5 & {\em Dyn.} VIb  \\\hline
      &       &       & 88.6 &                 \\\hline\hline
 2002 & 01/10 & 02/05 & 25.4 & {\em Dyn.} VIIa \\
 2002 & 02/14 & 03/09 & 22.2 & {\em Dyn.} VIIb \\
 2002 & 03/18 & 05/22 & 65.3 & {\em Dyn.} VIIc \\
 2002 & 05/24 & 06/03 &  9.3 & {\em Dyn.} VIIe \\\hline
      &       &       &122.2 &                 \\\hline\hline
\end{tabular}%
~~~
\begin{tabular}{ccc|c|l}
 Year & Start & End  & Length & Name \\%
      & \multicolumn{2}{c|}{month/day} & [day]  &      \\\hline
 2003 & 01/18 & 02/16 & 28.3 & {\em Dyn.} VIII \\\hline\hline
 2003 & 10/18 & 11/17 & 29.7 & {\em Dyn.} IX   \\\hline\hline
 2004 & 07/04 & 09/06 & 63.4 & {\em Dyn.} X    \\\hline\hline
 2005 & 06/25 & 08/15 & 51.0 & {\em Dyn.} XIa  \\\hline
 2005 & 08/17 & 08/31 & 13.5 & {\em Dyn.} XIb  \\\hline\hline
 2006 & 03/24 & 05/24 & 60.5 & {\em Dyn.} XII  \\\hline\hline
 2007 & 12/09 & 02/02 & 55.3 & {\em Dyn.} XIIIa \\\hline\hline
 2008 & 03/03 & 05/04 & 62.2 & {\em Dyn.} XIIIb \\\hline\hline
 2009 & 05/18 & 06/18 & 31.0 & {\em Dyn.} XIVa \\
 2009 & 07/02 & 07/12 &  9.7 & {\em Dyn.} XIVb \\
 2009 & 07/14 & 07/20 &  5.4 & {\em Dyn.} XIVc \\\hline
      &       &       & 46.1 &                 \\\hline\hline
 2010 & 05/08 & 06/07 & 30.3 & {\em Dyn.} XVa  \\ 
 2010 & 06/07 & 06/15 &  7.3 & {\em Dyn.} XVb  \\ 
 2010 & 06/15 & 06/15 &  0.6 & {\em Dyn.} XVc  \\ 
 2010 & 06/15 & 06/28 & 13.0 & {\em Dyn.} XVd  \\ 
 2010 & 06/28 & 07/01 &  2.3 & {\em Dyn.} XVe  \\ 
 2010 & 07/01 & 07/01 &  0.6 & {\em Dyn.} XVf  \\ 
 2010 & 07/03 & 07/10 &  6.5 & {\em Dyn.} XVh  \\ 
 2010 & 07/10 & 07/12 &  2.0 & {\em Dyn.} XVi  \\\hline 
      &       &       & 62.6 &                 \\\hline\hline
\multicolumn{5}{c}{~}\\*[2.3em]
\end{tabular}
}
\caption{Epochs when full-disk MDI Dopplergrams are available.
\label{tab:FullDiskEpochs}} 
\end{table}

%% file: tableSAll.tex
\begin{center}
\begin{longtable}{||c|ccc|ccc|ccc||} 
\caption[Selected subset of singlets]{Selected subset of singlets.
\label{tab:singlets}} \\
$m$ & 
 $\nu_{n,\ell,m}$ & $\sigma_{\nu_{n,\ell,m}}$  & $\Delta^{\nu}_{n,\ell,m}$ &
 $\nu_{n,\ell,m}$ & $\sigma_{\nu_{n,\ell,m}}$  & $\Delta^{\nu}_{n,\ell,m}$ &
 $\nu_{n,\ell,m}$ & $\sigma_{\nu_{n,\ell,m}}$  & $\Delta^{\nu}_{n,\ell,m}$
\\ \hline
\endfirsthead
\multicolumn{10}{c}%
{{\bfseries \tablename\ \thetable{} -- continued from previous page}} \\ \hline 
$m$ & 
 $\nu_{n,\ell,m}$ & $\sigma_{\nu_{n,\ell,m}}$  & $\Delta^{\nu}_{n,\ell,m}$ &
 $\nu_{n,\ell,m}$ & $\sigma_{\nu_{n,\ell,m}}$  & $\Delta^{\nu}_{n,\ell,m}$ &
 $\nu_{n,\ell,m}$ & $\sigma_{\nu_{n,\ell,m}}$  & $\Delta^{\nu}_{n,\ell,m}$
\\ \hline
\endhead
\hline \multicolumn{10}{|r|}{{Continued on next page}} \\ \hline
\endfoot
\hline \hline
\endlastfoot
%
%
%
 $\ell=100$
 & \multicolumn{3}{c|}{$n=1$}
 & \multicolumn{3}{c|}{$n=8$}
 & \multicolumn{3}{c||}{$n=14$}
\\
 -100 &       ~~~ &     ~~~ &     ~~~ &   3360.08 &    2.66 &    0.90 &   4665.92 &     ~~~ &    0.37 \\
  -80 &   1463.21 &    0.98 &    3.66 &   3366.17 &    2.22 &    0.67 &   4655.25 &    2.70 &    0.16 \\
  -60 &   1464.67 &    0.66 &    3.27 &   3372.48 &    3.48 &    0.42 &   4651.68 &    4.59 &   -0.36 \\
  -40 &       ~~~ &     ~~~ &     ~~~ &   3382.38 &    2.11 &   -0.82 &   4677.26 &    3.54 &   -0.64 \\
  -20 &   1486.21 &    1.85 &    3.44 &   3391.81 &    2.91 &    0.14 &   4671.52 &    4.06 &   -0.33 \\
    0 &       ~~~ &     ~~~ &     ~~~ &   3396.76 &    3.36 &    1.04 &   4677.05 &    3.73 &    0.41 \\
   20 &   1493.28 &    0.05 &    5.03 &   3405.98 &    2.65 &    1.60 &   4689.71 &    3.92 &    1.18 \\
   40 &       ~~~ &     ~~~ &     ~~~ &   3415.19 &    3.54 &    2.01 &   4696.68 &    3.25 &    1.50 \\
   60 &   1509.91 &    1.81 &    5.31 &   3421.46 &    3.17 &    1.65 &   4706.72 &    3.38 &    1.05 \\
   80 &       ~~~ &     ~~~ &     ~~~ &   3429.23 &    2.31 &    1.09 &   4716.58 &    3.48 &    0.57 \\
  100 &       ~~~ &     ~~~ &     ~~~ &   3438.31 &    3.10 &    0.91 &   4712.56 &    5.67 &    0.38 \\
\hline
 $\ell=200$
 & \multicolumn{3}{c|}{$n=0$}
 & \multicolumn{3}{c|}{$n=5$}
 & \multicolumn{3}{c||}{$n=9$}
\\
 -200 &   1340.61 &    0.23 &    3.17 &   3387.44 &    0.37 &    1.32 &   4628.16 &    2.30 &    1.21 \\
 -160 &   1358.98 &    0.66 &    2.80 &   3409.92 &    0.69 &    0.90 &   4651.74 &    2.73 &    0.82 \\
 -120 &   1377.24 &    1.36 &    1.80 &   3426.47 &    1.66 &   -0.12 &   4671.43 &    1.82 &   -0.11 \\
  -80 &       ~~~ &     ~~~ &     ~~~ &   3441.62 &    0.86 &   -0.82 &   4685.10 &    1.97 &   -0.81 \\
  -40 &   1412.51 &    0.94 &    2.04 &   3459.80 &    0.91 &   -0.14 &   4701.94 &    2.23 &   -0.27 \\
    0 &   1422.94 &    0.53 &    3.64 &   3475.29 &    0.54 &    1.39 &   4719.60 &    4.79 &    1.20 \\
   40 &   1440.79 &    1.42 &    5.23 &   3487.78 &    0.57 &    2.91 &   4731.66 &    1.99 &    2.68 \\
   80 &   1460.71 &    0.69 &    6.09 &   3502.31 &    1.73 &    3.66 &   4749.17 &    2.02 &    3.21 \\
  120 &   1474.25 &    1.32 &    5.72 &   3523.42 &    1.32 &    3.00 &   4767.42 &    2.14 &    2.47 \\
  160 &   1495.95 &    0.54 &    4.61 &   3540.49 &    1.12 &    1.92 &   4787.40 &    3.89 &    1.61 \\
  200 &   1511.24 &    0.31 &    4.06 &   3556.95 &    0.88 &    1.49 &   4799.06 &    1.86 &    1.28 \\
\hline
 $\ell=300$
 & \multicolumn{3}{c|}{$n=0$}
 & \multicolumn{3}{c|}{$n=4$}
 & \multicolumn{3}{c||}{$n=7$}
\\
 -300 &   1615.85 &    0.47 &    2.50 &   3524.09 &    0.26 &    1.31 &   4706.00 &    3.44 &    1.55 \\
 -240 &   1644.04 &    0.43 &    1.91 &   3551.80 &    0.38 &    0.72 &   4735.39 &    4.34 &    1.03 \\
 -180 &   1671.84 &    0.44 &    0.08 &   3580.30 &    0.76 &   -0.83 &   4763.52 &    3.70 &   -0.27 \\
 -120 &   1696.78 &    0.37 &   -0.58 &   3604.94 &    0.55 &   -1.87 &   4784.36 &    2.70 &   -1.29 \\
  -60 &   1720.06 &    0.35 &    0.61 &   3629.08 &    0.51 &   -0.91 &   4807.83 &    2.82 &   -0.56 \\
    0 &   1742.65 &    0.47 &    2.95 &   3653.42 &    1.44 &    1.38 &   4831.62 &    3.32 &    1.55 \\
   60 &   1763.32 &    0.52 &    5.28 &   3676.21 &    1.00 &    3.67 &   4855.95 &    3.97 &    3.66 \\
  120 &   1787.29 &    0.41 &    6.56 &   3696.90 &    0.83 &    4.67 &   4886.94 &    6.23 &    4.34 \\
  180 &   1813.92 &    0.32 &    6.03 &   3720.38 &    0.55 &    3.64 &   4904.75 &    3.31 &    3.29 \\
  240 &   1839.47 &    0.40 &    4.18 &   3752.99 &    0.39 &    2.13 &   4930.28 &    3.72 &    2.13 \\
  300 &   1867.36 &    0.50 &    3.36 &   3776.91 &    0.44 &    1.54 &   4957.49 &    4.56 &    1.68 \\
\hline
 $\ell=400$
 & \multicolumn{3}{c|}{$n=0$}
 & \multicolumn{3}{c|}{$n=3$}
 & \multicolumn{3}{c||}{$n=6$}
\\
 -400 &   1839.45 &    0.34 &    2.11 &   3439.71 &    1.13 &    1.31 &   4816.28 &    8.92 &    1.96 \\
 -320 &   1878.15 &    0.25 &    1.27 &   3481.20 &    0.78 &    0.55 &   4872.71 &    5.22 &    1.27 \\
 -240 &   1914.15 &    0.61 &   -1.56 &   3516.83 &    0.59 &   -1.56 &   4891.44 &    9.35 &   -0.43 \\
 -160 &   1948.60 &    0.23 &   -2.11 &   3550.95 &    0.42 &   -2.80 &   4937.41 &    5.64 &   -1.75 \\
  -80 &   1978.15 &    0.33 &   -0.49 &   3581.83 &    0.48 &   -1.56 &   4966.79 &    4.67 &   -0.80 \\
    0 &   2007.35 &    0.28 &    2.55 &   3617.48 &    0.66 &    1.43 &   5010.96 &    7.36 &    1.97 \\
   80 &   2037.30 &    0.34 &    5.59 &   3644.29 &    0.50 &    4.39 &   5031.16 &    5.16 &    4.71 \\
  160 &   2069.90 &    0.26 &    7.28 &   3674.82 &    0.49 &    5.62 &   5058.36 &    4.26 &    5.53 \\
  240 &   2103.62 &    0.35 &    6.78 &   3706.99 &    0.53 &    4.34 &   5095.35 &    6.32 &    4.14 \\
  320 &   2138.66 &    0.41 &    4.10 &   3741.46 &    1.07 &    2.40 &   5122.64 &    6.07 &    2.70 \\
  400 &   2176.16 &    0.27 &    2.97 &   3779.39 &    0.56 &    1.63 &   5158.84 &    5.39 &    2.12 \\
\hline
 $\ell=500$
 & \multicolumn{3}{c|}{$n=0$}
 & \multicolumn{3}{c|}{$n=3$}
 & \multicolumn{3}{c||}{$n=5$}
\\
 -500 &   2032.10 &    0.32 &    1.86 &   3721.59 &    1.21 &    1.40 &   4779.20 &    6.69 &    2.12 \\
 -400 &   2079.39 &    0.41 &    0.78 &   3776.78 &    1.62 &    0.51 &   4850.66 &    8.85 &    1.29 \\
 -300 &   2126.49 &    0.85 &   -2.80 &   3819.99 &    1.16 &   -1.89 &   4892.13 &    5.98 &   -0.83 \\
 -200 &   2168.71 &    0.37 &   -3.39 &   3864.13 &    1.22 &   -3.45 &   4944.70 &    7.41 &   -2.49 \\
 -100 &   2206.01 &    0.27 &   -1.35 &   3900.53 &    1.19 &   -2.04 &   4971.54 &    6.02 &   -1.30 \\
    0 &   2241.99 &    0.22 &    2.29 &   3936.12 &    1.30 &    1.53 &   5010.78 &    7.64 &    2.17 \\
  100 &   2280.01 &    0.27 &    5.96 &   3974.21 &    1.22 &    5.06 &   5044.81 &    5.69 &    5.63 \\
  200 &   2317.49 &    0.49 &    8.05 &   4011.97 &    0.86 &    6.30 &   5088.53 &    6.62 &    6.65 \\
  300 &   2358.19 &    0.72 &    7.48 &   4052.67 &    1.14 &    4.62 &   5121.74 &    7.40 &    4.92 \\
  400 &   2404.15 &    1.22 &    4.11 &   4096.97 &    1.84 &    2.57 &   5168.51 &    5.48 &    3.10 \\
  500 &   2451.90 &    0.26 &    2.71 &   4145.79 &    0.96 &    1.74 &   5212.01 &    7.12 &    2.33 \\
\hline
 $\ell=600$
 & \multicolumn{3}{c|}{$n=0$}
 & \multicolumn{3}{c|}{$n=2$}
 & \multicolumn{3}{c||}{$n=4$}
\\
 -600 &   2200.85 &    0.33 &    1.69 &   3375.48 &    0.59 &    1.45 &   4569.96 &    5.64 &    1.98 \\
 -480 &   2257.50 &    0.48 &    0.50 &   3434.14 &    0.48 &    0.30 &   4639.09 &    5.27 &    0.96 \\
 -360 &   2314.26 &    1.17 &   -3.44 &   3489.62 &    1.31 &   -2.90 &   4695.63 &    6.61 &   -1.64 \\
 -240 &   2363.97 &    1.75 &   -4.38 &   3541.13 &    0.76 &   -4.57 &   4746.18 &    5.60 &   -3.63 \\
 -120 &   2410.97 &    0.56 &   -2.05 &   3588.47 &    0.94 &   -2.68 &   4780.55 &    4.67 &   -2.16 \\
    0 &   2454.05 &    0.35 &    2.12 &   3627.75 &    0.79 &    1.64 &   4825.35 &    4.58 &    2.07 \\
  120 &   2498.71 &    0.56 &    6.32 &   3670.88 &    1.29 &    5.92 &   4878.69 &    6.53 &    6.28 \\
  240 &   2541.67 &    1.75 &    8.65 &   3723.76 &    1.30 &    7.58 &   4933.80 &    6.08 &    7.50 \\
  360 &       ~~~ &     ~~~ &     ~~~ &   3772.76 &    1.00 &    5.73 &   4963.75 &    4.89 &    5.41 \\
  480 &   2649.34 &    1.08 &    4.04 &   3825.62 &    1.10 &    3.01 &   5028.84 &    4.75 &    3.22 \\
  600 &   2704.88 &    0.42 &    2.54 &   3881.58 &    0.65 &    1.91 &   5084.24 &    4.74 &    2.25 \\
\hline
 $\ell=700$
 & \multicolumn{3}{c|}{$n=0$}
 & \multicolumn{3}{c|}{$n=2$}
 & \multicolumn{3}{c||}{$n=4$}
\\
 -700 &   2355.80 &    0.67 &    1.59 &   3585.34 &    1.18 &    1.52 &   4859.54 &   17.31 &    2.51 \\
 -560 &   2422.35 &    0.76 &    0.32 &   3655.52 &    1.27 &    0.18 &   4907.81 &   10.03 &    1.27 \\
 -420 &   2484.73 &    1.92 &   -3.75 &   3721.56 &    2.49 &   -3.38 &   4978.88 &   10.58 &   -1.82 \\
 -280 &   2545.93 &    0.68 &   -5.24 &   3780.02 &    2.10 &   -5.43 &   5035.02 &   11.13 &   -4.16 \\
 -140 &   2598.77 &    0.35 &   -2.71 &   3834.58 &    1.20 &   -3.33 &   5099.94 &    9.38 &   -2.44 \\
    0 &   2650.02 &    0.36 &    2.00 &   3886.56 &    1.52 &    1.76 &   5143.83 &   11.42 &    2.66 \\
  140 &   2700.27 &    0.88 &    6.73 &   3932.49 &    1.61 &    6.75 &   5193.52 &    8.96 &    7.67 \\
  280 &   2754.29 &    0.97 &    9.10 &   3990.67 &    1.69 &    8.46 &   5257.80 &    9.74 &    9.03 \\
  420 &   2810.04 &    2.09 &    7.60 &   4042.58 &    1.67 &    6.16 &   5299.17 &   11.27 &    6.55 \\
  560 &   2877.00 &    0.97 &    3.93 &   4114.32 &    1.71 &    3.24 &   5383.13 &   10.16 &    3.98 \\
  700 &   2941.49 &    0.40 &    2.42 &   4175.79 &    1.34 &    1.99 &   5433.74 &    7.97 &    2.79 \\
\hline
 $\ell=800$
 & \multicolumn{3}{c|}{$n=0$}
 & \multicolumn{3}{c|}{$n=2$}
 & \multicolumn{3}{c||}{$n=3$}
\\
 -800 &   2490.31 &    1.00 &    1.55 &   3786.28 &    2.67 &    1.64 &   4455.38 &    5.41 &    2.07 \\
 -640 &   2570.26 &    1.06 &    0.12 &   3871.52 &    1.92 &    0.03 &   4534.01 &    4.57 &    0.51 \\
 -480 &   2642.63 &    1.79 &   -4.16 &   3947.30 &    2.58 &   -3.95 &   4610.64 &    5.26 &   -3.19 \\
 -320 &   2708.33 &    1.17 &   -6.07 &   4011.97 &    1.79 &   -6.36 &   4667.91 &    5.59 &   -5.80 \\
 -160 &   2769.17 &    0.87 &   -3.48 &   4072.83 &    2.29 &   -4.05 &   4739.06 &    6.38 &   -3.71 \\
    0 &   2830.75 &    0.86 &    1.98 &   4132.14 &    1.81 &    1.82 &   4792.00 &    5.11 &    2.17 \\
  160 &   2890.48 &    0.84 &    7.37 &   4185.19 &    2.18 &    7.71 &   4849.83 &    5.39 &    8.13 \\
  320 &   2948.68 &    1.21 &    9.67 &   4256.81 &    2.41 &    9.51 &   4918.18 &    5.76 &    9.73 \\
  480 &   3020.75 &    1.53 &    7.62 &   4320.01 &    2.91 &    6.77 &   4986.92 &    5.49 &    6.87 \\
  640 &   3088.17 &    1.00 &    3.97 &       ~~~ &     ~~~ &     ~~~ &   5065.60 &    6.78 &    3.85 \\
  800 &   3165.50 &    1.01 &    2.37 &   4464.78 &    1.62 &    2.11 &   5134.08 &    5.31 &    2.42 \\
\hline
 $\ell=900$
 & \multicolumn{3}{c|}{$n=0$}
 & \multicolumn{3}{c|}{$n=2$}
 & \multicolumn{3}{c||}{$n=3$}
\\
 -900 &   2620.96 &    1.43 &    1.55 &   3997.98 &    3.27 &    1.81 &   4688.00 &   11.95 &    2.49 \\
 -720 &   2705.26 &    1.35 &   -0.15 &   4077.75 &    3.79 &   -0.12 &   4765.47 &    8.70 &    0.60 \\
 -540 &   2790.34 &    2.28 &   -4.84 &   4159.38 &    4.18 &   -4.60 &   4845.57 &    8.97 &   -3.67 \\
 -360 &   2862.60 &    1.30 &   -7.20 &   4234.49 &    3.13 &   -7.41 &   4913.34 &   11.78 &   -6.63 \\
 -180 &   2935.24 &    1.60 &   -4.34 &   4309.82 &    3.42 &   -4.83 &   5004.43 &   10.60 &   -4.24 \\
    0 &   3003.06 &    1.37 &    1.98 &   4374.78 &    3.11 &    2.01 &   5062.39 &    9.84 &    2.62 \\
  180 &   3061.50 &    1.52 &    8.21 &   4442.40 &    3.57 &    8.85 &   5125.96 &   10.22 &    9.54 \\
  360 &   3128.93 &    1.86 &   10.55 &   4506.11 &    3.86 &   10.77 &   5187.44 &    8.42 &   11.31 \\
  540 &   3203.81 &    2.29 &    7.95 &   4582.98 &    3.89 &    7.60 &   5274.33 &    9.25 &    8.04 \\
  720 &   3291.54 &    1.53 &    4.15 &   4661.43 &    3.29 &    3.99 &   5351.52 &    9.20 &    4.55 \\
  900 &   3372.62 &    1.52 &    2.38 &   4749.11 &    3.30 &    2.27 &   5439.97 &    8.28 &    2.84 \\
\hline
 $\ell=1000$
 & \multicolumn{3}{c|}{$n=0$}
 & \multicolumn{3}{c|}{$n=1$}
 & \multicolumn{3}{c||}{$n=2$}
\\
-1000 &       ~~~ &     ~~~ &     ~~~ &   3454.34 &    2.11 &    1.75 &   4183.54 &    4.48 &    2.04 \\
 -800 &   2837.09 &    2.54 &   -0.33 &   3555.08 &    4.40 &   -0.61 &   4280.47 &    4.26 &   -0.22 \\
 -600 &   2921.95 &    3.30 &   -5.48 &   3649.84 &    3.14 &   -6.06 &   4386.49 &    4.47 &   -5.31 \\
 -400 &   3006.85 &    2.90 &   -8.26 &   3730.87 &    2.60 &   -8.87 &   4454.42 &    4.30 &   -8.52 \\
 -200 &   3083.96 &    2.10 &   -5.36 &   3804.57 &    2.11 &   -5.77 &   4538.02 &    4.35 &   -5.69 \\
    0 &   3156.62 &    3.21 &    2.07 &   3879.78 &    2.21 &    2.03 &   4613.72 &    4.22 &    2.24 \\
  200 &   3225.57 &    2.16 &    9.34 &   3951.35 &    3.21 &    9.66 &   4683.41 &    5.59 &    9.99 \\
  400 &   3308.10 &    2.25 &   11.68 &   4030.11 &    2.58 &   12.08 &   4758.74 &    5.90 &   12.31 \\
  600 &   3387.98 &    3.00 &    8.61 &   4111.34 &    2.87 &    8.67 &   4841.74 &    4.19 &    8.74 \\
  800 &   3483.79 &    2.44 &    4.59 &   4198.87 &    2.31 &    4.45 &       ~~~ &     ~~~ &     ~~~ \\
 1000 &   3574.75 &    2.12 &    2.50 &   4296.71 &    1.80 &    2.36 &   5029.08 &    4.80 &    2.49 \\

\end{longtable}
\end{center}

%% file: tableMAll.tex
\begin{landscape}
\begin{center}
\begin{longtable}{||cc|ccc||ccc||cc|cc|cc||}
\caption[Selected subset of multiplets]{Selected subset of multiplets.
\label{tab:multiplets}} \\
 $n$ & $\ell$ & 
 $\nu_{n,\ell}$ & $\sigma_{\nu_{n,\ell}}$  & $\Delta^{\nu}_{n,\ell}$ &
 $\tilde{\Gamma}_{n,\ell}$ & $\tilde{\alpha}_{n,\ell}$ & $\tilde{A}_{n,\ell}$ &
 $\Gamma_{n,\ell}$ & $\sigma_{\Gamma_{n,\ell}}$ & $\alpha_{n,\ell}$ & $\sigma_{\alpha_{n,\ell}}$ & $A_{n,\ell}$ & $\sigma_{A_{n,\ell}}$       \\\hline
\endfirsthead
\multicolumn{14}{c}%
{{\bfseries \tablename\ \thetable{} -- continued from previous page}} \\ \hline 
 $n$ & $\ell$ & 
 $\nu_{n,\ell}$ & $\sigma_{\nu_{n,\ell}}$  & $\Delta^{\nu}_{n,\ell}$ &
 $\tilde{\Gamma}_{n,\ell}$ & $\tilde{\alpha}_{n,\ell}$ & $\tilde{A}_{n,\ell}$ &
 $\Gamma_{n,\ell}$ & $\sigma_{\Gamma_{n,\ell}}$ & $\alpha_{n,\ell}$ & $\sigma_{\alpha_{n,\ell}}$ & $A_{n,\ell}$ & $\sigma_{A_{n,\ell}}$       \\\hline
\endhead
\hline \multicolumn{14}{|r|}{{Continued on next page}} \\ \hline
\endfoot
\hline \hline
\endlastfoot
%
%
%
  1 &  100 &   1490.27 &    2.03 &    4.21 &   15.52 &  -0.086 &  5.67e-04 & $\le 1.96$ &    2.23 &  -0.095 &   0.068 &  7.66e-03 &  7.26e-04 \\
  2 &  100 &   1845.44 &    0.33 &    3.37 &   14.74 &  -0.132 &  4.78e-03 & $\le 1.96$ &    0.80 &  -0.134 &   0.023 &  4.89e-02 &  1.59e-03 \\
  3 &  100 &   2146.16 &    0.39 &    3.40 &   17.93 &  -0.100 &  1.85e-02 & $\le 1.96$ &    0.97 &  -0.107 &   0.022 &  1.47e-01 &  4.14e-03 \\
  4 &  100 &   2423.24 &    0.37 &    2.25 &   19.84 &  -0.026 &  5.48e-02 & $\le 1.96$ &    0.98 &  -0.039 &   0.018 &  4.11e-01 &  9.46e-03 \\
  5 &  100 &   2678.75 &    0.32 &    2.16 &   21.52 &   0.004 &  1.26e-01 & $\le 1.96$ &    1.35 &  -0.019 &   0.015 &  1.11e+00 &  2.75e-02 \\
  6 &  100 &   2928.89 &    0.43 &    1.80 &   23.76 &  -0.005 &  2.28e-01 & $\le 1.96$ &    1.58 &  -0.031 &   0.020 &  2.13e+00 &  4.38e-02 \\
  7 &  100 &   3167.81 &    0.41 &    1.27 &   25.77 &  -0.031 &  2.50e-01 & $\le 1.96$ &    1.70 &  -0.048 &   0.017 &  2.27e+00 &  4.53e-02 \\
  8 &  100 &   3398.58 &    0.29 &    1.04 &   28.70 &  -0.041 &  1.40e-01 & $\le 1.96$ &    1.69 &  -0.050 &   0.011 &  1.18e+00 &  2.44e-02 \\
  9 &  100 &   3623.47 &    0.29 &    0.80 &   31.36 &  -0.021 &  6.20e-02 & $\le 1.96$ &    1.93 &  -0.025 &   0.010 &  4.58e-01 &  8.19e-03 \\
 10 &  100 &   3845.15 &    0.32 &    0.67 &   36.51 &  -0.038 &  2.69e-02 &   12.68 &    1.86 &  -0.037 &   0.010 &  1.63e-01 &  2.80e-03 \\
 11 &  100 &   4061.31 &    0.38 &    0.53 &   41.39 &  -0.025 &  1.35e-02 &   17.45 &    2.32 &  -0.025 &   0.011 &  6.66e-02 &  1.00e-03 \\
 12 &  100 &   4272.21 &    0.47 &    0.48 &   50.24 &   0.013 &  7.06e-03 &   29.43 &    2.55 &   0.012 &   0.014 &  2.96e-02 &  4.81e-04 \\
 13 &  100 &   4479.18 &    0.55 &    0.49 &   58.60 &   0.096 &  3.85e-03 &   38.57 &    3.51 &   0.092 &   0.019 &  1.39e-02 &  2.41e-04 \\
 14 &  100 &   4667.48 &    1.04 &    0.41 &   68.51 &   0.134 &  2.31e-03 &   48.72 &    5.21 &   0.130 &   0.034 &  7.35e-03 &  2.10e-04 \\
 & & & & & & & & & & & & & \\
  0 &  200 &   1429.52 &    0.61 &    3.64 &    9.67 &  -0.077 &  5.12e-04 & $\le 1.96$ &    0.42 &  -0.085 &   0.044 &  3.82e-03 &  1.61e-04 \\
  1 &  200 &   1967.68 &    0.09 &    2.43 &   10.01 &  -0.119 &  1.32e-02 & $\le 1.96$ &    0.15 &  -0.127 &   0.008 &  7.28e-02 &  1.41e-03 \\
  2 &  200 &   2392.48 &    0.11 &    2.20 &   12.94 &  -0.080 &  7.03e-02 &    4.94 &    0.21 &  -0.085 &   0.007 &  3.77e-01 &  6.41e-03 \\
  3 &  200 &   2765.51 &    0.13 &    1.91 &   13.77 &  -0.040 &  2.40e-01 &    4.05 &    0.25 &  -0.042 &   0.008 &  1.36e+00 &  2.11e-02 \\
  4 &  200 &   3131.93 &    0.13 &    1.67 &   15.47 &  -0.039 &  3.85e-01 &    4.54 &    0.32 &  -0.040 &   0.008 &  2.18e+00 &  3.46e-02 \\
  5 &  200 &   3474.16 &    0.14 &    1.39 &   19.42 &  -0.054 &  1.74e-01 &    8.43 &    0.40 &  -0.057 &   0.007 &  8.01e-01 &  1.24e-02 \\
  6 &  200 &   3804.14 &    0.18 &    1.18 &   26.00 &  -0.077 &  5.01e-02 &   15.23 &    0.54 &  -0.081 &   0.006 &  1.73e-01 &  2.56e-03 \\
  7 &  200 &   4119.44 &    0.21 &    1.12 &   39.31 &  -0.080 &  1.39e-02 &   30.45 &    0.63 &  -0.083 &   0.005 &  3.86e-02 &  4.47e-04 \\
  8 &  200 &   4423.91 &    0.23 &    1.16 &   52.94 &  -0.077 &  5.12e-03 &   44.71 &    1.11 &  -0.079 &   0.005 &  1.30e-02 &  1.29e-04 \\
  9 &  200 &   4717.78 &    0.34 &    1.20 &   73.88 &  -0.072 &  2.11e-03 &   66.73 &    2.01 &  -0.073 &   0.008 &  5.13e-03 &  5.38e-05 \\
 & & & & & & & & & & & & & \\
  0 &  300 &   1742.04 &    0.06 &    2.95 &    8.42 &  -0.099 &  3.27e-03 & $\le 1.96$ &    0.10 &  -0.102 &   0.008 &  1.38e-02 &  2.46e-04 \\
  1 &  300 &   2286.96 &    0.07 &    2.02 &   11.14 &  -0.096 &  3.84e-02 &    5.80 &    0.13 &  -0.102 &   0.005 &  1.16e-01 &  1.74e-03 \\
  2 &  300 &   2779.88 &    0.08 &    1.71 &   13.43 &  -0.067 &  1.77e-01 &    7.91 &    0.19 &  -0.072 &   0.005 &  5.44e-01 &  7.96e-03 \\
  3 &  300 &   3234.66 &    0.10 &    1.58 &   14.97 &  -0.046 &  2.70e-01 &    7.73 &    0.24 &  -0.050 &   0.005 &  9.16e-01 &  1.23e-02 \\
  4 &  300 &   3652.31 &    0.11 &    1.38 &   21.80 &  -0.075 &  8.39e-02 &   15.28 &    0.32 &  -0.078 &   0.005 &  2.35e-01 &  2.92e-03 \\
  5 &  300 &   4062.44 &    0.15 &    1.36 &   39.85 &  -0.083 &  1.59e-02 &   35.14 &    0.42 &  -0.085 &   0.004 &  3.85e-02 &  3.76e-04 \\
  6 &  300 &   4456.84 &    0.21 &    1.46 &   64.60 &  -0.088 &  4.08e-03 &   60.84 &    0.81 &  -0.089 &   0.004 &  9.50e-03 &  7.66e-05 \\
  7 &  300 &   4834.19 &    0.33 &    1.55 &  101.35 &  -0.077 &  1.32e-03 &   98.18 &    1.75 &  -0.078 &   0.006 &  3.03e-03 &  2.20e-05 \\
 & & & & & & & & & & & & & \\
  0 &  400 &   2008.28 &    0.05 &    2.55 &    9.18 &  -0.119 &  8.54e-03 &    3.54 &    0.09 &  -0.120 &   0.005 &  2.31e-02 &  3.25e-04 \\
  1 &  400 &   2556.36 &    0.08 &    1.91 &   14.12 &  -0.105 &  5.37e-02 &   10.15 &    0.16 &  -0.108 &   0.004 &  1.27e-01 &  1.42e-03 \\
  2 &  400 &   3088.70 &    0.08 &    1.59 &   16.25 &  -0.054 &  1.60e-01 &   12.11 &    0.19 &  -0.057 &   0.004 &  4.02e-01 &  4.55e-03 \\
  3 &  400 &   3613.61 &    0.11 &    1.43 &   25.32 &  -0.066 &  6.71e-02 &   21.80 &    0.26 &  -0.068 &   0.004 &  1.64e-01 &  1.80e-03 \\
  4 &  400 &   4105.68 &    0.14 &    1.52 &   47.15 &  -0.086 &  1.07e-02 &   44.14 &    0.40 &  -0.087 &   0.003 &  2.49e-02 &  2.06e-04 \\
  5 &  400 &   4561.20 &    0.23 &    1.67 &   84.02 &  -0.069 &  2.29e-03 &   81.82 &    0.78 &  -0.070 &   0.004 &  5.27e-03 &  3.53e-05 \\
  6 &  400 &   5000.85 &    0.41 &    1.97 &  147.63 &  -0.026 &  6.65e-04 &  145.72 &    1.86 &  -0.027 &   0.007 &  1.52e-03 &  8.75e-06 \\
 & & & & & & & & & & & & & \\
  0 &  500 &   2242.80 &    0.07 &    2.29 &   11.41 &  -0.134 &  1.23e-02 &    7.03 &    0.13 &  -0.133 &   0.004 &  2.63e-02 &  2.75e-04 \\
  1 &  500 &   2801.33 &    0.09 &    1.87 &   19.34 &  -0.083 &  4.81e-02 &   16.38 &    0.19 &  -0.085 &   0.004 &  1.06e-01 &  1.01e-03 \\
  2 &  500 &   3367.17 &    0.09 &    1.59 &   23.55 &  -0.058 &  6.97e-02 &   20.68 &    0.21 &  -0.059 &   0.003 &  1.64e-01 &  1.55e-03 \\
  3 &  500 &   3936.25 &    0.14 &    1.53 &   44.75 &  -0.071 &  1.39e-02 &   42.71 &    0.32 &  -0.072 &   0.003 &  3.25e-02 &  2.55e-04 \\
  4 &  500 &   4493.43 &    0.19 &    1.72 &   88.71 &  -0.075 &  2.13e-03 &   87.27 &    0.60 &  -0.076 &   0.003 &  4.97e-03 &  2.78e-05 \\
  5 &  500 &   5009.82 &    0.40 &    2.17 &  162.85 &  -0.035 &  4.86e-04 &  161.52 &    1.57 &  -0.035 &   0.006 &  1.14e-03 &  5.18e-06 \\
 & & & & & & & & & & & & & \\
  0 &  600 &   2454.82 &    0.08 &    2.12 &   16.35 &  -0.140 &  1.15e-02 &   13.26 &    0.18 &  -0.139 &   0.004 &  2.29e-02 &  2.13e-04 \\
  1 &  600 &   3031.40 &    0.09 &    1.85 &   25.17 &  -0.077 &  3.36e-02 &   22.94 &    0.21 &  -0.078 &   0.003 &  7.52e-02 &  6.17e-04 \\
  2 &  600 &   3630.70 &    0.11 &    1.64 &   34.98 &  -0.069 &  2.28e-02 &   32.93 &    0.25 &  -0.070 &   0.003 &  5.43e-02 &  4.18e-04 \\
  3 &  600 &   4235.20 &    0.16 &    1.69 &   72.06 &  -0.077 &  3.45e-03 &   70.76 &    0.39 &  -0.077 &   0.002 &  8.35e-03 &  4.99e-05 \\
  4 &  600 &   4832.45 &    0.31 &    2.07 &  149.45 &  -0.033 &  5.67e-04 &  148.53 &    0.90 &  -0.033 &   0.005 &  1.39e-03 &  5.88e-06 \\
 & & & & & & & & & & & & & \\
  0 &  700 &   2649.22 &    0.09 &    2.00 &   24.14 &  -0.136 &  8.61e-03 &   22.02 &    0.20 &  -0.136 &   0.003 &  1.79e-02 &  1.43e-04 \\
  1 &  700 &   3251.42 &    0.10 &    1.87 &   34.92 &  -0.080 &  1.69e-02 &   33.19 &    0.23 &  -0.080 &   0.003 &  4.04e-02 &  2.76e-04 \\
  2 &  700 &   3884.80 &    0.12 &    1.76 &   51.83 &  -0.076 &  6.54e-03 &   50.47 &    0.30 &  -0.076 &   0.002 &  1.66e-02 &  1.08e-04 \\
  3 &  700 &   4520.80 &    0.19 &    1.95 &  107.58 &  -0.062 &  9.60e-04 &  106.75 &    0.51 &  -0.062 &   0.002 &  2.50e-03 &  1.20e-05 \\
  4 &  700 &   5145.12 &    0.48 &    2.66 &  229.68 &   0.041 &  1.76e-04 &  229.02 &    1.67 &   0.041 &   0.005 &  4.64e-04 &  1.72e-06 \\
 & & & & & & & & & & & & & \\
  0 &  800 &   2829.18 &    0.10 &    1.98 &   36.11 &  -0.128 &  5.28e-03 &   34.69 &    0.24 &  -0.128 &   0.003 &  1.22e-02 &  8.65e-05 \\
  1 &  800 &   3464.52 &    0.11 &    1.87 &   46.83 &  -0.080 &  7.05e-03 &   45.49 &    0.26 &  -0.080 &   0.002 &  1.87e-02 &  1.19e-04 \\
  2 &  800 &   4131.67 &    0.15 &    1.82 &   74.24 &  -0.068 &  1.92e-03 &   73.14 &    0.33 &  -0.068 &   0.002 &  5.38e-03 &  2.91e-05 \\
  3 &  800 &   4795.55 &    0.26 &    2.17 &  152.07 &  -0.019 &  2.97e-04 &  151.31 &    0.69 &  -0.019 &   0.003 &  8.51e-04 &  3.28e-06 \\
 & & & & & & & & & & & & & \\
  0 &  900 &   2997.09 &    0.12 &    1.98 &   52.14 &  -0.129 &  2.84e-03 &   51.07 &    0.28 &  -0.129 &   0.002 &  7.60e-03 &  4.52e-05 \\
  1 &  900 &   3672.76 &    0.13 &    1.95 &   61.91 &  -0.087 &  2.62e-03 &   60.89 &    0.30 &  -0.087 &   0.002 &  7.95e-03 &  4.78e-05 \\
  2 &  900 &   4373.29 &    0.17 &    2.01 &  102.63 &  -0.060 &  5.84e-04 &  101.71 &    0.41 &  -0.061 &   0.002 &  1.86e-03 &  9.06e-06 \\
  3 &  900 &   5062.73 &    0.39 &    2.62 &  221.50 &   0.022 &  9.63e-05 &  220.86 &    1.15 &   0.022 &   0.004 &  3.14e-04 &  1.10e-06 \\
 & & & & & & & & & & & & & \\
  0 & 1000 &   3154.77 &    0.13 &    2.07 &   75.30 &  -0.138 &  1.35e-03 &   74.84 &    0.36 &  -0.138 &   0.002 &  4.31e-03 &  2.47e-05 \\
  1 & 1000 &   3877.95 &    0.13 &    2.03 &   81.52 &  -0.091 &  9.07e-04 &   80.89 &    0.34 &  -0.091 &   0.002 &  3.26e-03 &  1.85e-05 \\
  2 & 1000 &   4610.80 &    0.20 &    2.24 &  135.00 &  -0.040 &  1.83e-04 &  134.58 &    0.53 &  -0.040 &   0.002 &  6.88e-04 &  3.04e-06 \\

\end{longtable}
\end{center}
\end{landscape}

%% file: tableCGAll.tex
\begin{center}
\begin{longtable}{||ccc|ccc|ccc||}
\caption[Selected subset of Clebsch-Gordan coefficients]{Selected subset of Clebsch-Gordan coefficients.
\label{tab:cgCoefs}} \\
$n$ & $\ell$ &   $\nu_{n,\ell}$ & 
$cg_{n,\ell}(1)$ & $cg_{n,\ell}(3)$ & $cg_{n,\ell}(5)$ & 
$\sigma_{cg_{n,\ell}(1)}$ & $\sigma_{cg_{n,\ell}(3)}$ &
$\sigma_{cg_{n,\ell}(5)}$ \\ \hline
\endfirsthead
\multicolumn{9}{c}%
{{\bfseries \tablename\ \thetable{} -- continued from previous page}} \\ \hline 
$n$ & $\ell$ &   $\nu_{n,\ell}$ & 
$cg_{n,\ell}(1)$ & $cg_{n,\ell}(3)$ & $cg_{n,\ell}(5)$ & 
$\sigma_{cg_{n,\ell}(1)}$ & $\sigma_{cg_{n,\ell}(3)}$ &
$\sigma_{cg_{n,\ell}(5)}$ \\ \hline
\endhead
\hline \multicolumn{9}{|r|}{{Continued on next page}} \\ \hline
\endfoot
\hline \hline
\endlastfoot
%
%
%
%
  1 &  100 &    1490.3 &  359.10 &    9.53 &   -5.98 &   11.17 &    9.33 &    7.30 \\
  2 &  100 &    1845.4 &  361.01 &   -4.29 &    1.75 &    2.33 &    1.61 &    1.19 \\
  3 &  100 &    2146.2 &  357.59 &   -7.02 &   -0.07 &    2.48 &    1.74 &    1.32 \\
  4 &  100 &    2423.2 &  361.39 &   -4.51 &   -1.67 &    2.46 &    1.65 &    1.29 \\
  5 &  100 &    2678.7 &  357.92 &  -10.23 &   -3.23 &    2.06 &    1.40 &    1.09 \\
  6 &  100 &    2928.9 &  360.87 &   -5.56 &   -0.47 &    2.75 &    1.83 &    1.51 \\
  7 &  100 &    3167.8 &  357.69 &  -11.04 &   -0.33 &    2.70 &    1.71 &    1.41 \\
  8 &  100 &    3398.6 &  357.18 &   -9.05 &   -1.71 &    1.85 &    1.25 &    0.98 \\
  9 &  100 &    3623.5 &  360.31 &  -10.61 &   -2.20 &    1.85 &    1.25 &    1.02 \\
 10 &  100 &    3845.1 &  355.93 &   -9.65 &    0.01 &    2.00 &    1.36 &    1.10 \\
 11 &  100 &    4061.3 &  359.91 &   -7.78 &   -0.17 &    2.46 &    1.66 &    1.31 \\
 12 &  100 &    4272.2 &  359.63 &   -9.27 &   -1.66 &    2.97 &    2.07 &    1.63 \\
 13 &  100 &    4479.2 &  358.03 &   -6.85 &   -3.21 &    3.62 &    2.41 &    1.86 \\
 14 &  100 &    4667.5 &  370.72 &  -15.02 &  -12.33 &   10.51 &    5.29 &    1.53 \\
 & & & & & & & & \\
  0 &  200 &    1429.5 &  348.78 &   -1.59 &   -0.11 &    1.94 &    1.31 &    1.17 \\
  1 &  200 &    1967.7 &  356.67 &   -8.22 &   -0.86 &    0.29 &    0.18 &    0.15 \\
  2 &  200 &    2392.5 &  359.23 &   -7.53 &   -0.86 &    0.35 &    0.22 &    0.18 \\
  3 &  200 &    2765.5 &  358.37 &   -7.23 &   -0.37 &    0.40 &    0.27 &    0.22 \\
  4 &  200 &    3131.9 &  360.23 &   -7.62 &   -0.66 &    0.42 &    0.28 &    0.23 \\
  5 &  200 &    3474.2 &  360.72 &   -7.16 &   -0.38 &    0.45 &    0.30 &    0.24 \\
  6 &  200 &    3804.1 &  359.57 &   -7.94 &   -0.19 &    0.56 &    0.39 &    0.30 \\
  7 &  200 &    4119.4 &  361.26 &   -8.11 &   -0.70 &    0.66 &    0.43 &    0.34 \\
  8 &  200 &    4423.9 &  360.44 &   -7.15 &   -0.08 &    0.75 &    0.49 &    0.38 \\
  9 &  200 &    4717.8 &  362.64 &   -6.11 &   -1.17 &    1.08 &    0.73 &    0.57 \\
 & & & & & & & & \\
  0 &  300 &    1742.0 &  356.22 &   -7.48 &   -0.70 &    0.14 &    0.09 &    0.07 \\
  1 &  300 &    2287.0 &  356.46 &   -7.73 &   -0.52 &    0.16 &    0.10 &    0.08 \\
  2 &  300 &    2779.9 &  356.75 &   -7.88 &   -0.73 &    0.19 &    0.12 &    0.10 \\
  3 &  300 &    3234.7 &  357.20 &   -8.02 &   -0.63 &    0.21 &    0.14 &    0.11 \\
  4 &  300 &    3652.3 &  357.63 &   -7.68 &   -0.78 &    0.25 &    0.16 &    0.13 \\
  5 &  300 &    4062.4 &  358.19 &   -7.93 &   -0.83 &    0.32 &    0.21 &    0.17 \\
  6 &  300 &    4456.8 &  359.71 &   -7.34 &   -0.38 &    0.46 &    0.31 &    0.25 \\
  7 &  300 &    4834.2 &  361.43 &   -5.55 &   -0.58 &    0.71 &    0.48 &    0.37 \\
 & & & & & & & & \\
  0 &  400 &    2008.3 &  356.08 &   -7.42 &   -0.60 &    0.09 &    0.06 &    0.04 \\
  1 &  400 &    2556.4 &  355.37 &   -7.78 &   -0.49 &    0.13 &    0.08 &    0.07 \\
  2 &  400 &    3088.7 &  355.87 &   -8.15 &   -0.52 &    0.14 &    0.09 &    0.07 \\
  3 &  400 &    3613.6 &  356.42 &   -8.20 &   -0.39 &    0.18 &    0.11 &    0.09 \\
  4 &  400 &    4105.7 &  357.62 &   -8.18 &   -0.77 &    0.23 &    0.16 &    0.12 \\
  5 &  400 &    4561.2 &  359.08 &   -7.17 &   -0.48 &    0.37 &    0.25 &    0.20 \\
  6 &  400 &    5000.9 &  360.31 &   -5.78 &   -0.30 &    0.65 &    0.44 &    0.35 \\
 & & & & & & & & \\
  0 &  500 &    2242.8 &  355.21 &   -7.50 &   -0.43 &    0.10 &    0.05 &    0.04 \\
  1 &  500 &    2801.3 &  355.03 &   -8.00 &   -0.47 &    0.12 &    0.07 &    0.06 \\
  2 &  500 &    3367.2 &  355.24 &   -8.33 &   -0.53 &    0.13 &    0.08 &    0.06 \\
  3 &  500 &    3936.3 &  355.95 &   -7.98 &   -0.42 &    0.19 &    0.12 &    0.10 \\
  4 &  500 &    4493.4 &  356.84 &   -8.23 &   -0.72 &    0.25 &    0.17 &    0.13 \\
  5 &  500 &    5009.8 &  357.52 &   -7.45 &   -0.61 &    0.51 &    0.34 &    0.28 \\
 & & & & & & & & \\
  0 &  600 &    2454.8 &  354.61 &   -7.77 &   -0.46 &    0.10 &    0.06 &    0.04 \\
  1 &  600 &    3031.4 &  354.45 &   -8.11 &   -0.42 &    0.11 &    0.07 &    0.05 \\
  2 &  600 &    3630.7 &  355.22 &   -8.01 &   -0.51 &    0.12 &    0.08 &    0.06 \\
  3 &  600 &    4235.2 &  356.39 &   -7.88 &   -0.59 &    0.17 &    0.11 &    0.09 \\
  4 &  600 &    4832.5 &  357.64 &   -7.47 &   -0.75 &    0.34 &    0.22 &    0.18 \\
 & & & & & & & & \\
  0 &  700 &    2649.2 &  354.23 &   -7.89 &   -0.45 &    0.10 &    0.06 &    0.04 \\
  1 &  700 &    3251.4 &  354.49 &   -8.10 &   -0.55 &    0.10 &    0.06 &    0.05 \\
  2 &  700 &    3884.8 &  355.17 &   -8.16 &   -0.49 &    0.11 &    0.07 &    0.06 \\
  3 &  700 &    4520.8 &  356.03 &   -7.93 &   -0.74 &    0.17 &    0.11 &    0.09 \\
  4 &  700 &    5145.1 &  356.71 &   -7.23 &   -0.50 &    0.44 &    0.29 &    0.24 \\
 & & & & & & & & \\
  0 &  800 &    2829.2 &  353.78 &   -8.12 &   -0.44 &    0.09 &    0.06 &    0.04 \\
  1 &  800 &    3464.5 &  354.34 &   -8.15 &   -0.48 &    0.09 &    0.06 &    0.05 \\
  2 &  800 &    4131.7 &  355.22 &   -8.07 &   -0.50 &    0.12 &    0.08 &    0.06 \\
  3 &  800 &    4795.6 &  356.77 &   -7.71 &   -0.55 &    0.21 &    0.14 &    0.11 \\
 & & & & & & & & \\
  0 &  900 &    2997.1 &  353.58 &   -8.20 &   -0.43 &    0.09 &    0.06 &    0.04 \\
  1 &  900 &    3672.8 &  354.33 &   -8.27 &   -0.56 &    0.09 &    0.06 &    0.05 \\
  2 &  900 &    4373.3 &  355.42 &   -8.29 &   -0.69 &    0.12 &    0.08 &    0.06 \\
  3 &  900 &    5062.7 &  357.41 &   -7.60 &   -0.72 &    0.28 &    0.19 &    0.15 \\
 & & & & & & & & \\
  0 & 1000 &    3154.8 &  353.46 &   -8.29 &   -0.57 &    0.08 &    0.06 &    0.04 \\
  1 & 1000 &    3878.0 &  354.16 &   -8.38 &   -0.55 &    0.09 &    0.06 &    0.05 \\
  2 & 1000 &    4610.8 &  354.82 &   -8.44 &   -0.71 &    0.13 &    0.09 &    0.07 \\

\end{longtable}
\end{center}